\documentclass[12pt,a4paper]{book}
\usepackage{a4}
\usepackage{exscale}
\usepackage{fancyhdr}
\usepackage{graphicx}
\usepackage[latin1]{inputenc}
\usepackage{ngerman}

\usepackage[
  pdftitle={Neural Synchronization and Cryptography},
  pdfauthor={Andreas Ruttor},
  pdfsubject={Neural Cryptography},
  pdfkeywords={neural networks, synchronization, cryptography,
               statistical physics, nonlinear dynamics}
]{hyperref}

\usepackage[numbers,sort&compress]{natbib}
\usepackage{hypernat}

\fancypagestyle{plain}
{%
  \fancyhf{}
  \fancyfoot[LE,RO]{\thepage}
  
}

\begin{document}
\begin{center}
  \thispagestyle{empty}
  \vspace*{1cm}
  \textbf{\Huge Neural Synchronization \\[1ex] and Cryptography}
  \vfill
  \begin{minipage}[][0.5 \textwidth][c]{\textwidth}
    \centering
    \includegraphics[width=0.9 \textwidth]{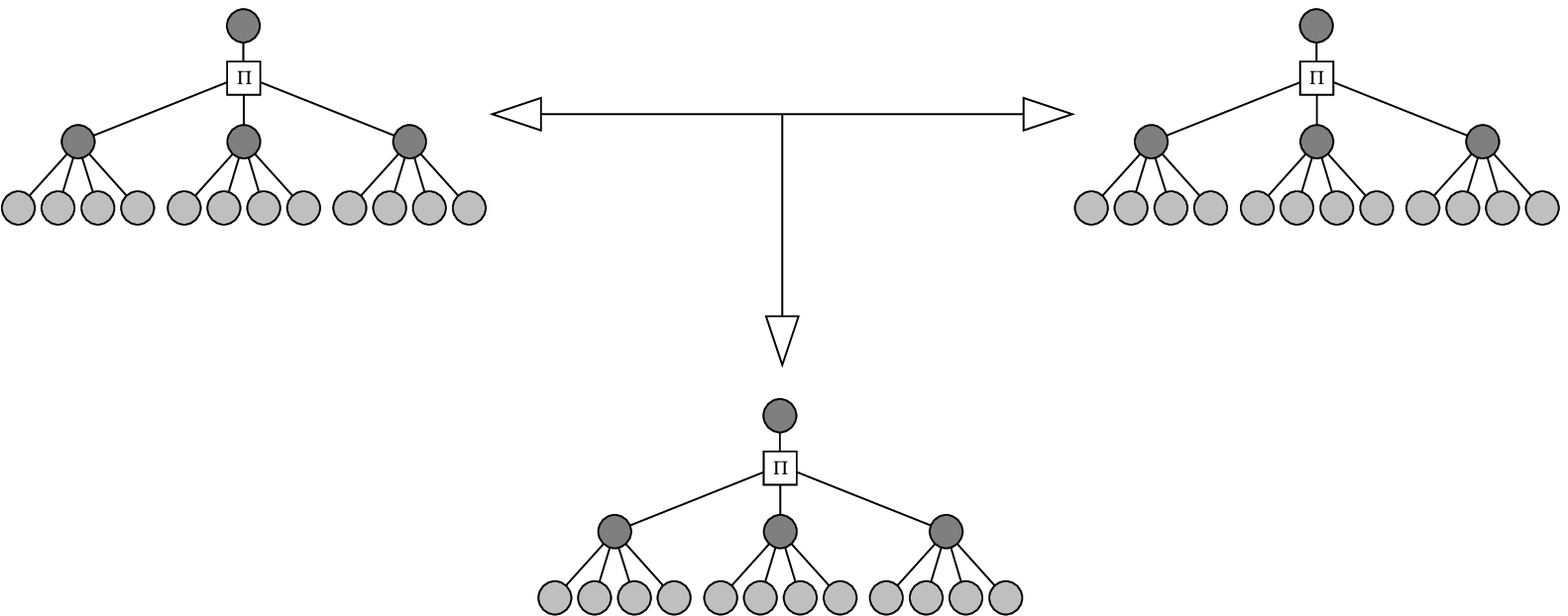}
  \end{minipage}
  \vfill
  Dissertation zur Erlangung des\\
  naturwissenschaftlichen Doktorgrades\\
  der Bayerischen Julius-Maximilians-Universit"at W"urzburg\\[4ex]
  vorgelegt von\\[2ex]
  {\Large Andreas Ruttor}\\[1ex]
  aus W"urzburg\\
  \vfill
  \textsc{W"urzburg 2006}
  \vspace*{1cm}
\end{center}

\clearpage

\begin{flushleft}
  \thispagestyle{empty}
  \vspace*{1cm}
  \vfill
  \noindent
  Eingereicht am 24.11.2006\\
  bei der Fakult"at f"ur Physik und Astronomie\\
  \vspace{0.5cm}
  \noindent \\
  1. Gutachter: Prof. Dr. W. Kinzel\\
  2. Gutachter: Prof. Dr. F. Assaad\\
  der Dissertation.\\
  \ \\
  1. Pr"ufer: Prof. Dr. W. Kinzel\\
  2. Pr"ufer: Prof. Dr. F. Assaad\\
  3. Pr"ufer: Prof. Dr. P. Jakob\\
  im Promotionskolloquium\\
  \ \\
  Tag des Promotionskolloquiums: 18.05.2007\\
  \ \\
  Doktorurkunde ausgeh"andigt am: 20.07.2007\\
  \vspace{1cm}
\end{flushleft}

\cleardoublepage

\chapter*{Abstract}

Neural networks can synchronize by learning from each other. For that
purpose they receive common inputs and exchange their outputs.
Adjusting discrete weights according to a suitable learning rule then
leads to full synchronization in a finite number of steps. It is also
possible to train additional neural networks by using the inputs and
outputs generated during this process as examples. Several algorithms
for both tasks are presented and analyzed.

In the case of Tree Parity Machines the dynamics of both processes is
driven by attractive and repulsive stochastic forces. Thus it can be
described well by models based on random walks, which represent either
the weights themselves or order parameters of their distribution.
However, synchronization is much faster than learning. This effect is
caused by different frequencies of attractive and repulsive steps, as
only neural networks interacting with each other are able to skip
unsuitable inputs. Scaling laws for the number of steps needed for
full synchronization and successful learning are derived using
analytical models. They indicate that the difference between both
processes can be controlled by changing the synaptic depth. In the
case of bidirectional interaction the synchronization time increases
proportional to the square of this parameter, but it grows
exponentially, if information is transmitted in one direction only.

Because of this effect neural synchronization can be used to construct
a cryptographic key-exchange protocol. Here the partners benefit from
mutual interaction, so that a passive attacker is usually unable to
learn the generated key in time. The success probabilities of
different attack methods are determined by numerical simulations and
scaling laws are derived from the data. If the synaptic depth is
increased, the complexity of a successful attack grows exponentially,
but there is only a polynomial increase of the effort needed to
generate a key. Therefore the partners can reach any desired level of
security by choosing suitable parameters. In addition, the entropy of
the weight distribution is used to determine the effective number of
keys, which are generated in different runs of the key-exchange
protocol using the same sequence of input vectors.

If the common random inputs are replaced with queries, synchronization
is possible, too. However, the partners have more control over the
difficulty of the key exchange and the attacks. Therefore they can
improve the security without increasing the average synchronization
time.

\clearpage

\thispagestyle{empty}

\cleardoublepage

\selectlanguage{ngerman}
\chapter*{Zusammenfassung}

Neuronale Netze, die die gleichen Eingaben erhalten und ihre Ausgaben
austauschen, k"onnen voneinander lernen und auf diese Weise
synchronisieren. Wenn diskrete Gewichte und eine geeignete Lernregel
verwendet werden, kommt es in endlich vielen Schritten zur
vollst"andigen Synchronisation. Mit den dabei erzeugten Beispielen
lassen sich weitere neuronale Netze trainieren. Es werden mehrere
Algorithmen f"ur beide Aufgaben vorgestellt und untersucht.

Attraktive und repulsive Zufallskr"afte treiben bei Tree Parity
Machines sowohl den Synchronisationsvorgang als auch die Lernprozesse
an, so dass sich alle Abl"aufe gut durch Random-Walk-Modelle
beschreiben lassen. Dabei sind die Random Walks entweder die Gewichte
selbst oder Ordnungsparameter ihrer Verteilung. Allerdings sind
miteinander wechselwirkende neuronale Netze in der Lage, ungeeignete
Eingaben zu "uberspringen und so repulsive Schritte teilweise zu
vermeiden. Deshalb k"onnen Tree Parity Machines schneller
synchronisieren als lernen. Aus analytischen Modellen abgeleitete
Skalengesetze zeigen, dass der Unterschied zwischen beiden Vorg"angen
von der synaptischen Tiefe abh"angt. Wenn die beiden neuronalen Netze
sich gegenseitig beeinflussen k"onnen, steigt die Synchronisationszeit
nur proportional zu diesem Parameter an; sie w"achst jedoch
exponentiell, sobald die Informationen nur in eine Richtung flie"sen.

Deswegen l"asst sich mittels neuronaler Synchronisation ein
kryptographisches Schl"usselaustauschprotokoll realisieren. Da die
Partner sich gegenseitig beeinflussen, der Angreifer diese
M"oglichkeit aber nicht hat, gelingt es ihm meistens nicht, den
erzeugten Schl"ussel rechtzeitig zu finden. Die
Erfolgswahrscheinlichkeiten der verschiedenen Angriffe werden mittels
numerischer Simulationen bestimmt. Die dabei gefundenen Skalengesetze
zeigen, dass die Komplexit"at eines erfolgreichen Angriffs
exponentiell mit der synaptischen Tiefe ansteigt, aber der Aufwand
f"ur den Schl"usselaustausch selbst nur polynomial anw"achst. Somit
k"onnen die Partner jedes beliebige Sicherheitsniveau durch geeignete
Wahl der Parameter erreichen. Au"serdem wird die effektive Zahl der
Schl"ussel berechnet, die das Schl"usselaustauschprotokoll bei
vorgegebener Zeitreihe der Eingaben erzeugen kann.

Der neuronale Schl"usselaustausch funktioniert auch dann, wenn die
Zufalls"-eingaben durch Queries ersetzt werden. Jedoch haben die
Partner in diesem Fall mehr Kontrolle "uber die Komplexit"at der
Synchronisation und der Angriffe. Deshalb gelingt es, die Sicherheit
zu verbessern, ohne den Aufwand zu erh"ohen.

\clearpage

\thispagestyle{empty}

\cleardoublepage

\selectlanguage{USenglish}
\tableofcontents
\chapter{Introduction}
\label{chap:introduction}

Synchronization is an interesting phenomenon, which can be observed in
a lot of physical and also biological systems \cite{Pikovsky:2001:S}.
It has been first discovered for weakly coupled oscillators, which
develop a constant phase relation to each other. While a lot of
systems show this type of synchronization, a periodic time evolution
is not required. This is clearly visible in the case of chaotic
systems. These can be synchronized by a common source of noise
\cite{Kim:2002:SSC, Cuomo:1993:CIS} or by interaction
\cite{Pecora:1990:SCS, Argyris:2005:CBC}.

As soon as \emph{full synchronization} is achieved, one observes two
or more systems with identical dynamics. But sometimes only parts
synchronize. And it is even possible that one finds a fixed relation
between the states of the systems instead of identical dynamics. Thus
these phenomena look very different, although they are all some kind
of synchronization. In most situations it does not matter, if the
interaction is unidirectional or bidirectional. So there is usually no
difference between components, which influence each other actively and
those which are passively influenced by the dynamics of other systems.

Recently it has been discovered that artificial neural networks can
synchronize, too \cite{Metzler:2000:INN, Kinzel:2000:DIN}. These
mathematical models have been first developed to study and simulate
the behavior of biological neurons. But it was soon discovered that
complex problems in computer science can be solved by using neural
networks. This is especially true if there is little information about
the problem available. In this case the development of a conventional
algorithm is very difficult or even impossible. In contrast, neural
networks have the ability to learn from examples. That is why one does
not have to know the exact rule in order to train a neural network. In
fact, it is sufficient to give some examples of the desired
classification and the network takes care of the generalization.
Several methods and applications of neural networks can be found in
\cite{Hertz:1991:ITN}.

A feed-forward neural network defines a mapping between its input
vector $\mathbf{x}$ and one or more output values $\sigma_i$. Of
course, this mapping is not fixed, but can be changed by adjusting the
weight vector $\mathbf{w}$, which defines the influence of each input
value on the output. For the update of the weights there are two basic
algorithms possible: In batch learning all examples are presented at
the same time and then an optimal weight vector is calculated.
Obviously, this only works for static rules. But in online learning
only one example is used in each time step. Therefore it is possible
to train a neural network using dynamical rules, which change over
time. Thus the examples can be generated by another neural network,
which adjusts its weights, too.

This approach leads to interacting neural feed-forward networks, which
synchronize by mutual learning \cite{Metzler:2000:INN}. They receive
common input vectors and are trained using the outputs of the other
networks. After a short time full synchronization is reached and one
observes either parallel or anti-parallel weight vectors, which stay
synchronized, although they move in time. Similar to other systems
there is no obvious difference between unidirectional and
bidirectional interaction in the case of simple perceptrons
\cite{Kinzel:2003:DGI}.

But Tree Parity Machines, which are more complex neural networks with
a special structure, show a new phenomenon. Synchronization by mutual
learning is much faster than learning by adapting to examples
generated by other networks \cite{Kinzel:2002:INN, Kinzel:2002:TIN,
  Kinzel:2003:DGI, Kinzel:2002:NC}. Therefore one can distinguish
active and passive participants in such a communication. This allows
for new applications, which are not possible with the systems known
before. Especially the idea to use neural synchronization for a
cryptographic key-exchange protocol, which has been first proposed in
\cite{Kanter:2002:SEI}, has stimulated most research in this area
\cite{Kinzel:2002:INN, Kinzel:2003:DGI, Kinzel:2002:NC,
  Kinzel:2002:TIN, Kanter:2003:TNN, Klein:2005:SNN,
  Mislovaty:2002:SKE, Rosen-Zvi:2002:CBN, Rosen-Zvi:2002:MLT,
  Ruttor:2004:NCF, Klimov:2003:ANC, Shacham:2004:CAN, Ruttor:2006:GAN,
  Ruttor:2005:NCQ, Mislovaty:2003:PCC}.

\begin{figure}
  \centering
  \includegraphics{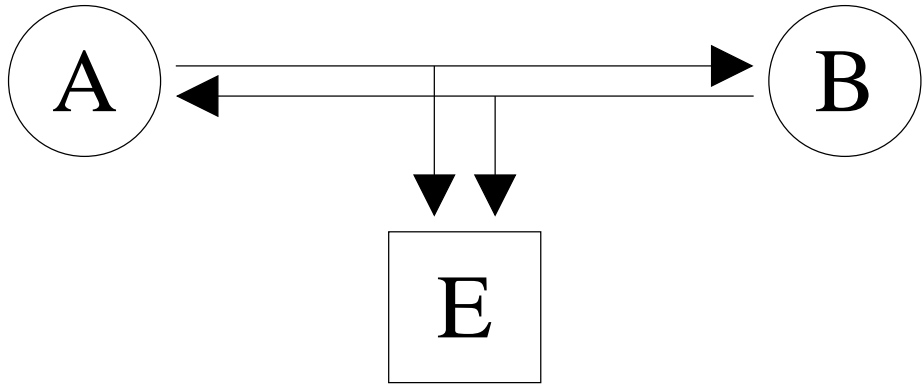}
  \caption{Key exchange between two partners with a passive attacker
    listening to the communication.}
  \label{fig:communication}
\end{figure}

Such an algorithm can be used to solve a common cryptographic
problem~\cite{Stinson:1995:CTP}: Two partners \textbf{A}lice and
\textbf{B}ob want to exchange secret messages over a public channel.
In order to protect the content against an opponent \textbf{E}ve, A
encrypts her message using a fast symmetric encryption algorithm. But
now B needs to know A's key for reading her message. This situation is
depicted in figure~\ref{fig:communication}.

In fact, there are three possible solutions for this key-exchange
problem \cite{Beutelspacher:2002:K}. First A and B could use a second
private channel to transmit the key, e.~g.~they could meet in person
for this purpose. But usually this is very difficult or just
impossible. Alternatively, the partners can use public-key
cryptography. Here an asymmetric encryption algorithm is employed so
that the public keys of A's and B's key pair can be exchanged between
the partners without the need to keep them secret. But asymmetric
encryption is much slower than symmetric algorithms. That is why it is
only used to transmit a symmetric session key. However, one can
achieve the same result by using a key-exchange protocol. In this case
messages are transmitted over the public channel and afterwards A and
B generate a secret key based on the exchanged information. But E is
unable to discover the key because listening to the communication is
not sufficient.

Such a protocol can be constructed using neural synchronization
\cite{Kanter:2002:SEI}. Two Tree Parity Machines, one for A and one
for B respectively, start with random initial weights, which are kept
secret. In each step a new random input vector is generated publicly.
Then the partners calculate the output of their neural networks and
send it to each other. Afterwards the weight vectors are updated
according to a suitable learning rule. Because both inputs and weights
are discrete, this procedure leads to full synchronization,
$\mathbf{w}_i^A = \mathbf{w}_i^B$, after a finite number of steps.
Then A and B can use the weight vectors as a common secret key.

In this case the difference between unidirectional learning and
bidirectional synchronization is essential for the security of the
cryptographic application. As E cannot influence A and B, she is
usually not able to achieve synchronization by the time A and B finish
generating the key and stop the transmission of the output bits
\cite{Kinzel:2002:INN}. Consequently, attacks based on learning have
only a small probability of success \cite{Mislovaty:2002:SKE}. But
using other methods is difficult, too. After all the attacker does not
know the internal representation of the multi-layer neural networks.
In contrast, it is easy to reconstruct the learning process of a
perceptron exactly due to the lack of hidden units. This corresponds
with the observation that E is nearly always successful, if these
simple networks are used \cite{Kinzel:2003:DGI}.

Of course, one wants to compare the level of security achieved by the
neural key-exchange protocol with other algorithms for key exchange.
For that purpose some assumptions are necessary, which are standard
for all cryptographic systems:
\begin{itemize}
\item The attacker E knows all the messages exchanged between A and B.
  Thus each participant has the same amount of information about all
  the others. Furthermore the security of the neural key-exchange
  protocol does not depend on some special properties of the
  transmission channel.
\item E is unable to change the messages, so that only passive attacks
  are considered. In order to achieve security against active methods,
  e.~g.~man-in-the-middle attacks, one has to implement additional
  provisions for authentication.
\item The algorithm is public, because keeping it secret does not
  improve the security at all, but prevents cryptographic analysis.
  Although vulnerabilities may not be revealed, if one uses
  \emph{security by obscurity}, an attacker can find them
  nevertheless.
\end{itemize}

In chapter~\ref{chap:neurosync} the basic algorithm for neural
synchronization is explained. Definitions of the order parameters used
to analyze this effect can be found there, too. Additionally, it
contains descriptions of all known methods for E's attacks on the
neural key-exchange protocol.

Then the dynamics of neural synchronization is discussed in
chapter~\ref{chap:dynamics}. It is shown that it is, in fact, a
complex process driven by stochastic attractive and repulsive forces,
whose properties depend on the chosen parameters. Looking especially
at the average change of the overlap between corresponding hidden
units in A's, B's and E's Tree Parity Machine reveals the differences
between bidirectional and unidirectional interaction clearly.

Chapter~\ref{chap:security} focuses on the security of the neural
key-exchange protocol, which is essential for this application of
neural synchronization. Of course, simulations of cryptographic useful
systems do not show successful attacks and the other way round. That
is why finding scaling laws in regard to effort and security is very
important. As these relations can be used to extrapolate reliably,
they play a major role here.

Finally, chapter~\ref{chap:queries} presents a modification of the
neural key-exchange protocol: Queries generated by A and B replace the
random sequence of input vectors. Thus the partners have more
influence on the process of synchronization, because they are able to
control the frequency of repulsive steps as a function of the overlap.
In doing so, A and B can improve the security of the neural
key-exchange protocol without increasing the synchronization time.

\chapter{Neural synchronization}
\label{chap:neurosync}

Synchronization of neural networks \cite{Metzler:2000:INN,
  Kinzel:2000:DIN, Kinzel:2002:INN, Kinzel:2002:TIN, Kinzel:2003:DGI}
is a special case of an online learning situation.  Two neural
networks start with randomly chosen weight vectors. In each time step
they receive a common input vector, calculate their outputs, and
communicate them to each other. If they agree on the mapping between
the current input and the output, their weights are updated according
to a suitable learning rule.

In the case of discrete weight values this process leads to full
synchronization in a finite number of steps \cite{Kinzel:2003:DGI,
  Kinzel:2002:INN, Kinzel:2002:NC, Kinzel:2002:TIN, Ruttor:2004:SRW}.
Afterwards corresponding weights in both networks have the same value,
even if they are updated by further applications of the learning rule.
Thus full synchronization is an absorbing state.

Additionally, a third neural network can be trained using the
examples, input vectors and output values, generated by the process of
synchronization. As this neural network cannot influence the others,
it corresponds to a student network which tries to learn a time
dependent mapping between inputs and outputs.

In the case of perceptrons, which are simple neural networks, one
cannot find any significant difference between these two situations:
the average number of steps needed for synchronization and learning is
the same \cite{Metzler:2000:INN, Kinzel:2000:DIN}. But in the case of
the more complex Tree Parity Machines an interesting phenomenon can be
observed: two neural networks learning from each other synchronize
faster than a third network only listening to the communication
\cite{Kinzel:2002:INN, Kinzel:2003:DGI, Kinzel:2002:NC,
  Kinzel:2002:TIN}.

This difference between bidirectional and unidirectional interaction
can be used to solve the cryptographic key-exchange problem
\cite{Kanter:2002:SEI}. For that purpose the partners A and B
synchronize their Tree Parity Machines. In doing so they generate
their common session key faster than an attacker is able to discover
it by training another neural network. Consequently, the difference
between synchronization and learning is essential for the security of
the neural key-exchange protocol.

In this chapter the basic framework for neural synchronization is
presented. This includes the structure of the networks, the learning
rules, and the quantities used to describe the process of
synchronization.

\section{Tree Parity Machines}

\begin{figure}[h]
  \centering
  \includegraphics{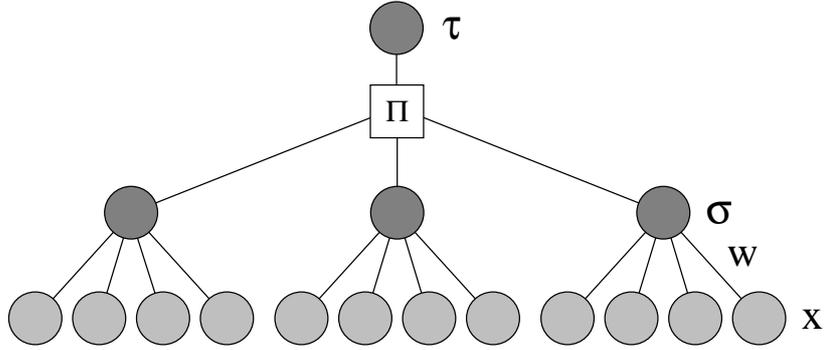}
  \caption{A Tree Parity Machine with $K=3$ and $N=4$.}
  \label{fig:tpm}
\end{figure}

Tree Parity Machines, which are used by partners and attackers in
neural cryptography, are multi-layer feed-forward networks. Their
general structure is shown in figure~\ref{fig:tpm}.

Such a neural network consists of $K$ hidden units, which are
perceptrons with independent receptive fields. Each one has $N$ input
neurons and one output neuron. All input values are binary,
\begin{equation}
  x_{i,j} \in \{-1, +1\} \,,
\end{equation}
and the weights, which define the mapping from input to output, are
discrete numbers between $-L$ and $+L$,
\begin{equation}
  w_{i,j} \in \{-L, -L+1, \dots, +L\} \,.
\end{equation}
Here the index $i=1,\dots,K$ denotes the $i$-th hidden unit of the
Tree Parity Machine and $j=1, \dots, N$ the elements of the vector.

As in other neural networks the weighted sum over the current input
values is used to determine the output of the hidden units. Therefore
the full state of each hidden neuron is given by its local field
\begin{equation}
  \label{eq:field}
  h_i = \frac{1}{\sqrt{N}} \mathbf{w}_i \cdot \mathbf{x}_i =
  \frac{1}{\sqrt{N}} \sum_{j=1}^N w_{i,j} \, x_{i,j} \,.
\end{equation}
The output $\sigma_i$ of the $i$-th hidden unit is then defined as the
sign of $h_i$,
\begin{equation}
  \sigma_i = \mathrm{sgn}( h_i ) \,,
\end{equation}
but the special case $h_i = 0$ is mapped to $\sigma_i = -1$ in order
to ensure a binary output value. Thus a hidden unit is only active,
$\sigma_i = +1$, if the weighted sum over its inputs is positive,
otherwise it is inactive, $\sigma_i = -1$.

Then the total output $\tau$ of a Tree Parity Machine is given by the
product (parity) of the hidden units,
\begin{equation}
  \tau = \prod_{i=1}^K \sigma_i \,,
\end{equation}
so that $\tau$ only indicates, if the number of inactive hidden units,
with $\sigma_i = -1$, is even ($\tau = +1$) or odd ($\tau = -1$).
Consequently, there are $2^{K-1}$ different internal representations
$(\sigma_1, \sigma_2, \dots, \sigma_K)$, which lead to the same output
value $\tau$.

If there is only one hidden unit, $\tau$ is equal to $\sigma_1$.
Consequently, a Tree Parity Machine with $K=1$ shows the same behavior
as a perceptron, which can be regarded as a special case of the more
complex neural network.

\section{Learning rules}
\label{sec:rules}

At the beginning of the synchronization process A's and B's Tree
Parity Machines start with randomly chosen and therefore uncorrelated
weight vectors $\mathbf{w}^{A/B}_i$. In each time step $K$ public
input vectors $\mathbf{x}_i$ are generated randomly and the
corresponding output bits $\tau^{A/B}$ are calculated.

Afterwards A and B communicate their output bits to each other. If
they disagree, $\tau^A \not= \tau^B$, the weights are not changed.
Otherwise one of the following learning rules suitable for
synchronization is applied:
\begin{itemize}
\item In the case of the \emph{Hebbian learning rule}
  \cite{Mislovaty:2002:SKE} both neural networks learn from each
  other:
  \begin{equation}
    \label{eq:hebb_plus}
    w_{i,j}^{+} = g(w_{i,j} + x_{i,j} \tau \Theta(\sigma_i \tau)
    \Theta(\tau^A \tau^B)) \,.
  \end{equation}
\item It is also possible that both networks are trained with the
  opposite of their own output. This is achieved by using the
  \emph{anti-Hebbian learning rule} \cite{Kinzel:2002:TIN}:
  \begin{equation}
    \label{eq:hebb_minus}
    w_{i,j}^{+} = g(w_{i,j} - x_{i,j} \tau \Theta(\sigma_i \tau)
    \Theta(\tau^A \tau^B)) \,.
  \end{equation}
\item But the set value of the output is not important for
  synchronization as long as it is the same for all participating
  neural networks. That is why one can use the \emph{random-walk
    learning rule} \cite{Kinzel:2002:NC}, too:
  \begin{equation}
    \label{eq:random_walk}
    w_{i,j}^{+} = g(w_{i,j} + x_{i,j} \Theta(\sigma_i \tau) \Theta(\tau^A
    \tau^B)) \,.
  \end{equation}
\end{itemize}
In any way only weights are changed by these learning rules, which are
in hidden units with $\sigma_i = \tau$. By doing so it is impossible
to tell which weights are updated without knowing the internal
representation $(\sigma_1, \sigma_2, \dots, \sigma_K)$. This feature
is especially needed for the cryptographic application of neural
synchronization.

Of course, the learning rules have to assure that the weights stay in
the allowed range between $-L$ and $+L$. If any weight moves outside
this region, it is reset to the nearest boundary value $\pm L$. This
is achieved by the function $g(w)$ in each learning rule:
\begin{equation}
  \label{eq:g}
  g(w) = \left\{
    \begin{array}{cl}
      \mathrm{sgn}(w) \, L & \mbox{ for $|w|>L$ } \\
      w                    & \mbox{ otherwise }   \\
    \end{array}
  \right.\,.
\end{equation}

Afterwards the current synchronization step is finished. This process
can be repeated until corresponding weights in A's and B's Tree Parity
Machine have equal values, $\mathbf{w}_i^A = \mathbf{w}_i^B$. Further
applications of the learning rule are unable to destroy this
synchronization, because the movements of the weights depend only on
the inputs and weights, which are then identical in A's and B's neural
networks.

\section{Order parameters}

In order to describe the correlations between two Tree Parity Machines
caused by the synchronization process, one can look at the probability
distribution of the weight values in each hidden unit. It is given by
$(2 L + 1)$ variables
\begin{equation}
  p_{a,b}^i = P(w_{i,j}^A = a \wedge w_{i,j}^B = b) \,,
\end{equation}
which are defined as the probability to find a weight with $w_{i,j}^A
= a$ in A's Tree Parity Machine and $w_{i,j}^B = b$ in B's neural
network.

While these probabilities are only approximately given as relative
frequencies in simulations with finite $N$, their development can be
calculated using exact equations of motion in the limit $N \rightarrow
\infty$ \cite{Rosen-Zvi:2002:CBN, Rosen-Zvi:2002:MLT,
  Ruttor:2004:NCF}. This method is explained in detail in
appendix~\ref{chap:icalc}.

In both cases, simulation and iterative calculation, the standard
order parameters \cite{Engel:2001:SML}, which are also used for the
analysis of online learning, can be calculated as functions of
$p_{a,b}^i$:
\begin{eqnarray}
  \label{eq:qa}
  Q_i^A =& \displaystyle \frac{1}{N} \mathbf{w}_i^A \mathbf{w}_i^A
  &= \sum_{a=-L}^{L} \sum_{b=-L}^{L} a^2 \, p_{a,b}^i \, ,\\
  \label{eq:qb}
  Q_i^B =& \displaystyle \frac{1}{N} \mathbf{w}_i^B \mathbf{w}_i^B
  &= \sum_{a=-L}^{L} \sum_{b=-L}^{L} b^2 \, p_{a,b}^i \, ,\\
  \label{eq:rab}
  R_i^{AB} =& \displaystyle \frac{1}{N} \mathbf{w}_i^A \mathbf{w}_i^B
  &= \sum_{a=-L}^{L} \sum_{b=-L}^{L} a \, b \, p_{a,b}^i \,.
\end{eqnarray}
Then the level of synchronization is given by the normalized overlap
\cite{Engel:2001:SML} between two corresponding hidden units:
\begin{equation}
  \rho_i^{AB} = \frac{\mathbf{w}_i^A \cdot
    \mathbf{w}_i^B}{\sqrt{\mathbf{w}_i^A \cdot \mathbf{w}_i^A}
    \sqrt{\mathbf{w}_i^B \cdot \mathbf{w}_i^B}} =
  \frac{R_i^{AB}}{\sqrt{Q_i^A Q_i^B}} \,.
\end{equation}
Uncorrelated hidden units, e.~g.~at the beginning of the
synchronization process, have $\rho_i = 0$, while the maximum value
$\rho_i = 1$ is reached for fully synchronized weights. Consequently,
$\rho_i$ is the most important quantity for analyzing the process of
synchronization.

But it is also interesting to estimate the mutual information gained
by the partners during the process of synchronization. For this
purpose one has to calculate the entropy \cite{Cover:1991:EIT}
\begin{equation}
  \label{eq:entropy}
  S_i^{AB} = - N \sum_{a=-L}^L \sum_{b=-L}^L p_{a,b}^i \ln p_{a,b}^i
\end{equation}
of the joint weight distribution of A's and B's neural networks.
Similarly the entropy of the weights in a single hidden unit is given
by
\begin{eqnarray}
  \label{eq:entropy_a}
  S_i^A &=& - N \sum_{a=-L}^L \left( \sum_{b=-L}^L p_{a,b}^i \right)
  \ln \left( \sum_{b=-L}^L p_{a,b}^i \right) \,, \\
  \label{eq:entropy_b}
  S_i^B &=& - N \sum_{b=-L}^L \left( \sum_{a=-L}^L p_{a,b}^i \right)
  \ln \left( \sum_{a=-L}^L p_{a,b}^i \right) \,.
\end{eqnarray}
Of course, these equations assume that there are no correlations
between different weights in one hidden unit. This is correct in the
limit $N \rightarrow \infty$, but not necessarily for small systems.

Using (\ref{eq:entropy}), (\ref{eq:entropy_a}), and
(\ref{eq:entropy_b}) the mutual information \cite{Cover:1991:EIT} of
A's and B's Tree Parity Machines can be calculated as
\begin{equation}
  \label{eq:minfo}
  I^{AB} = \sum_{i=1}^K \left( S_i^A + S_i^B - S_i^{AB} \right) \,.
\end{equation}
At the beginning of the synchronization process, the partners only
know the weight configuration of their own neural network, so that
$I^{AB} = 0$. But for fully synchronized weight vectors this quantity
is equal to the entropy of a single Tree Parity Machine, which is
given by
\begin{equation}
  \label{eq:s0}
  S_0 = K N \ln ( 2 L + 1 )
\end{equation}
in the case of uniformly distributed weights.

\section{Neural cryptography}

The neural key-exchange protocol \cite{Kanter:2002:SEI} is an
application of neural synchronization. Both partners A and B use a
Tree Parity Machine with the same structure. The parameters $K$, $L$
and $N$ are public. Each neural network starts with randomly chosen
weight vectors. These initial conditions are kept secret. During the
synchronization process, which is described in
section~\ref{sec:rules}, only the input vectors $\mathbf{x}_i$ and the
total outputs $\tau^A$, $\tau^B$ are transmitted over the public
channel. Therefore each participant just knows the internal
representation $(\sigma_1, \sigma_2, \dots, \sigma_K)$ of his own Tree
Parity Machine. Keeping this information secret is essential for the
security of the key-exchange protocol. After achieving full
synchronization A and B use the weight vectors as common secret key.

The main problem of the attacker E is that the internal
representations $(\sigma_1, \sigma_2, \dots, \sigma_K)$ of A's and B's
Tree Parity Machines are not known to her. As the movement of the
weights depends on $\sigma_i$, it is important for a successful attack
to guess the state of the hidden units correctly. Of course, most
known attacks use this approach. But there are other possibilities and
it is indeed possible that a clever attack method will be found, which
breaks the security of neural cryptography completely. However, this
risk exists for all cryptographic algorithms except the one-time pad.

\subsection{Simple attack}

For the \emph{simple attack} \cite{Kanter:2002:SEI} E just trains a
third Tree Parity Machine with the examples consisting of input
vectors $\mathbf{x}_i$ and output bits $\tau^A$. These can be obtained
easily by intercepting the messages transmitted by the partners over
the public channel. E's neural network has the same structure as A's
and B's and starts with random initial weights, too.

In each time step the attacker calculates the output of her neural
network. Afterwards E uses the same learning rule as the partners, but
$\tau^E$ is replaced by $\tau^A$. Thus the update of the weights is
given by one of the following equations:
\begin{itemize}
\item Hebbian learning rule:
  \begin{equation}
    w_{i,j}^{E+} = g(w_{i,j}^E + x_{i,j} \tau^A \Theta(\sigma_i^E
    \tau^A) \Theta(\tau^A \tau^B)) \,.
  \end{equation}
\item Anti-Hebbian learning rule:
  \begin{equation}
    w_{i,j}^{E+} = g(w_{i,j}^E - x_{i,j} \tau^A \Theta(\sigma_i^E
    \tau^A) \Theta(\tau^A \tau^B)) \,.
  \end{equation}
\item Random walk learning rule:
  \begin{equation}
    w_{i,j}^{E+} = g(w_{i,j}^E + x_{i,j} \Theta(\sigma_i^E \tau^A)
    \Theta(\tau^A \tau^B)) \,.
  \end{equation}
\end{itemize}
So E uses the internal representation $(\sigma_1^E, \sigma_2^E, \dots,
\sigma_K^E)$ of her own network in order to estimate A's, even if the
total output is different. As $\tau^A \not= \tau^E$ indicates that
there is at least one hidden unit with $\sigma_i^A \not= \sigma_i^E$,
this is certainly not the best algorithm available for an attacker.

\subsection{Geometric attack}

The \emph{geometric attack} \cite{Klimov:2003:ANC} performs better
than the simple attack, because E takes $\tau^E$ and the local fields
of her hidden units into account. In fact, it is the most successful
method for an attacker using only a single Tree Parity Machine.

Similar to the simple attack E tries to imitate B without being able
to interact with A. As long as $\tau^A = \tau^E$, this can be done by
just applying the same learning rule as the partners A and B. But in
the case of $\tau^E \not= \tau^A$ E cannot stop A's update of the
weights. Instead the attacker tries to correct the internal
representation of her own Tree Parity Machine using the local fields
$h_1^E$, $h_2^E$, \dots, $h_K^E$ as additional information. These
quantities can be used to determine the level of confidence associated
with the output of each hidden unit \cite{Ein-Dor:1999:CPN}. As a low
absolute value $|h_i^E|$ indicates a high probability of $\sigma_i^A
\not= \sigma_i^E$, the attacker changes the output $\sigma_i^E$ of the
hidden unit with minimal $|h_i^E|$ and the total output $\tau^E$
before applying the learning rule.

Of course, the geometric attack does not always succeed in estimating
the internal representation of A's Tree Parity Machine correctly.
Sometimes there are several hidden units with $\sigma_i^A \not=
\sigma_i^E$. In this case the change of one output bit is not enough.
It is also possible that $\sigma_i^A = \sigma_i^E$ for the hidden unit
with minimal $|h_i^E|$, so that the geometric correction makes the
result worse than before.

\subsection{Majority attack}

With the \emph{majority attack} \cite{Shacham:2004:CAN} E can improve
her ability to predict the internal representation of A's neural
network. For that purpose the attacker uses an ensemble of $M$ Tree
Parity Machines instead of a single neural network. At the beginning
of the synchronization process the weight vectors of all attacking
networks are chosen randomly, so that their average overlap is zero.

Similar to other attacks, E does not change the weights in time steps
with $\tau^A \not= \tau^B$, because the partners skip these input
vectors, too. But for $\tau^A = \tau^B$ an update is necessary and the
attacker calculates the output bits $\tau^{E,m}$ of her Tree Parity
Machines. If the output bit $\tau^{E,m}$ of the $m$-th attacking
network disagrees with $\tau^A$, E searches the hidden unit $i$ with
minimal absolute local field $|h_i^{E,m}|$. Then the output bits
$\sigma_i^{E,m}$ and $\tau^{E,m}$ are inverted similarly to the
geometric attack. Afterwards the attacker counts the internal
representations $(\sigma^{E,m}_1, \dots, \sigma^{E,m}_K)$ of her Tree
Parity Machines and selects the most common one. This majority vote is
then adopted by all attacking networks for the application of the
learning rule.

But these identical updates create and amplify correlations between
E's Tree Parity Machines, which reduce the efficiency of the majority
attack. Especially if the attacking neural networks become fully
synchronized, this method is reduced to a geometric attack.

In order to keep the Tree Parity Machines as uncorrelated as possible,
majority attack and geometric attack are used alternately
\cite{Shacham:2004:CAN}. In even time steps the majority vote is used
for learning, but otherwise E only applies the geometric correction.
Therefore not all updates of the weight vectors are identical, so that
the overlap between them is reduced.  Additionally, E replaces the
majority attack by the geometric attack in the first $100$ time steps
of the synchronization process.

\subsection{Genetic attack}

The \emph{genetic attack} \cite{Ruttor:2006:GAN} offers an alternative
approach for the opponent, which is not based on optimizing the
prediction of the internal representation, but on an evolutionary
algorithm. E starts with only one randomly initialized Tree Parity
Machine, but she can use up to $M$ neural networks.

Whenever the partners update the weights because of $\tau^A = \tau^B$
in a time step, the following genetic algorithm is applied:
\begin{itemize}
\item As long as E has at most $M / 2^{K-1}$ Tree Parity Machines, she
  determines all $2^{K-1}$ internal representations
  $(\sigma_1^E,\dots,\sigma_K^E)$ which reproduce the output $\tau^A$.
  Afterwards these are used to update the weights in the attacking
  networks according to the learning rule. By doing so E creates
  $2^{K-1}$ variants of each Tree Parity Machine in this
  \emph{mutation step}.
\item But if E already has more than $M / 2^{K-1}$ neural networks,
  only the fittest Tree Parity Machines should be kept. This is
  achieved by discarding all networks which predicted less than $U$
  outputs $\tau^A$ in the last $V$ learning steps, with $\tau^A =
  \tau^B$, successfully. A limit of $U=10$ and a history of $V=20$ are
  used as default values for the \emph{selection step}. Additionally,
  E keeps at least $20$ of her Tree Parity Machines.
\end{itemize}

The efficiency of the genetic attack mostly depends on the algorithm
which selects the fittest neural networks. In the ideal case the Tree
Parity Machine, which has the same sequence of internal
representations as A is never discarded. Then the problem of the
opponent E would be reduced to the synchronization of $K$ perceptrons
and the genetic attack would succeed certainly. However, this
algorithm as well as other methods available for the opponent E are
not perfect, which is clearly shown in chapter~\ref{chap:security}.

\chapter{Dynamics of the neural synchronization process}
\label{chap:dynamics}

Neural synchronization is a stochastic process consisting of discrete
steps, in which the weights of participating neural networks are
adjusted according to the algorithms presented in
chapter~\ref{chap:neurosync}. In order to understand why
unidirectional learning and bidirectional synchronization show
different effects, it is reasonable to take a closer look at the
dynamics of these processes.

Although both are completely determined by the initial weight vectors
$\mathbf{w}_i$ of the Tree Parity Machines and the sequence of random
input vectors $\mathbf{x}_i$, one cannot calculate the result of each
initial condition, as there are too many except for very small
systems. Instead of that the effect of the synchronization steps on
the overlap $\rho_i$ of two corresponding hidden units is analyzed.
This order parameter is defined as the cosine of the angle between the
weight vectors \cite{Engel:2001:SML}. Attractive steps increase the
overlap, while repulsive steps decrease it \cite{Ruttor:2004:NCF}.

As the probabilities for both types of steps as well as the average
step sizes $\langle \Delta \rho_\mathrm{a} \rangle$, $\langle \Delta
\rho_\mathrm{r} \rangle$ depend on the current overlap, neural
synchronization can be regarded as a random walk in $\rho$-space.
Hence the average change of the overlap $\langle \Delta \rho(\rho)
\rangle$ shows the most important properties of the dynamics.
Especially the difference between bidirectional and unidirectional
interaction is clearly visible.

As long as two Tree Parity Machines influence each other, repulsive
steps have only little effect on the process of synchronization.
Therefore it is possible to neglect this type of step in order to
determine the scaling of the synchronization time $t_\mathrm{sync}$.
For that purpose a random walk model consisting of two corresponding
weights is analyzed \cite{Ruttor:2004:SRW}.

But in the case of unidirectional interaction the higher frequency of
repulsive steps leads to a completely different dynamics of the
system, so that synchronization is only possible by fluctuations.
Hence the scaling of $\langle t_\mathrm{sync} \rangle$ changes to an
exponential increase with $L$. This effect is important for the
cryptographic application of neural synchronization, as it is
essential for the security of the neural key-exchange protocol.

\section{Effect of the learning rules}
\label{sec:effect}

The learning rules used for synchronizing Tree Parity Machines, which
have been presented in section~\ref{sec:rules}, share a common
structure. That is why they can be described by a single equation
\begin{equation}
  \label{eq:update}
  w_{i,j}^{+} = g(w_{i,j} + f(\sigma_i, \tau^A, \tau^B) x_{i,j})
\end{equation}
with a function $f(\sigma, \tau^A, \tau^B)$, which can take the values
$-1$, $0$, or $+1$. In the case of bidirectional interaction it is
given by
\begin{equation}
  \label{eq:f}
  f(\sigma, \tau^A, \tau^B) = \Theta(\sigma \tau^A) \Theta(\tau^A
  \tau^B)
  \left\{
    \begin{array}{cl}
      \sigma  & \mbox{ Hebbian learning rule }      \\
      -\sigma & \mbox{ anti-Hebbian learning rule } \\
      1       & \mbox{ random walk learning rule }
    \end{array}
  \right. \,.
\end{equation}
The common part $\Theta(\sigma \tau^A) \Theta(\tau^A \tau^B)$ of
$f(\sigma, \tau^A, \tau^B)$ controls, when the weight vector of a
hidden unit is adjusted. Because it is responsible for the occurrence
of attractive and repulsive steps as shown in section~\ref{sec:steps},
all three learning rules have similar effects on the overlap. But the
second part, which influences the direction of the movements, changes
the distribution of the weights in the case of Hebbian and
anti-Hebbian learning. This results in deviations, especially for
small system sizes, which is the topic of section~\ref{sec:wdist}.

Equation (\ref{eq:update}) together with (\ref{eq:f}) also describes
the update of the weights for unidirectional interaction, after the
output $\tau^E$ and the internal representation $(\sigma_1^E,
\sigma_2^E, \dots, \sigma_K^E)$ have been adjusted by the learning
algorithm. That is why one observes the same types of steps in this
case.

\subsection{Distribution of the weights}
\label{sec:wdist}

According to (\ref{eq:f}) the only difference between the learning
rules is, whether and how the output $\sigma_i$ of a hidden unit
affects $\Delta w_{i,j} = w_{i,j}^{+} - w_{i,j}$. Although this does not
change the qualitative effect of an update step, it influences the
distribution of the weights \cite{Ruttor:2006:GAN}.

In the case of the Hebbian rule (\ref{eq:hebb_plus}), A's and B's Tree
Parity Machines learn their own output. Therefore the direction in
which the weight $w_{i,j}$ moves is determined by the product $\sigma_i
x_{i,j}$. As the output $\sigma_i$ is a function of all input values,
$x_{i,j}$ and $\sigma_i$ are correlated random variables. Thus the
probabilities to observe $\sigma_i x_{i,j} = +1$ or $\sigma_i x_{i,j} =
-1$ are not equal, but depend on the value of the corresponding weight
$w_{i,j}$:
\begin{equation}
  \label{eq:wcorr}
  P(\sigma_i x_{i,j} = 1) = \frac{1}{2} \left[ 1 + \mathrm{erf} \left(
      \frac{w_{i,j}}{\sqrt{N Q_i - w_{i,j}^2}} \right) \right] \,.
\end{equation}
According to this equation, $\sigma_i x_{i,j} = \mathrm{sgn}(w_{i,j})$
occurs more often than the opposite, $\sigma_i x_{i,j} =
-\mathrm{sgn}(w_{i,j})$. Consequently, the Hebbian learning rule
(\ref{eq:hebb_plus}) pushes the weights towards the boundaries at $-L$
and $+L$.

In order to quantify this effect the stationary probability
distribution of the weights for $t \rightarrow \infty$ is calculated
using (\ref{eq:wcorr}) for the transition probabilities. This leads to
\cite{Ruttor:2006:GAN}
\begin{equation}
  \label{eq:wdplus}
  P(w_{i,j} = w) = p_0 \prod_{m=1}^{|w|} \frac{1 + \mathrm{erf} \left(
      \frac{m - 1}{\sqrt{N Q_i - (m - 1)^2}} \right)}{1 - \mathrm{erf}
    \left( \frac{m}{\sqrt{N Q_i - m^2}} \right)} \,.
\end{equation}
Here the normalization constant $p_0$ is given by
\begin{equation}
  p_0 = \left( \sum_{w = -L}^{L} \prod_{m=1}^{|w|} \frac{1 +
      \mathrm{erf} \left( \frac{m - 1}{\sqrt{N Q_i - (m - 1)^2}}
      \right)}{1 - \mathrm{erf} \left( \frac{m}{\sqrt{N Q_i - m^2}}
      \right)} \right)^{-1} \,.
\end{equation}
In the limit $N \rightarrow \infty$ the argument of the error
functions vanishes, so that the weights stay uniformly distributed. In
this case the initial length
\begin{equation}
  \label{eq:rwlength}
  \sqrt{Q_i(t = 0)} = \sqrt{\frac{L(L+1)}{3}}
\end{equation}
of the weight vectors is not changed by the process of
synchronization.

But, for finite $N$ the probability distribution (\ref{eq:wdplus})
itself depends on the order parameter $Q_i$. Therefore its expectation
value is given by the solution of the following equation:
\begin{equation}
  Q_i = \sum_{w = -L}^{L} w^2 P(w_{i,j} = w) \,.
\end{equation}
Expanding it in terms of $N^{-1/2}$ results in \cite{Ruttor:2006:GAN}
\begin{eqnarray}
  \label{eq:qplus}
  Q_i = \frac{L (L + 1)}{3} + \frac{8 L^4 + 16 L^3 -10 L^2 - 18 L +
    9}{15 \sqrt{3 \pi L (L + 1)}} \frac{1}{\sqrt{N}} + \mathrm{O}
  \left( \frac{L^4}{N} \right)
\end{eqnarray}
as a first-order approximation of $Q_i$ for large system sizes. The
asymptotic behavior of this order parameter in the case of $1 \ll L
\ll \sqrt{N}$ is given by
\begin{equation}
  Q_i \sim \frac{L (L + 1)}{3} \left( 1 + \frac{8}{5 \sqrt{3 \pi}}
    \frac{L}{\sqrt{N}} \right) \,.
\end{equation}
Thus each application of the Hebbian learning rule increases the
length of the weight vectors $\mathbf{w}_i$ until a steady state is
reached. The size of this effect depends on $L / \sqrt{N}$ and
disappears in the limit $L / \sqrt{N} \rightarrow 0$.

\begin{figure}
  \centering
  \includegraphics[scale=0.5]{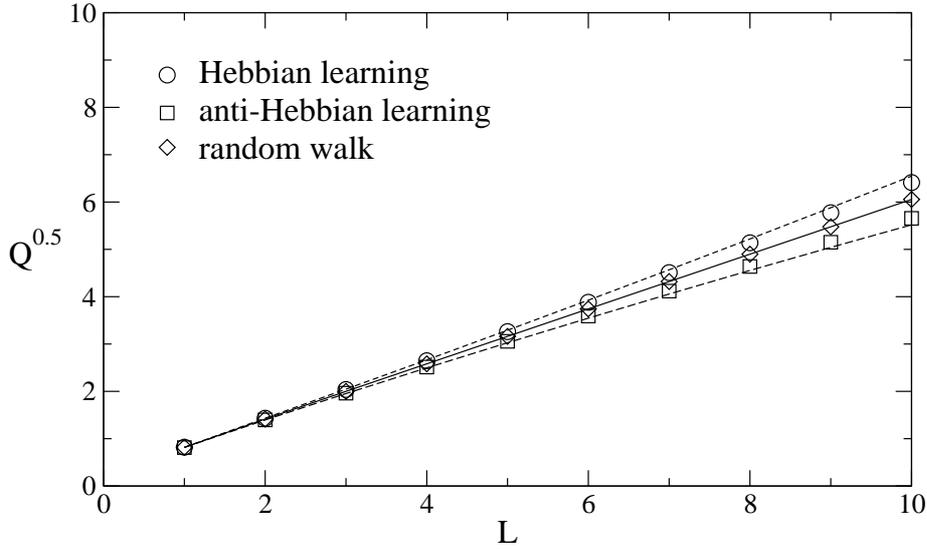}
  \caption{Length of the weight vectors in the steady state for $K=3$
    and $N=1000$. Symbols denote results averaged over $1000$
    simulations and lines show the first-order approximation given in
    (\ref{eq:qplus}) and (\ref{eq:qminus}).}
  \label{fig:length}
\end{figure}

In the case of the anti-Hebbian rule (\ref{eq:hebb_minus}) A's and B's
Tree Parity Machines learn the opposite of their own outputs.
Therefore the weights are pulled away from the boundaries instead of
being pushed towards $\pm L$. Here the first-order approximation of
$Q_i$ is given by \cite{Ruttor:2006:GAN}
\begin{equation}
  \label{eq:qminus}
  Q_i = \frac{L (L + 1)}{3} - \frac{8 L^4 + 16 L^3 - 10 L^2 - 18 L +
    9}{15 \sqrt{3 \pi L (L + 1)}} \frac{1}{\sqrt{N}} + \mathrm{O}
  \left( \frac{L^4}{N} \right) \,,
\end{equation}
which asymptotically converges to
\begin{equation}
  Q_i \sim \frac{L (L + 1)}{3} \left( 1 - \frac{8}{5 \sqrt{3 \pi}}
    \frac{L}{\sqrt{N}} \right) 
\end{equation}
in the case $1 \ll L \ll \sqrt{N}$. Hence applying the anti-Hebbian
learning rule decreases the length of the weight vectors
$\mathbf{w}_i$ until a steady state is reached. As before, $L /
\sqrt{N}$ determines the size of this effect.

In contrast, the random walk rule (\ref{eq:random_walk}) always uses a
fixed set output. Here the weights stay uniformly distributed, as the
random input values $x_{i,j}$ alone determine the direction of the
movements. Consequently, the length of the weight vectors is always
given by (\ref{eq:rwlength}).

Figure~\ref{fig:length} shows that the theoretical predictions are in
good quantitative agreement with simulation results as long as $L^2$
is small compared to the system size $N$. The deviations for large $L$
are caused by higher-order terms which are ignored in (\ref{eq:qplus})
and (\ref{eq:qminus}).

\begin{figure}
  \centering
  \includegraphics[scale=0.5]{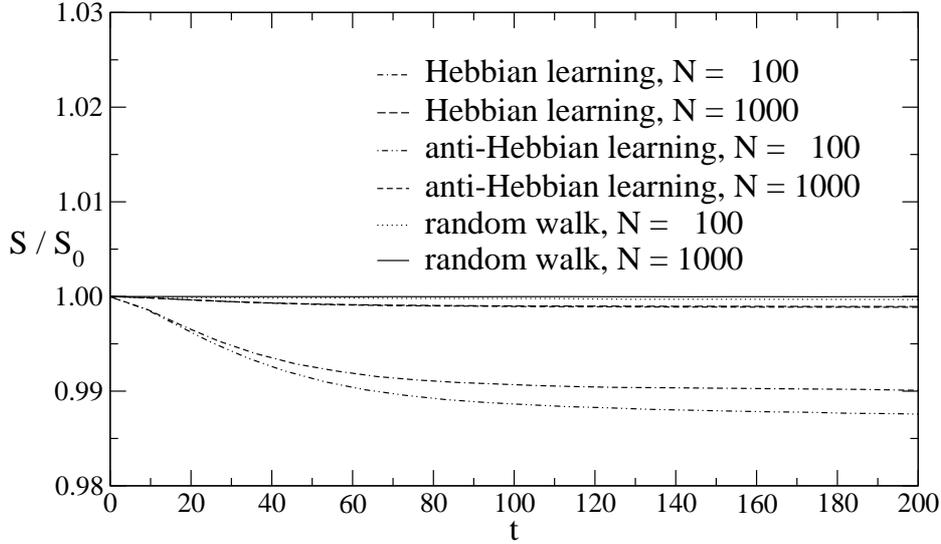}
  \caption{Time evolution of the weight distribution in the case of
    synchronization with $K=3$ and $L=5$, obtained in $100$
    simulations consisting of $100$ pairs of Tree Parity Machines.}
  \label{fig:entropy}
\end{figure}

Of course, the change of the weight distribution is also directly
visible in the relative entropy $S^A / S_0$ as shown in
figure~\ref{fig:entropy}. While the weights are always uniformly
distributed at the beginning of the synchronization process, so that
$S^A = S_0$, only the random walk learning rule preserves this
property. Otherwise $S^A$ decreases until the length of the weight
vectors reaches its stationary state after a few steps. Therefore the
transient has only little influence on the process of synchronization
and one can assume a constant value of both $Q_i$ and $S^A$.

In the limit $N \rightarrow \infty$, however, a system using Hebbian
or anti-Hebbian learning exhibits the same dynamics as observed in the
case of the random walk rule for all system sizes. Consequently, there
are two possibilities to determine the properties of neural
synchronization without interfering finite-size effects. First, one
can run simulations for the random walk learning rule and moderate
system sizes. Second, the evolution of the probabilities $p_{a,b}^i$,
which describe the distribution of the weights in two corresponding
hidden units, can be calculated iteratively for $N \rightarrow
\infty$. Both methods have been used in order to obtain the results
presented in this thesis.

\subsection{Attractive and repulsive steps}
\label{sec:steps}

As the internal representation $(\sigma_1, \sigma_2, \dots, \sigma_K)$
is not visible to other neural networks, two types of synchronization
steps are possible:
\begin{itemize}
\item For $\tau^A = \sigma_i^A = \sigma_i^B = \tau^B$ the weights of
  both corresponding hidden units are moved in the same direction. As
  long as both weights, $w_{i,j}^A$ and $w_{i,j}^B$, stay in the range
  between $-L$ and $+L$, their distance $d_{i,j} = |w_{i,j}^A -
  w_{i,j}^B|$ remains unchanged. But if one of them hits the boundary
  at $\pm L$, it is reflected, so that $d_{i,j}$ decreases by one,
  until $d_{i,j} = 0$ is reached. Therefore a sequence of these
  \emph{attractive steps} leads to full synchronization eventually.
\item If $\tau^A = \tau^B$, but $\sigma_i^A \not= \sigma_i^B$, only
  the weight vector of one hidden unit is changed. Two corresponding
  weights which have been already synchronized before, $w_{i,j}^A =
  w_{i,j}^B$, are separated by this movement, unless this is prevented
  by the boundary conditions. Consequently, this \emph{repulsive step}
  reduces the correlations between corresponding weights and impedes
  the process of synchronization.
\end{itemize}
In all other situations the weights of the $i$-th hidden unit in A's
and B's Tree Parity Machines are not modified at all.

In the limit $N \rightarrow \infty$ the effects of attractive and
repulsive steps can be described by the following equations of motion
for the probability distribution of the weights
\cite{Rosen-Zvi:2002:CBN, Rosen-Zvi:2002:MLT, Ruttor:2004:NCF}. In
attractive steps the weights perform an anisotropic diffusion
\begin{equation}
  \label{eq:pplus}
  p_{a,b}^{i+} = \frac{1}{2} \left( p^i_{a+1,b+1} + p^i_{a-1,b-1}
  \right)
\end{equation}
and move on the diagonals of a $(2 L + 1) \times (2 L + 1)$ square
lattice. Repulsive steps, instead, are equal to normal diffusion steps
\begin{equation}
  \label{eq:pminus}
  p_{a,b}^{i+} = \frac{1}{4} \left( p^i_{a+1,b} + p^i_{a-1,b} +
    p^i_{a,b+1} + p^i_{a,b-1} \right) 
\end{equation}
on the same lattice. However, one has to take the reflecting boundary
conditions into account. Therefore (\ref{eq:pplus}) and
(\ref{eq:pminus}) are only defined for $-L < a,b < +L$. Similar
equations for the weights on the boundary can be found in
appendix~\ref{chap:icalc}.

Starting from the development of the variables $p_{a,b}$ one can
calculate the change of the overlap in both types of steps. In
general, the results
\begin{equation}
  \Delta \rho_\mathrm{a} = \frac{3}{L (L + 1)} \left( 1 -
    \sum_{j=-L}^L (2 j + 2) p_{L,j} + p_{L,L} \right)
\end{equation}
for attractive steps and
\begin{equation}
  \Delta \rho_\mathrm{r} = - \frac{3}{L (L + 1)} \sum_{j=-L}^L
  \frac{j}{2} (p_{L,j} - p_{-L,j})
\end{equation}
for repulsive steps are not only functions of the current overlap, but
also depend explicitly on the probability distribution of the weights.
That is why $\Delta \rho_\mathrm{a}(\rho)$ and $\Delta
\rho_\mathrm{r}(\rho)$ are random variables, whose properties have to
be determined in simulations of finite systems or iterative
calculations for $N \rightarrow \infty$.

\begin{figure}
  \centering
  \includegraphics[scale=0.5]{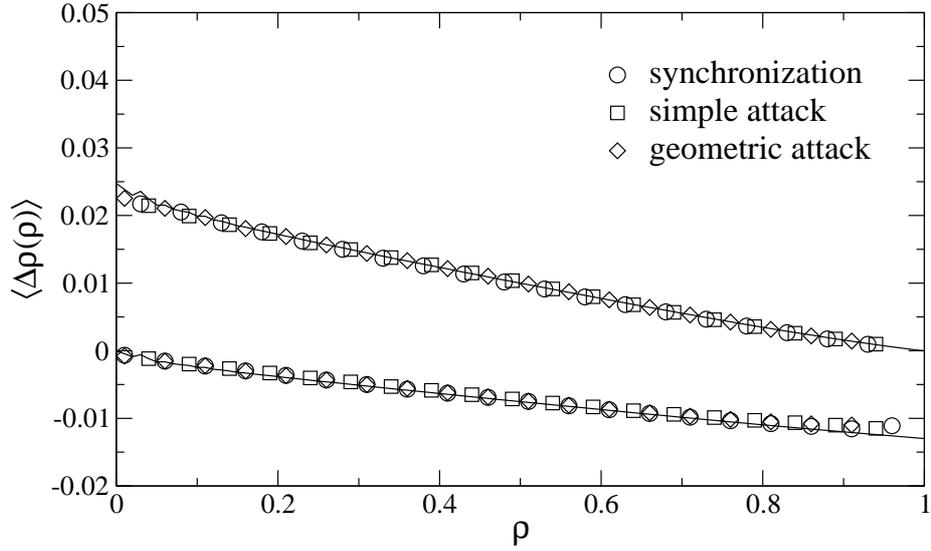}
  \caption{Effect of attractive (upper curve) and repulsive steps
    (lower curve) for $K=3$ and $L=10$. Symbols represent averages
    over $1000$ simulations using $N=100$ and the random walk learning
    rule. The line shows the corresponding result of $1000$ iterative
    calculations for synchronization in the limit $N \rightarrow
    \infty$.}
  \label{fig:effect}
\end{figure}

Figure~\ref{fig:effect} shows that each attractive step increases the
overlap on average. At the beginning of the synchronization it has its
maximum effect \cite{Ruttor:2007:DNC},
\begin{equation}
  \label{eq:atscaling}
  \Delta \rho_\mathrm{a}(\rho = 0) = \frac{12 L}{(L + 1)(2 L + 1)^2}
  \sim \frac{3}{L^2} \,,
\end{equation}
as the weights are uncorrelated,
\begin{equation}
  p_{a,b}(\rho = 0) = \frac{1}{(2 L + 1)^2} \,.
\end{equation}
But as soon as full synchronization is reached, an attractive step
cannot increase the overlap further, so that $\Delta
\rho_\mathrm{a}(\rho = 1) = 0$. Thus $\rho = 1$ is a fixed point for a
sequence of these steps.

In contrast, a repulsive step reduces a previously gained positive
overlap on average. Its maximum effect \cite{Ruttor:2007:DNC},
\begin{equation}
  \label{eq:rpscaling}
  \Delta \rho_\mathrm{r}(\rho = 1) = -\frac{3}{(L + 1)(2 L + 1)} \sim
  -\frac{3}{2 L^2} \,,
\end{equation}
is reached in the case of fully synchronized weights,
\begin{equation}
  p_{a,b}(\rho = 1) = \left\{
    \begin{array}{cl}
      (2 L + 1)^{-1} & \mbox{ for $a=b$ } \\
      0              & \mbox{ for $a \not= b$ }
    \end{array}
  \right. \,.
\end{equation}
But if the weights are uncorrelated, $\rho=0$, a repulsive step has no
effect. Hence $\rho = 0$ is a fixed point for a sequence of these
steps.

It is clearly visible in figure~\ref{fig:effect} that the results
obtained by simulations with the random walk learning rule and
iterative calculations for $N \rightarrow \infty$ are in good
quantitative agreement. This shows that both $\langle \Delta
\rho_\mathrm{a}(\rho) \rangle$ and $\langle \Delta
\rho_\mathrm{r}(\rho) \rangle$ are independent of the system size $N$.
Additionally, the choice of the synchronization algorithm does not
matter, which indicates a similar distribution of the weights for both
unidirectional and bidirectional interaction. Consequently, the
differences observed between learning and synchronization are caused
by the probabilities of attractive and repulsive steps, but not their
effects.

\begin{figure}
  \centering
  \includegraphics[scale=0.5]{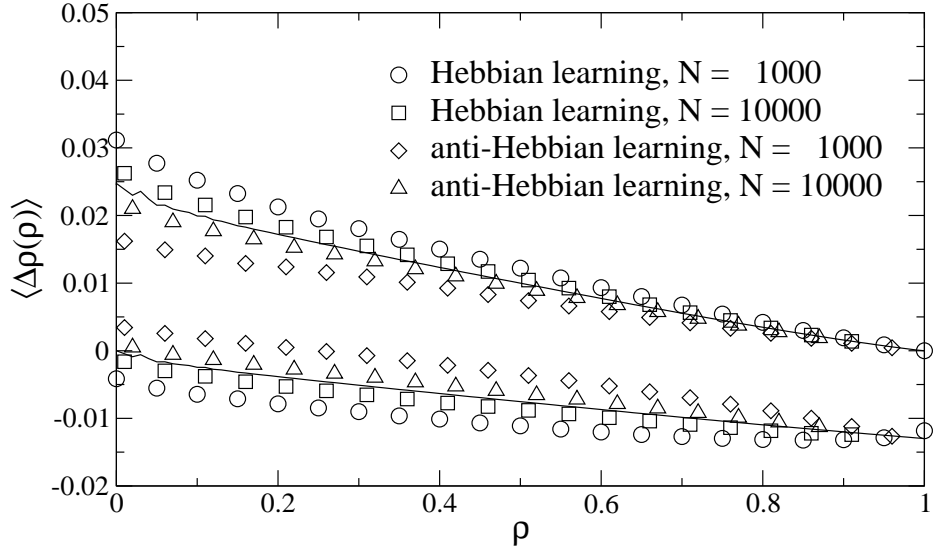}
  \caption{Effect of attractive (upper curves) and repulsive steps
    (lower curves) for different learning rules with $K=3$ and $L=10$.
    Symbols denote averages over $1000$ simulations, while the lines
    show the results of $1000$ iterative calculations.}
  \label{fig:qeffect}
\end{figure}

However, the distribution of the weights is obviously altered by
Hebbian and anti-Hebbian learning in finite systems, so that average
change of the overlap in attractive and repulsive steps is different
from the result for the random walk learning rule. This is clearly
visible in figure~\ref{fig:qeffect}. In the case of the Hebbian
learning rule the effect of both types of steps is enhanced, but for
anti-Hebbian learning it is reduced. It is even possible that an
repulsive step has an attractive effect on average, if the overlap
$\rho$ is small. This explains why one observes finite-size effects in
the case of large $L / \sqrt{N}$ \cite{Mislovaty:2002:SKE}.

\begin{figure}
  \centering
  \includegraphics[scale=0.5]{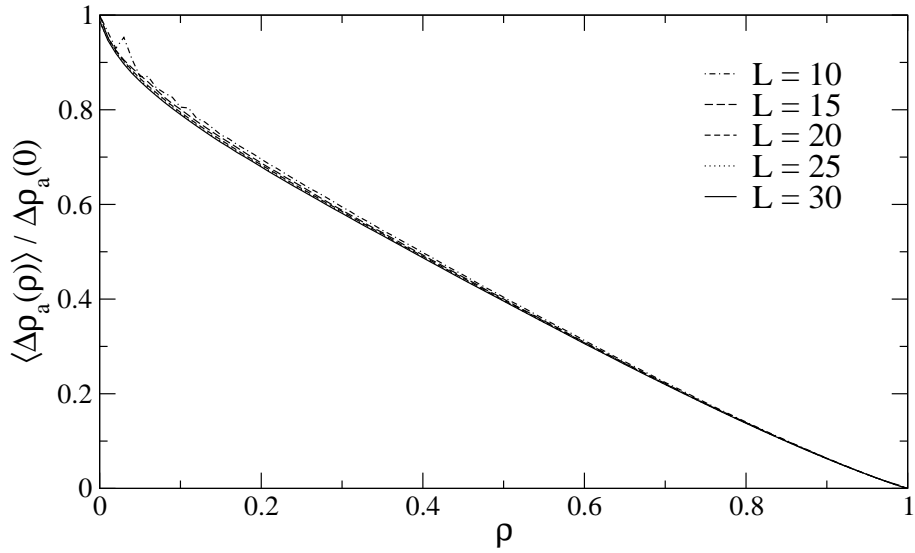}
  \caption{Scaling behavior of the average step size $\langle \Delta
    \rho_\mathrm{a} \rangle$ for attractive steps.  These results were
    obtained in $1000$ iterative calculations for $K=3$ and $N
    \rightarrow \infty$.}
  \label{fig:atscaling}
\end{figure}

\begin{figure}
  \centering
  \includegraphics[scale=0.5]{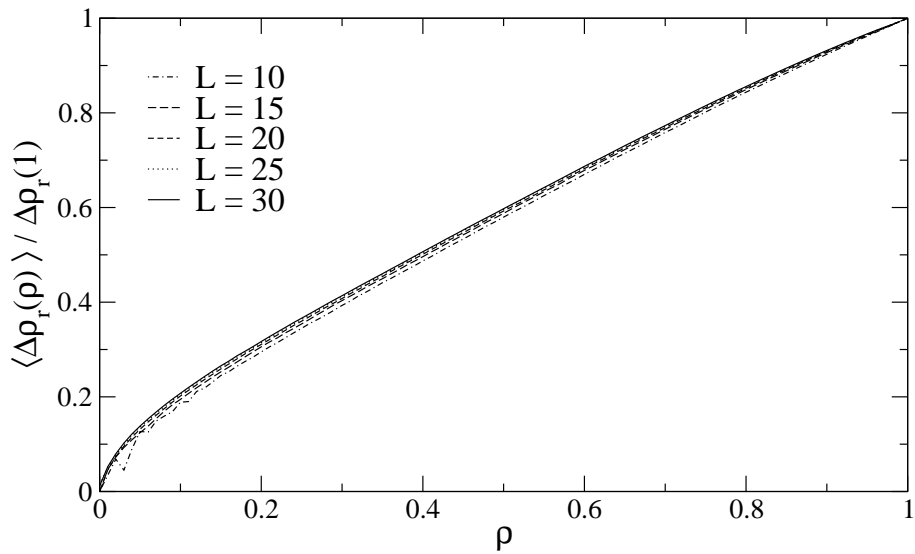}
  \caption{Scaling behavior of the average step size $\langle \Delta
    \rho_\mathrm{r} \rangle$ for repulsive steps. These results were
    obtained in $1000$ iterative calculations for $K=3$ and $N
    \rightarrow \infty$.}
  \label{fig:rpscaling}
\end{figure}

Using the equations (\ref{eq:atscaling}) and (\ref{eq:rpscaling}) one
can obtain the rescaled quantities $\langle \Delta
\rho_\mathrm{a}(\rho) \rangle / \Delta \rho_\mathrm{a}(0)$ and
$\langle \Delta \rho_\mathrm{r}(\rho) \rangle / \Delta
\rho_\mathrm{r}(1)$. They become asymptotically independent of the
synaptic depth $L$ in the limit $L \rightarrow \infty$ as shown in
figure~\ref{fig:atscaling} and figure~\ref{fig:rpscaling}. Therefore
these two scaling functions together with $\Delta \rho_\mathrm{a}(0)$
and $\Delta \rho_\mathrm{r}(1)$ are sufficient to describe the effect
of attractive and repulsive steps \cite{Ruttor:2007:DNC}.

\section{Transition probabilities}
\label{sec:prob}

While $\langle \Delta \rho_\mathrm{a} \rangle$ and $\langle \Delta
\rho_\mathrm{r} \rangle$ are identical for synchronization and
learning, the probabilities of attractive and repulsive steps depend
on the type of interaction between the neural networks. Therefore
these quantities are important for the differences between partners
and attackers in neural cryptography.

A repulsive step can only occur if two corresponding hidden units have
different $\sigma_i$. The probability for this event is given by the
well-known generalization error \cite{Engel:2001:SML}
\begin{equation}
  \label{eq:generr}
  \epsilon_i = \frac{1}{\pi} \arccos \rho_i
\end{equation}
of the perceptron. However, disagreeing hidden units alone are not
sufficient for a repulsive step, as the weights of all neural networks
are only changed if $\tau^A = \tau^B$. Therefore the probability of a
repulsive step is given by
\begin{equation}
  P_\mathrm{r} = P(\sigma_i^A \not= \sigma_i^{B/E} | \tau^A = \tau^B)
  \,,
\end{equation}
after possible corrections of the output bits have been applied in the
case of advanced learning algorithms. Similarly, one finds
\begin{equation}
  P_\mathrm{a} = P(\tau^A = \sigma_i^A = \sigma_i^{B/E} | \tau^A =
  \tau^B)
\end{equation}
for the probability of attractive steps.
 
\subsection{Simple attack}

In the case of the simple attack, the outputs $\sigma_i^E$ of E's Tree
Parity Machine are not corrected before the application of the
learning rule and the update of the weights occurs independent of
$\tau^E$, as mutual interaction is not possible. Therefore a repulsive
step in the $i$-th hidden unit occurs with probability
\cite{Ruttor:2004:NCF}
\begin{equation}
  P_\mathrm{r}^E = \epsilon_i \,.
\end{equation}
But if two corresponding hidden units agree on their output
$\sigma_i$, this does not always lead to an attractive step, because
$\sigma_i = \tau$ is another necessary condition for an update of the
weights. Thus the probability of an attractive step is given by
\cite{Ruttor:2007:DNC}
\begin{equation}
  P_\mathrm{a}^E = \frac{1}{2} (1 - \epsilon_i)
\end{equation}
for $K > 1$. In the special case $K=1$, however, $\sigma_i = \tau$ is
always true, so that this type of steps occurs with double frequency:
$P_\mathrm{a}^E = 1 - \epsilon_i$.

\subsection{Synchronization}

In contrast, mutual interaction is an integral part of bidirectional
synchronization. When an odd number of hidden units disagrees on the
output, $\tau^A \not= \tau^B$ signals that adjusting the weights would
have a repulsive effect on at least one of the weight vectors.
Therefore A and B skip this synchronization step.

But when an even number of hidden units disagrees on the output, the
partners cannot detect repulsive steps by comparing $\tau^A$ and
$\tau^B$. Additionally, identical internal representations in both
networks are more likely than two or more different output bits
$\sigma_i^A \not= \sigma_i^B$, if there are already some correlations
between the Tree Parity Machines. Consequently, the weights are
updated if $\tau^A = \tau^B$.

In the case of identical overlap in all $K$ hidden units, $\epsilon_i
= \epsilon$, the probability of this event is given by
\begin{equation}
  P_\mathrm{u} = P(\tau^A = \tau^B) = \sum_{i=0}^{K/2} {K \choose 2 i}
  (1 - \epsilon)^{K - 2 i} \, \epsilon^{2 i} \,.
\end{equation}
Of course, only attractive steps are possible if two perceptrons learn
from each other ($K=1$). But for synchronization of Tree Parity
Machines with $K > 1$, the probabilities of attractive and repulsive
are given by:
\begin{eqnarray}
  \label{eq:pab}
  P_\mathrm{a}^B &=& \frac{1}{2 P_\mathrm{u}} \sum_{i=0}^{(K-1)/2} {K
    - 1 \choose 2 i} (1 - \epsilon)^{K - 2 i} \, \epsilon^{2 i} \,, \\
  \label{eq:prb}
  P_\mathrm{r}^B &=& \frac{1}{P_\mathrm{u}} \sum_{i=1}^{K/2} {K - 1
    \choose 2 i - 1} (1 - \epsilon)^{K - 2 i} \, \epsilon^{2 i} \,.
\end{eqnarray}
In the case of three hidden units ($K = 3$), which is the usual choice
for the neural key-exchange protocol, this leads to
\cite{Ruttor:2004:NCF, Ruttor:2007:DNC}
\begin{eqnarray}
  P_\mathrm{a}^B &=& \frac{1}{2} \, \frac{(1 - \epsilon)^3 + (1 -
    \epsilon) \epsilon^2}{(1 - \epsilon)^3 + 3 (1 - \epsilon)
    \epsilon^2}\,, \\
  P_\mathrm{r}^B &=& \frac{2 (1 - \epsilon) \epsilon^2}{(1 -
    \epsilon)^3 + 3 (1 - \epsilon) \epsilon^2} \,.
\end{eqnarray}

\begin{figure}
  \centering
  \includegraphics[scale=0.5]{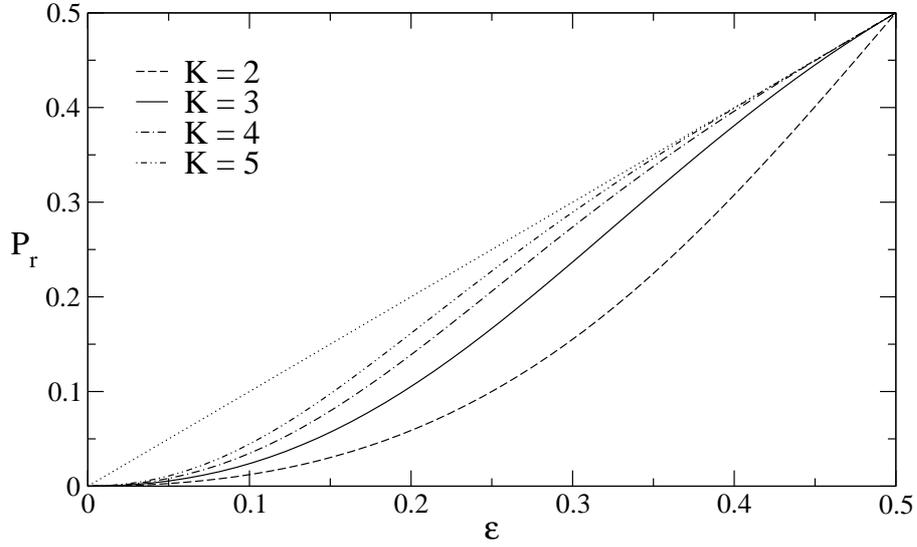}
  \caption{Probability $P_\mathrm{r}^B(\rho)$ of repulsive steps for
    synchronization with mutual interaction under the condition
    $\tau^A=\tau^B$. The dotted line shows $P_\mathrm{r}^E(\rho)$ for
    a simple attack.}
  \label{fig:prsync}
\end{figure}

Figure~\ref{fig:prsync} shows that repulsive steps occur more
frequently in E's Tree Parity Machine than in A's or B's for equal
overlap $0 < \rho < 1$. That is why the partners A and B have a clear
advantage over a simple attacker in neural cryptography. But this
difference becomes smaller and smaller with increasing $K$.
Consequently, a large number of hidden units is detrimental for the
security of the neural key-exchange protocol against the simple
attack.

\subsection{Geometric attack}

\begin{figure}
  \centering
  \includegraphics[scale=0.5]{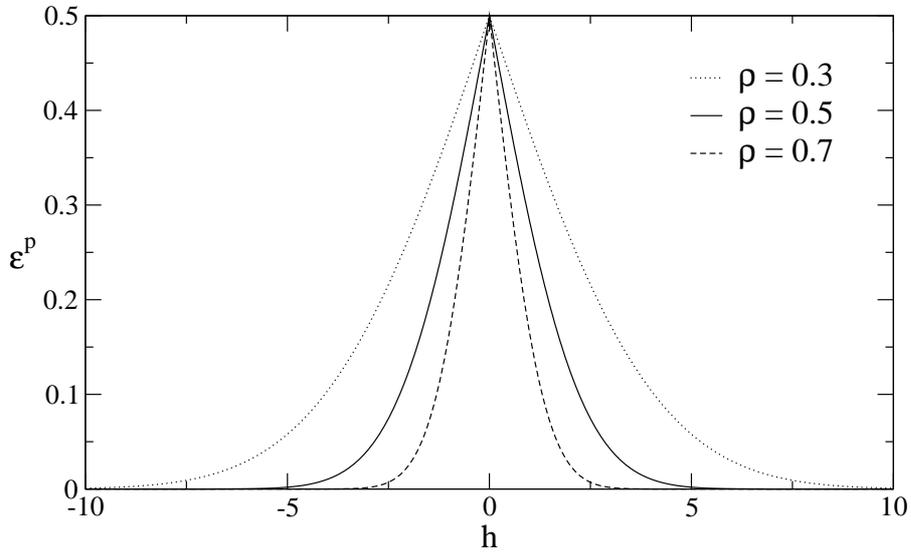}
  \caption{Prediction error $\epsilon_i^\mathrm{p}$ as a function of
    the local field $h_i^E$ for different values of the overlap
    $\rho_i^{AE}$ and $Q_i=1$.}
  \label{fig:prgeo}
\end{figure}

However, E can do better than simple learning by taking the local
field into account. Then the probability of $\sigma_i^E \not=
\sigma_i^A$ is given by the prediction error \cite{Ein-Dor:1999:CPN}
\begin{equation}
  \label{eq:perr}
  \epsilon^\mathrm{p}_i = \frac{1}{2} \left[ 1 - \mathrm{erf} \left(
      \frac{\rho_i}{\sqrt{2(1-\rho_i^2)}} \frac{|h_i|}{\sqrt{Q_i}}
    \right) \right]
\end{equation}
of the perceptron, which depends not only on the overlap $\rho_i$, but
also on the absolute value $|h_i^E|$ of the local field. This quantity
is a strictly decreasing function of $|h_i^E|$ as shown in
figure~\ref{fig:prgeo}. Therefore the geometric attack is often able
to find the hidden unit with $\sigma_i^E \not= \sigma_i^A$ by
searching for the minimum of $|h_i^E|$. If only the $i$-th hidden unit
disagrees and all other have $\sigma_j^E = \sigma_j^A$, the
probability for a successful correction of the internal representation
by using the geometric attack is given by \cite{Ruttor:2006:GAN}
\begin{equation}
  \label{eq:pg}
  P_\mathrm{g} = \int_{0}^{\infty} \prod_{j \neq i} \left(
    \int_{h_i}^{\infty} \frac{2}{\sqrt{2 \pi Q_j}} \frac{1 -
      \epsilon^\mathrm{p}_j}{1 - \epsilon_j} \, e^{-\frac{h_j^2}{2
        Q_j}} \, \mathrm{d}h_j \right) \frac{2}{\sqrt{2 \pi Q_i}}
  \frac{\epsilon^\mathrm{p}_i}{\epsilon_i} \, e^{-\frac{h_i^2}{2
      Q_i}}\, \mathrm{d}h_i \,.
\end{equation}
In the case of identical order parameters $Q=Q_j^E$ and $R=R_j^{AE}$
this equation can be easily extended to $k$ out of $K$ hidden units
with different outputs $\sigma_j^A \not= \sigma_j^E$. Then the
probability for successful correction of $\sigma_i^E \not= \sigma_i^A$
is given by
\begin{eqnarray}
  P_k^{+} &=& \int_{0}^{\infty} \left( \frac{2}{\sqrt{2 \pi Q}}
  \right)^K \left( \int_{h_i}^{\infty} \frac{1 -
      \epsilon^\mathrm{p}(h)}{1 - \epsilon} \, e^{-\frac{h^2}{2 Q}} \,
    \mathrm{d}h \right)^{K-k} \nonumber\\ &\times& 
  \left( \int_{h_i}^{\infty} \frac{\epsilon^\mathrm{p}(h)}{\epsilon}
    \, e^{-\frac{h^2}{2 Q}} \, \mathrm{d}h \right)^{k-1}
  \frac{\epsilon^\mathrm{p}(h_i)}{\epsilon} \, e^{-\frac{h_i^2}{2 Q}}
  \, \mathrm{d}h_i \,.
\end{eqnarray}
Using a similar equation the probability for an erroneous correction
of $\sigma_i^E = \sigma_i^A$ can be calculated, too:
\begin{eqnarray}
  P_k^{-} &=& \int_{0}^{\infty} \left( \frac{2}{\sqrt{2 \pi Q}}
  \right)^K \left( \int_{h_i}^{\infty} \frac{1 -
      \epsilon^\mathrm{p}(h)}{1 - \epsilon} \, e^{-\frac{h^2}{2 Q}} \,
    \mathrm{d}h \right)^{K-k-1} \nonumber\\ &\times& 
  \left( \int_{h_i}^{\infty} \frac{\epsilon^\mathrm{p}(h)}{\epsilon}
    \, e^{-\frac{h^2}{2 Q}} \, \mathrm{d}h \right)^{k}
  \frac{1 - \epsilon^\mathrm{p}(h_i)}{1 - \epsilon} \,
  e^{-\frac{h_i^2}{2 Q}} \, \mathrm{d}h_i \,.
\end{eqnarray}

Taking all possible internal representations of A's and E's neural
networks into account, the probability of repulsive steps consists of
three parts in the case of the geometric attack.
\begin{itemize}
\item If the number of hidden units with $\sigma_i^E \not= \sigma_i^A$
  is even, no geometric correction happens at all. This is similar to
  bidirectional synchronization, so that one finds
  \begin{equation}
    P_\mathrm{r,1}^E = \sum_{i=1}^{K/2} {K - 1 \choose 2 i - 1} (1 -
    \epsilon)^{K - 2 i} \, \epsilon^{2 i} \,.
  \end{equation}
\item It is possible that the hidden unit with the minimum $|h_i^E|$
  has the same output as its counterpart in A's Tree Parity Machine.
  Then the geometric correction increases the deviation of the
  internal representations. The second part of $P_\mathrm{r}^E$ takes
  this event into account:
  \begin{equation}
    P_\mathrm{r,2}^E = \sum_{i=1}^{K/2} {K - 1 \choose 2 i - 1}
    P_{2 i - 1}^{-} \, (1 - \epsilon)^{K - 2 i + 1} \, \epsilon^{2 i -
      1} \,.
  \end{equation}
\item Similarly the geometric attack does not fix a deviation in the
  $i$-th hidden unit, if the output of another one is flipped instead.
  Indeed, this causes a repulsive step with probability
  \begin{equation}
    P_\mathrm{r,3}^E = \sum_{i=0}^{(K-1)/2} {K - 1 \choose 2 i} (1 -
    P_{2 i + 1}^{+}) (1 - \epsilon)^{K - 2 i - 1} \, \epsilon^{2 i +
      1} \,.
  \end{equation}
\end{itemize}
Thus the probabilities of attractive and repulsive steps in the $i$-th
hidden unit for $K > 1$ and identical order parameters are given by
\begin{eqnarray}
  P_\mathrm{a}^E &=& \frac{1}{2} \left( 1 - \sum_{j=1}^3
    P_\mathrm{r,j}^E \right) \,, \\ 
  P_\mathrm{r}^E &=& \sum_{j=1}^3 P_\mathrm{r,j}^E \,.
\end{eqnarray}
In the case $K = 1$, however, only attractive steps occur, because the
algorithm of the geometric attack is then able to correct all
deviations. And especially for $K = 3$ one can calculate these
probabilities using (\ref{eq:pg}) instead of the general equations,
which yields \cite{Ruttor:2006:GAN}
\begin{eqnarray}
  P_\mathrm{a}^E &=& \frac{1}{2} (1 + 2 P_\mathrm{g}) (1 - \epsilon)^2
  \epsilon + \frac{1}{2} (1 - \epsilon)^3 \nonumber \\
  &+& \frac{1}{2} (1 - \epsilon) \epsilon^2 + \frac{1}{6} \epsilon^3
  \, \\
  P_\mathrm{r}^E &=& 2 (1 - P_\mathrm{g}) (1 - \epsilon)^2 \epsilon +
  2 (1 - \epsilon) \epsilon^2 + \frac{2}{3} \epsilon^3 \,.
\end{eqnarray}

\begin{figure}
  \centering
  \includegraphics[scale=0.5]{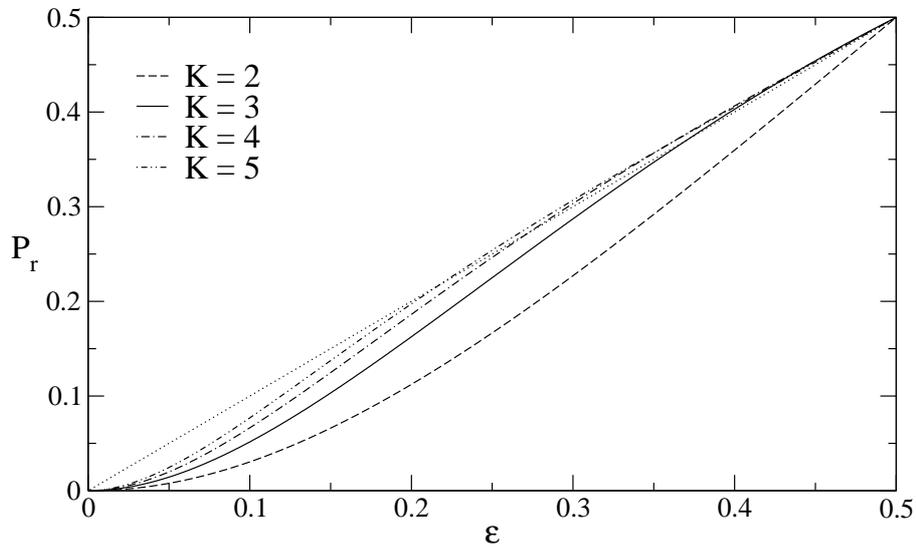}
  \caption{Probability of repulsive steps for an attacker using the
    geometric attack. The dotted line shows $P_\mathrm{r}$ for the
    simple attack.}
  \label{fig:prgeox}
\end{figure}

\begin{figure}
  \centering
  \includegraphics[scale=0.5]{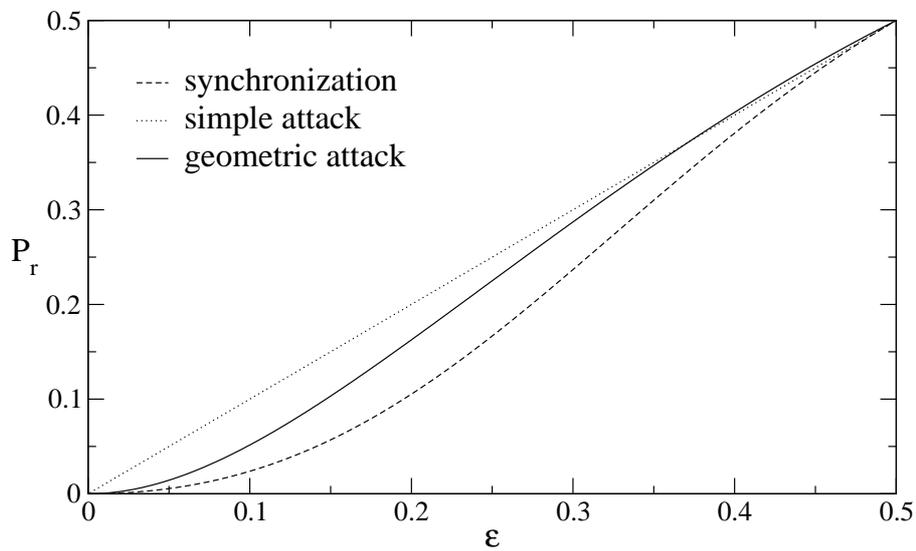}
  \caption{Probability of repulsive steps for Tree Parity Machines
    with $K=3$ hidden units and different types of interaction.}
  \label{fig:prcmp}
\end{figure}

As shown in figure~\ref{fig:prgeox} $P_\mathrm{r}^E$ grows, if the
number of hidden units is increased. It is even possible that the
geometric attack performs worse than the simple attack at the
beginning of the synchronization process ($\epsilon \approx 0.5$).
While this behavior is similar to that observed in
figure~\ref{fig:prsync}, $P_\mathrm{r}^E$ is still higher than
$P_\mathrm{r}^B$ for identical $K$. Consequently, even this advanced
algorithm for unidirectional learning has a disadvantage compared to
bidirectional synchronization, which is clearly visible in
figure~\ref{fig:prcmp}.

\section{Dynamics of the weights}
\label{sec:rwmodel}

In each attractive step corresponding weights of A's and B's Tree
Parity Machines move in the same direction, which is chosen with equal
probability in the case of the random walk learning rule. The same is
true for Hebbian and anti-Hebbian learning in the limit $N \rightarrow
\infty$ as shown in section~\ref{sec:wdist}. Of course, repulsive
steps disturb this synchronization process. But for small overlap they
have little effect, while they occur only seldom in the case of large
$\rho$. That is why one can neglect repulsive steps in some situations
and consequently describe neural synchronization as an ensemble of
random walks with reflecting boundaries, driven by pairwise identical
random signals \cite{Kinzel:2002:INN, Kinzel:2003:DGI}.

\begin{figure}[h]
  \centering
  \includegraphics[scale=0.5]{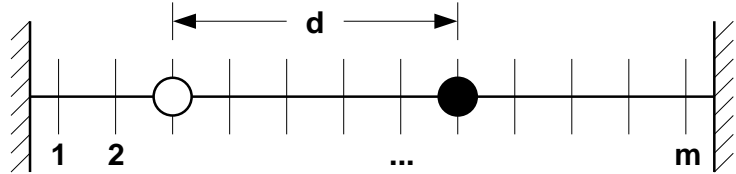}
  \caption{Random walks with reflecting boundaries.}
  \label{fig:rwalk}
\end{figure}

This leads to a simple model for a pair of weights, which is shown in
figure~\ref{fig:rwalk} \cite{Ruttor:2004:SRW}. Two random walks
corresponding to $w_{i,j}^A$ and $w_{i,j}^B$ can move on a
one-dimensional line with $m = 2 L + 1$ sites. In each step a
direction, either left or right, is chosen randomly. Then the random
walkers move in this direction. If one of them hits the boundary, it
is reflected, so that its own position remains unchanged. As this does
not affect the other random walker, which moves towards the first one,
the distance $d$ between them shrinks by $1$ at each reflection.
Otherwise $d$ remains constant.

The most important quantity of this model is the synchronization time
$T$ of the two random walkers, which is defined as the number of steps
needed to reach $d = 0$ starting with random initial positions. In
order to calculate the mean value $\langle T \rangle$ and analyze the
probability distribution $P(T=t)$, this process is divided into
independent parts, each of them with constant distance $d$. Their
duration $S_{d,z}$ is given by the time between two reflections. Of
course, this quantity depends not only on the distance $d$, but also
on the initial position $z = L + \mathrm{min}(w_{i,j}^A, w_{i,j}^B) + 1$
of the left random walker.

\subsection{Waiting time for a reflection}

If the first move is to the right, a reflection only occurs for $z = m
- d$. Otherwise, the synchronization process continues as if the
initial position had been $z + 1$. In this case the average number of
steps with distance $d$ is given by $\langle S_{d,z+1} + 1 \rangle$.
Similarly, if the two random walkers move to the left in the first
step, this quantity is equal to $\langle S_{d,z-1} + 1 \rangle$.
Averaging over both possibilities leads to the following difference
equation \cite{Ruttor:2004:SRW}:
\begin{equation}
  \label{eq:rwsync}
  \langle S_{d,z} \rangle = \frac{1}{2} \langle S_{d,z-1} \rangle +
  \frac{1}{2} \langle S_{d,z+1} \rangle + 1 \,.
\end{equation}

Reflections are only possible, if the current position $z$ is either
$1$ or $m - d$. In both situations $d$ changes with probability
$\frac{1}{2}$ in the next step, which is taken into account by using
the boundary conditions
\begin{equation}
  S_{d,0} = 0 \quad \mbox{and} \quad S_{d,m-d+1} = 0 \,.
\end{equation}

As (\ref{eq:rwsync}) is identical to the classical ruin problem
\cite{Feller:1968:IPT}, its solution is given by
\begin{equation}
  \label{eq:rwsol}
  \langle S_{d,z} \rangle = (m - d + 1)z - z^2 \,.
\end{equation}

In order to calculate the standard deviation of the synchronization
time an additional difference equation,
\begin{equation}
  \label{eq:rwvar}
  \langle S_{d,z}^2 \rangle = \frac{1}{2} \langle (S_{d,z-1} + 1)^2
  \rangle + \frac{1}{2} \langle (S_{d,z+1} + 1)^2 \rangle \,,
\end{equation}
is necessary, which can be obtained in a similar manner as equation
(\ref{eq:rwsync}). Using both (\ref{eq:rwsol}) and (\ref{eq:rwvar})
leads to the relation \cite{Ruttor:2004:SRW}
\begin{eqnarray}
  \langle S_{d,z}^2 \rangle - \langle S_{d,z} \rangle^2
  &=& \frac{\langle S_{d,z-1}^2 \rangle - \langle S_{d,z-1}
    \rangle^2}{2} + \frac{\langle S_{d,z+1}^2 \rangle - \langle
    S_{d,z+1} \rangle^2}{2} \nonumber \\
  &+& (m - d + 1 - 2 z)^2
\end{eqnarray}
for the variance of $S_{d,z}$. Applying a Z-transformation finally
yields the solution
\begin{equation}
  \langle S_{d,z}^2 \rangle - \langle S_{d,z} \rangle^2 = \frac{(m - d
    + 1 - z)^2 + z^2 - 2}{3} \, \langle S_{d,z} \rangle \,.
\end{equation}

While the first two moments of $S_{d,z}$ are sufficient to calculate
the mean value and the standard deviation of $T$, the probability
distribution $P(S_{d,z}=t)$ must be known in order to further analyze
$P(T=t)$. For that purpose a result known from the solution of the
classical ruin problem \cite{Feller:1968:IPT} is used: The probability
that a fair game ends with the ruin of one player in time step $t$ is
given by
\begin{equation}
  \label{eq:u}
  u(t) = \frac{1}{a} \sum_{k=1}^{a-1} \sin \left( \frac{k \pi
      z}{a} \right) \left[ \sin \left( \frac{k \pi}{a} \right) + \sin
    \left( k \pi - \frac{k \pi}{a} \right) \right] \left[ \cos \left(
      \frac{k \pi}{a} \right) \right]^{t-1} \,.
\end{equation}
In the random walk model $a - 1 = m - d$ denotes the number of
possible positions for two random walkers with distance $d$. And
$u(t)$ is the probability distribution of the random variable
$S_{d,z}$. As before, $z = L + \mathrm{min}(w_{i,j}^A, w_{i,j}^B) + 1$
denotes the initial position of the left random walker.

\subsection{Synchronization of two random walks}

With these results one can determine the properties of the
synchronization time $T_{d,z}$ for two random walks starting at
position $z$ and distance $d$. After the first reflection at time
$S_{d,z}$ one of the random walkers is located at the boundary. As the
model is symmetric, both possibilities $z = 1$ or $z = m - d$ are
equal. Hence the second reflection takes place after $S_{d,z} +
S_{d-1,1}$ steps and, consequently, the total synchronization time is
given by
\begin{equation}
  \label{eq:rwt}
  T_{d,z} = S_{d,z} + \sum_{j=1}^{d-1} S_{j,1} \,.
\end{equation}
Using (\ref{eq:rwsol}) leads to \cite{Ruttor:2004:SRW}
\begin{equation}
  \langle T_{d,z} \rangle = (m - d + 1) z - z^2 + \, \frac{1}{2} (d -
  1)(2 m - d)
\end{equation}
for the expectation value of this random variable. In a similar manner
one can calculate the variance of $T_{d,z}$, because the parts of the
synchronization process are mutually independent.

Finally, one has to average over all possible initial conditions in
order to determine the mean value and the standard deviation of the
synchronization time $T$ for randomly chosen starting positions of the
two random walkers \cite{Ruttor:2004:SRW}:
\begin{eqnarray}
  \label{eq:rwtime}
  \langle T \rangle &=& \frac{2}{m^2} \sum_{d=1}^{m-1}
  \sum_{z=1}^{m-d} \langle T_{d,z} \rangle = \frac{(m-1)^2}{3} +
  \frac{m-1}{3m} \,, \\
  \langle T^2 \rangle &=& \frac{2}{m^2} \sum_{d=1}^{m-1}
  \sum_{z=1}^{m-d} \langle T_{d,z}^2 \rangle \nonumber \\
  &=& \frac{17 m^5 - 51 m^4 + 65 m^3 - 45 m^2 + 8 m + 6}{90 m} \,.
\end{eqnarray}

\begin{figure}
  \centering
  \includegraphics[scale=0.5]{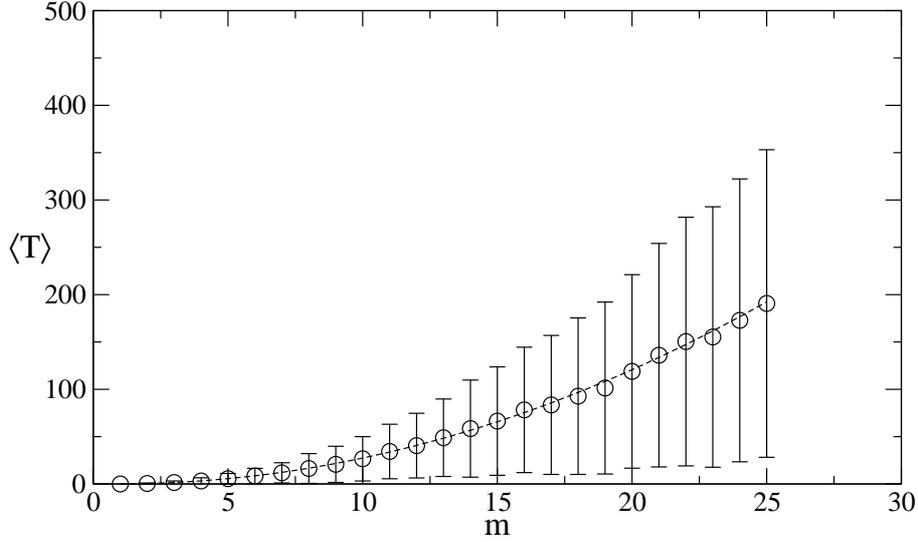}
  \caption{Synchronization time of two random walks as a function of
    the system size $m = 2 L + 1$. Error bars denote the standard
    deviation observed in $1000$ simulations. The analytical solution
    (\ref{eq:rwtime}) is plotted as dashed curve.}
  \label{fig:rwtime}
\end{figure}

Thus the average number of attractive steps required to reach a
synchronized state, which is shown in figure~\ref{fig:rwtime},
increases nearly proportional to $m^2$. In particular for large system
sizes $m$ the asymptotic behavior is given by
\begin{equation}
  \label{eq:rwmean}
  \langle T \rangle \sim \frac{1}{3} m^2 \sim \frac{4}{3} L^2 \,.
\end{equation}
As shown later in section~\ref{sec:async} this result is consistent
with the scaling behavior $\langle t_\mathrm{sync} \rangle \propto
L^2$ found in the case of neural synchronization
\cite{Mislovaty:2002:SKE}.

In numerical simulations, both for random walks and neural networks,
large fluctuations of the synchronization time are observed. The
reason for this effect is that not only the mean value but also the
standard deviation of $T$ \cite{Ruttor:2004:SRW},
\begin{equation}
  \label{eq:rwdev}
  \sigma_T = \sqrt{\frac{7 m^6 - 11 m^5 - 15 m^4 + 55 m^3 - 72 m^2 +
      46 m - 10}{90 m^2}} \,,
\end{equation}
increases with the extension $m$ of the random walks. A closer look at
(\ref{eq:rwmean}) and (\ref{eq:rwdev}) reveals that $\sigma_T$ is
asymptotically proportional to $\langle T \rangle$:
\begin{equation}
  \sigma_T \sim \sqrt{\frac{7}{10}} \, \langle T \rangle \,.
\end{equation}
Therefore the relative fluctuations $\sigma_T / \langle T \rangle$ are
nearly independent of $m$ and not negligible. Consequently, one cannot
assume a typical synchronization time, but has to take the full
distribution $P(T=t)$ into account.

\subsection{Probability distribution}

As $T_{d,z}$ is the sum over $S_{i,j}$ for each distance $i$ from $d$
to $1$ according to (\ref{eq:rwt}), its probability distribution
$\mathrm{P}(T_{d,z}=t)$ is a convolution of $d$ functions $u(t)$
defined in (\ref{eq:u}). The convolution of two different geometric
sequences $b_n = b^n$ and $c_n = c^n$ is itself a linear combination
of these sequences:
\begin{equation}
  \label{eq:convolution}
  b_n \ast c_n = \sum_{j=1}^{n-1} b^j c^{n-j} = \frac{c}{b-c} \, b_n +
  \frac{b}{c-b} \, c_n \,.
\end{equation}
Thus $\mathrm{P}(T_{d,z}=t)$ can be written as a sum over geometric
sequences, too:
\begin{equation}
  \mathrm{P}(T_{d,z}=t) = \sum_{a=m - d + 1}^{m} \sum_{k=1}^{a-1}
  c^{d,z}_{a,k} \left[ \cos \left( \frac{k \pi}{a} \right)
  \right]^{t-1} \,.
\end{equation}

In order to obtain $\mathrm{P}(T=t)$ for random initial conditions,
one has to average over all possible starting positions of both
random walkers. But even this result
\begin{equation}
  \mathrm{P}(T=t) = \frac{2}{m^2} \sum_{d=1}^{m-1} \sum_{z=1}^{m-d}
  \mathrm{P}(T_{d,z}=t)
\end{equation}
can be written as a sum over a lot of geometric sequences:
\begin{equation}
  \mathrm{P}(T=t) = \sum_{a=2}^{m} \sum_{k=1}^{a-1} c_{a,k} \left[
    \cos \left( \frac{k \pi}{a} \right) \right]^{t-1} \,.
\end{equation}

\begin{figure}
  \centering
  \includegraphics[scale=0.5]{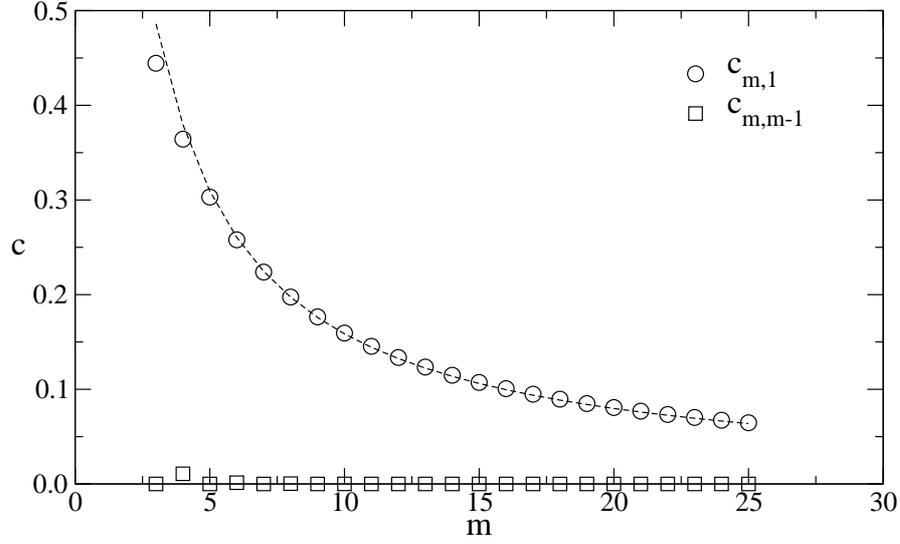}
  \caption{Value of the coefficients $c_{m,1}$ and $c_{m,m-1}$ as a
    function of $m$. The approximation given in (\ref{eq:rwapprox}) is
    shown as dashed curve.}
  \label{fig:rwcoeff}
\end{figure}

For long times, however, only the terms with the largest absolute
value of the coefficient $\cos (k \pi / a)$ are relevant, because the
others decline exponentially faster. Hence one can neglect them in the
limit $t \rightarrow \infty$, so that the asymptotic behavior of the
probability distribution is given by
\begin{equation} 
  \label{eq:rwdist}
  \mathrm{P}(T=t) \sim [ c_{m,1} + (-1)^{t-1} c_{m,m-1} ]
  \left[ \cos \left( \frac{\pi}{m} \right) \right]^{t-1} \,.
\end{equation}
The two coefficients $c_{m,1}$ and $c_{m,m-1}$ in this equation can be
calculated using (\ref{eq:convolution}). This leads to the following
result \cite{Ruttor:2004:SRW}, which is shown in
figure~\ref{fig:rwcoeff}:
\begin{eqnarray}
  c_{m,1} &\!\!=\!\!& \frac{\sin^2 (\pi / m)}{m^2 m!} \sum_{d=1}^{m-1}
  \frac{2^{d+1} (m - d)!}{1 - \delta_{d,1} \cos (\pi / m)} \nonumber \\
  &\!\!\times\!\!& \prod_{a=m-d+1}^{m-1} \sum_{k=1}^{a-1} \frac{\sin^2
    (k \pi / 2)}{\cos (\pi / m) - \cos (k \pi / a)} \, \frac{\sin^2 (k
    \pi / a)}{1 - \delta_{a,m-d+1} \cos (k \pi / a)} , \\
  c_{m,m-1} &\!\!=\!\!& \frac{\sin^2 (\pi / m) \cos^2 (m \pi / 2)}{m^2
    m!} \sum_{d=1}^{m-1} (-1)^{d-1} \frac{2^{d+1} (m - d)!}{1 +
    \delta_{d,1} \cos (\pi / m)} \nonumber \\
  &\!\!\times\!\!& \prod_{a=m-d+1}^{m-1} \sum_{k=1}^{a-1} \frac{\sin^2
    (k \pi / 2)}{\cos (\pi / m) + \cos (k \pi / a)} \, \frac{\sin^2 (k
    \pi / a)}{1 - \delta_{a,m-d+1} \cos (k \pi / a)} .
\end{eqnarray}
As the value of $c_{m,m-1}$ is given by an alternating sum, this
coefficient is much smaller than $c_{m,1}$. Additionally, it is
exactly zero for odd values of $m$ because of the factor $\cos^2 (m
\pi / 2)$. The other coefficient $c_{m,1}$, however, can be
approximated by \cite{Ruttor:2004:SRW}
\begin{equation}
  \label{eq:rwapprox}
  c_{m,1} \approx 0.324 \, m \left[ 1 - \cos \left( \frac{\pi}{m}
    \right) \right]
\end{equation}
for $m \gg 1$, which is clearly visible in figure~\ref{fig:rwcoeff},
too.

\begin{figure}
  \centering
  \includegraphics[scale=0.5]{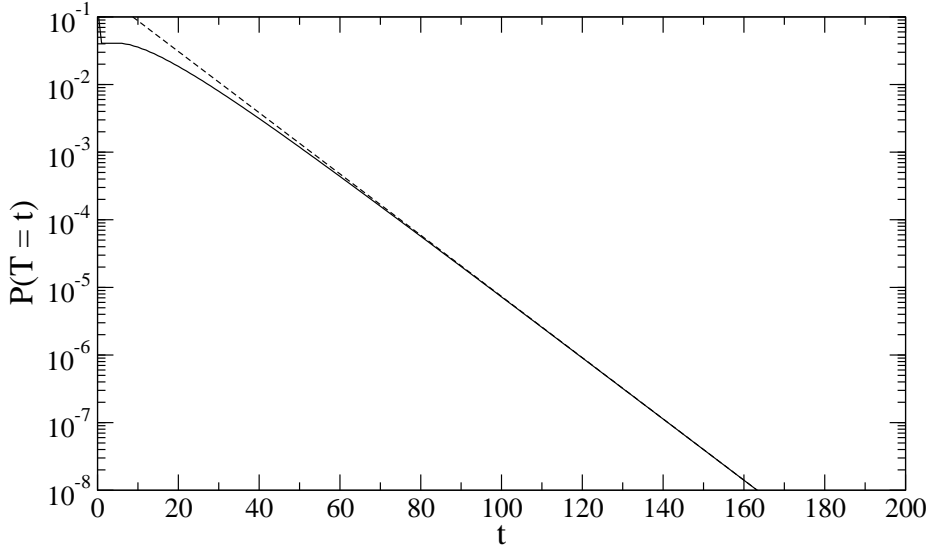}
  \caption{Probability distribution $\mathrm{P}(T=t)$ of the
    synchronization time for $m=7$ ($L=3$). The numerical result is
    plotted as full curve. The dashed line denotes the asymptotic
    function defined in (\ref{eq:rwasymp}).}
  \label{fig:rwasymp}
\end{figure}

In the case of neural synchronization, $m = 2 L + 1$ is always odd, so
that $c_{m,m-1} = 0$. Here $\mathrm{P}(T=t)$ asymptotically converges
to a geometric probability distribution for long synchronization
times:
\begin{equation}
  \label{eq:rwasymp}
  \mathrm{P}(T=t) \sim c_{m,1} \left[ \cos \left( \frac{\pi}{m}
    \right) \right]^{t-1} \,.
\end{equation}
Figure~\ref{fig:rwasymp} shows that this analytical solution describes
$P(T=t)$ well, except for some deviations at the beginning of the
synchronization process. But for small values of $t$ one can use the
equations of motion for $p_{a,b}$ in order to calculate $P(T=t)$
iteratively.

\subsection{Extreme order statistics}

In this section the model is extended to $N$ independent pairs of
random walks driven by identical random noise. This corresponds to two
hidden units with $N$ weights, which start uncorrelated and reach full
synchronization after $T_N$ attractive steps.

Although $\langle T \rangle$ is the mean value of the synchronization
time for a pair of weights, $w_{i,j}^A$ and $w_{i,j}^B$, it is not equal
to $\langle T_N \rangle$. The reason is that the weight vectors have
to be completely identical in the case of full synchronization.
Therefore $T_N$ is the maximum value of $T$ observed in $N$
independent samples corresponding to the different weights of a hidden
unit.

As the distribution function $\mathrm{P}(T \leq t)$ is known, the
probability distribution of $T_N$ is given by
\begin{equation}
  \label{eq:pmax}
  \mathrm{P}(T_N \leq t) = \mathrm{P}(T \leq t)^N \,.
\end{equation}

\begin{figure}
  \centering
  \includegraphics[scale=0.5]{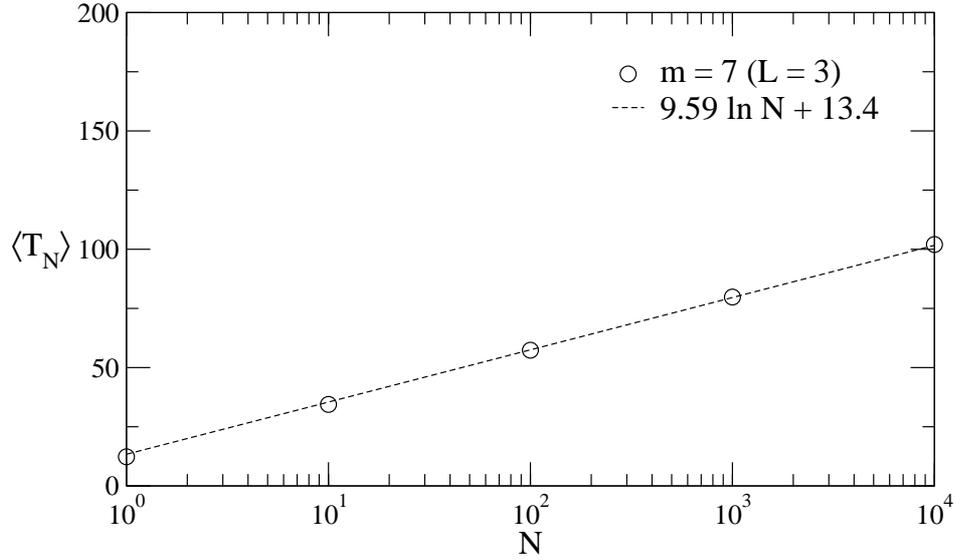}
  \caption{Average synchronization time $\langle T_N \rangle$ as a
    function of $N$ for $m=7$ ($L=3$). Results of the numerical
    calculation using (\ref{eq:pmax}) are represented by circles. The
    dashed line shows the expectation value of $T_N$ calculated in
    (\ref{eq:gumexp}).}
  \label{fig:maximum}
\end{figure}

Hence one can calculate the average value $\langle T_N \rangle$ using
the numerically computed distribution $\mathrm{P}(T_N \leq t)$. The
result, which is shown in figure~\ref{fig:maximum}, indicates that
$\langle T_N \rangle$ increases logarithmically with the number of
pairs of random walkers:
\begin{equation}
  \label{eq:maximum}
  \langle T_N \rangle - \langle T \rangle \propto \ln N \,.
\end{equation}

For large $N$ only the asymptotic behavior of $\mathrm{P}(T \leq t)$
is relevant for the distribution of $T_N$. The exponential decay of
$\mathrm{P}(T=t)$ according to (\ref{eq:rwdist}) yields a Gumbel
distribution for $\mathrm{P}(T_N \leq t)$ \cite{Galambos:1940:ATE},
\begin{equation}
  \label{eq:gum}
  G(t) = \exp \left( -e^\frac{t_\mathrm{a} - t}{t_\mathrm{b}} \right)
  \,,
\end{equation}
for $N \gg m$ with the parameters
\begin{equation}
  \label{eq:gumpar}
  t_\mathrm{a} = t_\mathrm{b} \ln \frac{N c_{m,1}}{1 - \cos (\pi / m)}
  \quad \mbox{and} \quad
  t_\mathrm{b} = - \frac{1}{\ln \cos (\pi / m)} \,.
\end{equation}
Substituting (\ref{eq:gumpar}) into (\ref{eq:gum}) yields
\cite{Ruttor:2004:SRW}
\begin{equation}
  \mathrm{P}(T_N \leq t) = \exp \left( - \frac{N c_{m,1}
      \cos^t (\pi / m)}{1 - \cos (\pi / m)} \right)
\end{equation}
as the distribution function for the total synchronization time of $N$
pairs of random walks ($N \gg m$). The expectation value of this
probability distribution is given by \cite{Galambos:1940:ATE}
\begin{equation}
  \label{eq:gumexp}
  \langle T_N \rangle = t_\mathrm{a} + t_\mathrm{b} \gamma
  = - \frac{1}{\ln \cos (\pi / m)} \left( \gamma + \ln N + \ln
    \frac{c_{m,1}}{1 - \cos (\pi / m)} \right) \,.
\end{equation}
Here $\gamma$ denotes the Euler-Mascheroni constant. For $N \gg m \gg
1$ the asymptotic behavior of the synchronization time is given by
\begin{equation}
  \langle T_N \rangle \sim \frac{2}{\pi^2} \, m^2 \left(
    \gamma + \ln N + \ln \frac{2 m^2 \, c_{m,1}}{\pi^2} \right) \,.
\end{equation}
Using (\ref{eq:rwapprox}) finally leads to the result
\cite{Ruttor:2004:SRW}
\begin{equation}
  \label{eq:rwresult}
  \langle T_N \rangle \approx \frac{2}{\pi^2} \, m^2 \left( \ln N
    + \ln (0.577 \, m) \right) \,,
\end{equation}
which shows that $\langle T_N \rangle$ increases proportional to $m^2
\ln N$.

Of course, neural synchronization is somewhat more complex than this
model using random walks driven by pairwise identical noise. Because
of the structure of the learning rules the weights are not changed in
each step. Including these idle steps certainly increases the
synchronization time $t_\mathrm{sync}$. Additionally, repulsive steps
destroying synchronization are possible, too. Nevertheless, a similar
scaling law $\langle t_\mathrm{sync} \rangle \propto L^2 \ln N$ can be
observed for the synchronization of two Tree Parity Machines as long
as repulsive effects have only little influence on the dynamics of the
system.

\section{Random walk of the overlap}

The most important order parameter of the synchronization process is
the overlap between the weight vectors of the participating neural
networks. The results of section~\ref{sec:effect} and
section~\ref{sec:prob} indicate that its change over time can be
described by a random walk with position dependent step sizes,
$\langle \Delta \rho_\mathrm{a} \rangle$, $\langle \Delta
\rho_\mathrm{r} \rangle$, and transition probabilities,
$P_\mathrm{a}$, $P_\mathrm{r}$ \cite{Ruttor:2007:DNC}. Of course,
only the transition probabilities are exact functions of $\rho$, while
the step sizes fluctuate randomly around their average values.
Consequently, this model is not suitable for quantitative predictions,
but nevertheless one can determine important properties regarding the
qualitative behavior of the system. For this purpose, the average
change of the overlap
\begin{equation}
  \label{eq:average}
  \langle \Delta \rho \rangle = P_\mathrm{a}(\rho) \langle \Delta
  \rho_\mathrm{a}(\rho) \rangle + P_\mathrm{r}(\rho) \langle \Delta
  \rho_\mathrm{r}(\rho) \rangle
\end{equation}
in one synchronization step as a function of $\rho$ is especially
useful.

\begin{figure}
  \centering
  \includegraphics[scale=0.5]{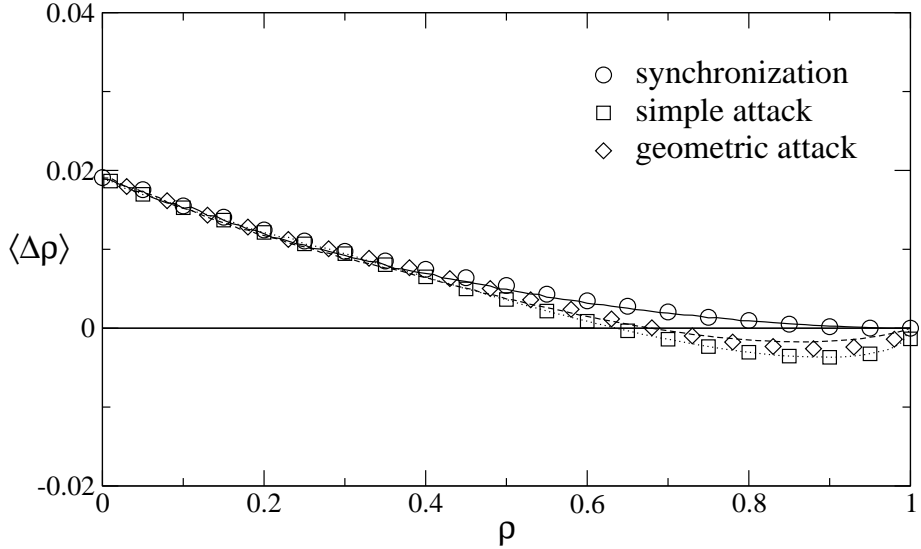}
  \caption{Average change of the overlap for $K=3$, $L=5$, and random
    walk learning rule. Symbols denote results obtained from $1000$
    simulations, while the lines have been calculated using
    (\ref{eq:average}).}
  \label{fig:average}
\end{figure}

Figure~\ref{fig:average} clearly shows the difference between
synchronization and learning for $K=3$. In the case of bidirectional
interaction, $\langle \Delta \rho \rangle$ is always positive until
the process reaches the absorbing state at $\rho=1$. But for
unidirectional interaction, there is a fixed point at $\rho_\mathrm{f}
< 1$. That is why a further increase of the overlap is only possible
by fluctuations. Consequently, there are two different types of
dynamics, which play a role in the process of synchronization.

\subsection{Synchronization on average}
\label{sec:async}

If $\langle \Delta \rho \rangle$ is always positive for $\rho < 1$,
each update of the weights has an attractive effect on average. In
this case repulsive steps delay the process of synchronization, but
the dynamics is dominated by the effect of attractive steps. Therefore
it is similar to that of random walks discussed in
section~\ref{sec:rwmodel}.

\begin{figure}
  \centering
  \includegraphics[scale=0.5]{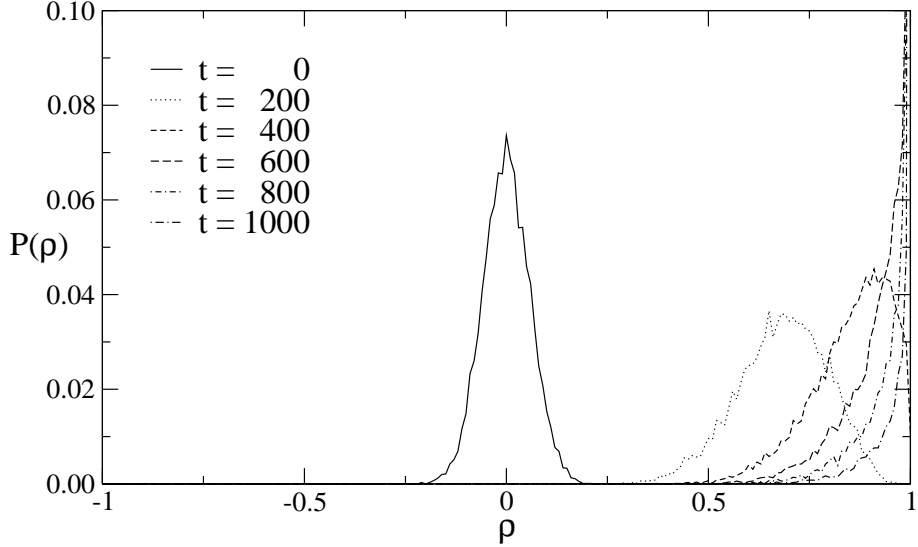}
  \caption{Distribution of the overlap in different time steps. These
    results were obtained in $100$ simulations for synchronization
    with $K=3$, $L=5$, $N=100$, and random walk learning rule.}
  \label{fig:fsync}
\end{figure}

As shown in figure~\ref{fig:fsync} the distribution of the overlap
gets closer to the absorbing state at $\rho=1$ in each time step. And
the velocity of this process is determined by $\langle \Delta \rho
\rangle$. That is why $\rho$ increases fast at the beginning of the
synchronization, but more slowly towards the end.

However, the average change of the overlap depends on the synaptic
depth $L$, too. While the transition probabilities $P_\mathrm{a}$ and
$P_\mathrm{r}$ are unaffected by a change of $L$, the step sizes
$\langle \Delta \rho_\mathrm{a} \rangle$ and $\langle \Delta
\rho_\mathrm{r} \rangle$ shrink proportional to $L^{-2}$ according to
(\ref{eq:atscaling}) and (\ref{eq:rpscaling}). Hence $\langle \Delta
\rho \rangle$ also decreases proportional to $L^{-2}$ so that a large
synaptic depth slows down the dynamics. That is why one expects
\begin{equation}
  \label{eq:synctime}
  \langle t_\mathrm{sync} \rangle \propto \frac{1}{\langle \Delta \rho
    \rangle} \propto L^2
\end{equation}
for the scaling of the synchronization time.

\begin{figure}
  \centering
  \includegraphics[scale=0.5]{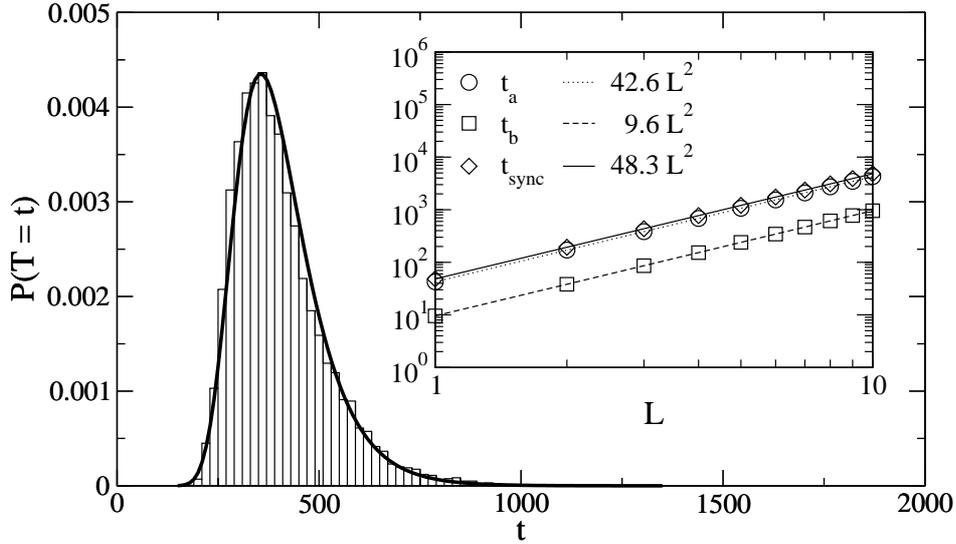}
  \caption{Probability distribution of the synchronization time for
    two Tree Parity Machines with $K=3$, $L=3$, $N=1000$, and random
    walk learning rule. The histogram shows the relative frequency of
    occurrence observed in $10\,000$ simulations and the thick curve
    represents a fit of the Gumbel distribution. Fit parameters for
    different values of $L$ are shown in the inset.}
  \label{fig:neurodist}
\end{figure}

In fact, the probability $P(t_\mathrm{sync} \leq t)$ to achieve
identical weight vectors in A's and B's neural networks in at most $t$
steps is described well by a Gumbel distribution (\ref{eq:gum}):
\begin{equation}
  \label{eq:async}
  P_\mathrm{sync}^B(t) = \exp \left( -e^\frac{t_\mathrm{a} -
      t}{t_\mathrm{b}} \right) \,.
\end{equation}
Similar to the model in section~\ref{sec:rwmodel} the parameters
$t_\mathrm{a}$ and $t_\mathrm{b}$ increase both proportional to $L^2$,
which is clearly visible in figure~\ref{fig:neurodist}. Consequently,
the average synchronization time scales like $\langle t_\mathrm{sync}
\rangle \propto L^2 \ln N$, in agreement with~(\ref{eq:rwresult}).

Additionally, figure~\ref{fig:fsync} indicates that large fluctuations
of the overlap can be observed during the process of neural
synchronization. For $t=0$ the width of the distribution is due to the
finite number of weights and vanishes in the limit $N \rightarrow
\infty$. But later fluctuations are mainly amplified by the interplay
of discrete attractive and repulsive steps. This effect cannot be
avoided by increasing $N$, because this does not change the step
sizes. Therefore the order parameter $\rho$ is not a self-averaging
quantity \cite{Reents:1998:SAL}: one cannot replace $\rho$ by $\langle
\rho \rangle$ in the equations of motion in order to calculate the
time evolution of the overlap analytically. Instead, the whole
probability distribution of the weights has to be taken into account.

\subsection{Synchronization by fluctuations}
\label{sec:fsync}

If there is a fixed point at $\rho_\mathrm{f} < 1$, then the dynamics
of neural synchronization changes drastically. As long as $\rho <
\rho_\mathrm{f}$ the overlap increases on average. But then a
quasi-stationary state is reached. Further synchronization is only
possible by fluctuations, which are caused by the discrete nature of
attractive and repulsive steps.

\begin{figure}
  \centering
  \includegraphics[scale=0.5]{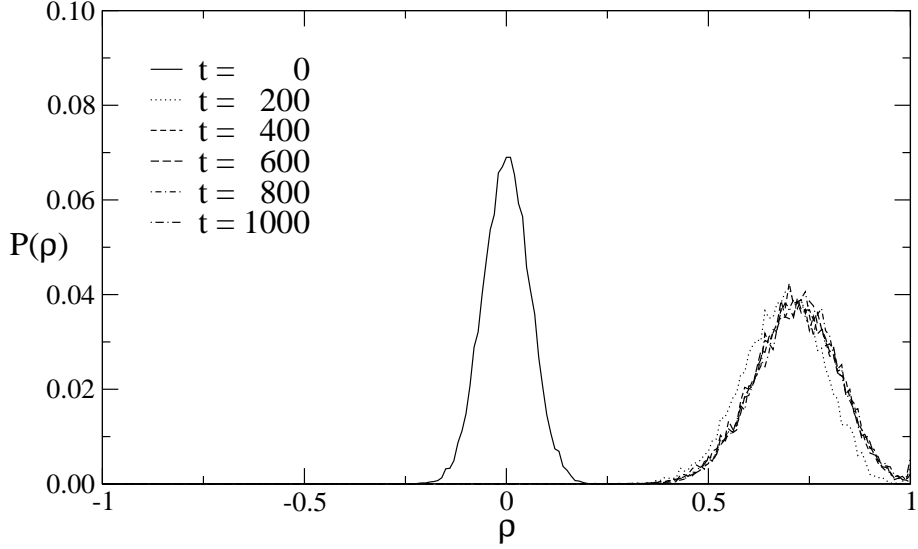}
  \caption{Distribution of the overlap in different time steps. These
    results were obtained in $100$ simulations for the geometric
    attack with $K=3$, $L=5$, $N=100$, and random walk learning rule.}
  \label{fig:fgeo}
\end{figure}

Figure~\ref{fig:fgeo} shows both the initial transient and the
quasi-stationary state. The latter can be described by a normal
distribution with average value $\rho_\mathrm{f}$ and a standard
deviation $\sigma_\mathrm{f}$.

In order to determine the scaling of the fluctuations, a linear
approximation of $\langle \Delta \rho(\rho) \rangle$ is used as a
simple model \cite{Ruttor:2007:DNC},
\begin{equation}
  \label{eq:diff}
  \Delta \rho(t) = - \alpha_\mathrm{f} (\rho(t) - \rho_\mathrm{f}) +
  \beta_\mathrm{f} \xi(t) \,, 
\end{equation}
without taking the boundary conditions into account. Here the $\xi(t)$
are random numbers with zero mean and unit variance. The two
parameters are defined as
\begin{eqnarray}
  \alpha_\mathrm{f} &=& - \left. \frac{\mathrm{d}}{\mathrm{d} \rho}
    \langle \Delta \rho(\rho) \rangle \right|_{\rho=\rho_\mathrm{f}}
  \,, \\ 
  \beta_\mathrm{f} &=& \sqrt{\langle (\Delta \rho(\rho_\mathrm{f}))^2
    \rangle} \,.
\end{eqnarray}
In this model, the solution of (\ref{eq:diff}),
\begin{equation}
  \rho(t + 1) - \rho_\mathrm{f} = \beta_\mathrm{f} \sum_{i=0}^{t} (1 -
  \alpha_\mathrm{f})^{t - i} \xi(i) \,,
\end{equation}
describes the time evolution of the overlap. Here the initial
condition $\rho(0) = \rho_\mathrm{f}$ was assumed, which is admittedly
irrelevant in the limit $t \rightarrow \infty$. Calculating the
variance of the overlap in the stationary state yields
\cite{Ruttor:2007:DNC}
\begin{equation}
  \sigma_\mathrm{f}^2 = \beta_\mathrm{f}^2 \sum_{t=0}^{\infty} (1 -
  \alpha_\mathrm{f})^{2 t}  = \frac{\beta_\mathrm{f}^2}{2
    \alpha_\mathrm{f} - \alpha_\mathrm{f}^2} \,.
\end{equation}

\begin{figure}
  \centering
  \includegraphics[scale=0.5]{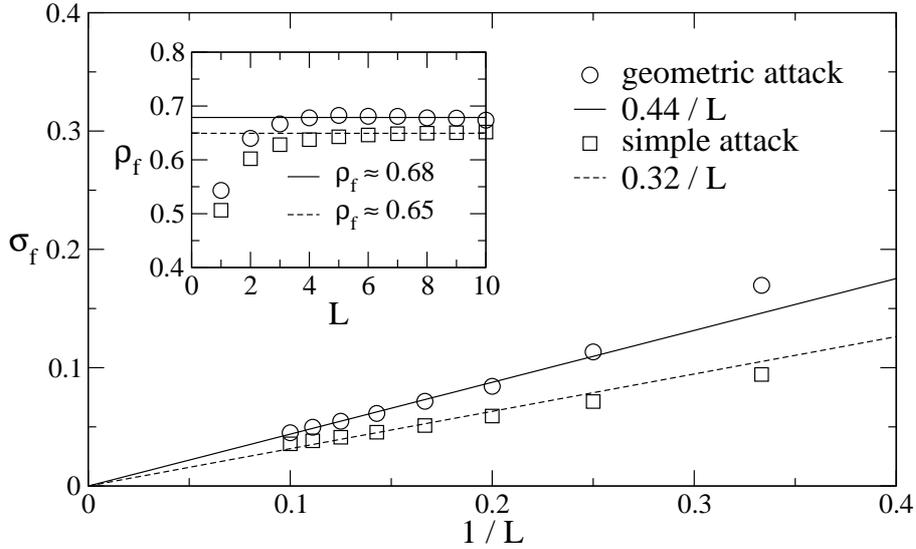}
  \caption{Standard deviation of $\rho$ at the fixed point for $K=3$,
    $N=1000$, random walk learning rule, and unidirectional
    synchronization, averaged over $10\,000$ simulations. The inset
    shows the position of the fixed point.}
  \label{fig:width}
\end{figure}

As the step sizes of the random walk in $\rho$-space decrease
proportional to $L^{-2}$ for $L \gg 1$ according to
(\ref{eq:atscaling}) and (\ref{eq:rpscaling}), this is also the
scaling behavior of the parameters $\alpha_\mathrm{f}$ and
$\beta_\mathrm{f}$. Thus one finds
\begin{equation}
  \sigma_\mathrm{f} \propto \frac{1}{L}
\end{equation}
for larger values of the synaptic depth. Although this simple model
does not include the more complex features of $\langle \Delta
\rho(\rho) \rangle$, its scaling behavior is clearly reproduced in
figure~\ref{fig:width}. Deviations for small values of $L$ are caused
by finite-size effects.

Consequently, E is unable to synchronize with A and B in the limit $L
\rightarrow \infty$, even if she uses the geometric attack. This is
also true for any other algorithm resulting in a dynamics of the
overlap, which has a fixed point at $\rho_\mathrm{f} < 1$.

\begin{figure}
  \centering
  \includegraphics[scale=0.5]{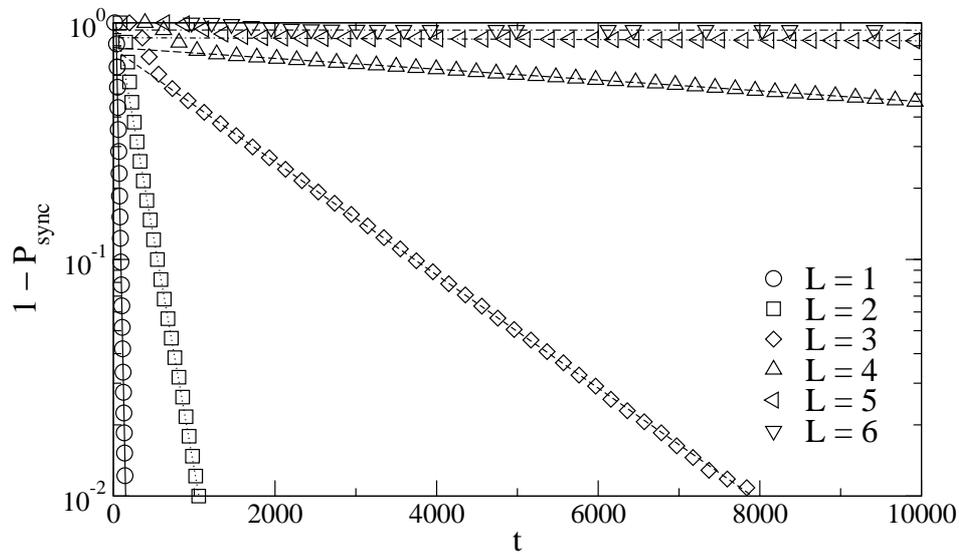}
  \caption{Probability distribution of $t_\mathrm{sync}$ for $K=3$,
    $N=1000$, random walk learning rule, and geometric attack. Symbols
    denote results averaged over $1000$ simulations and the lines show
    fits with (\ref{eq:fsync}).}
  \label{fig:tfdist}
\end{figure}

For finite synaptic depth, however, the attacker has a chance of
getting beyond the fixed point at $\rho_\mathrm{f}$ by fluctuations.
The probability that this event occurs in any given step is
independent of $t$, once the quasi-stationary state has been reached.
Thus $P^E_\mathrm{sync}(t)$ is not given by a Gumbel distribution
(\ref{eq:gum}), but described well for $t \gg t_0$ by an exponential
distribution,
\begin{equation}
  \label{eq:fsync}
  P^E_\mathrm{sync}(t) = 1 - e^{-\frac{t - t_0}{t_\mathrm{f}}} \,,
\end{equation}
with time constant $t_\mathrm{f}$. This is clearly visible in
figure~\ref{fig:tfdist}. Because of $t_\mathrm{f} \gg t_0$ one needs
\begin{equation}
  \langle t_\mathrm{sync} \rangle \approx \left\{
    \begin{array}{cl}
      t_\mathrm{f} \, e^{t_0 / t_\mathrm{f}} & \mbox{ for $t_0 < 0$}
      \\
      t_\mathrm{f} + t_0                     & \mbox{ for $t_0 \geq
        0$}
    \end{array}
  \right\} \approx t_\mathrm{f}
\end{equation}
steps on average to reach $\rho=1$ using unidirectional learning.

\begin{figure}
  \centering
  \includegraphics[scale=0.5]{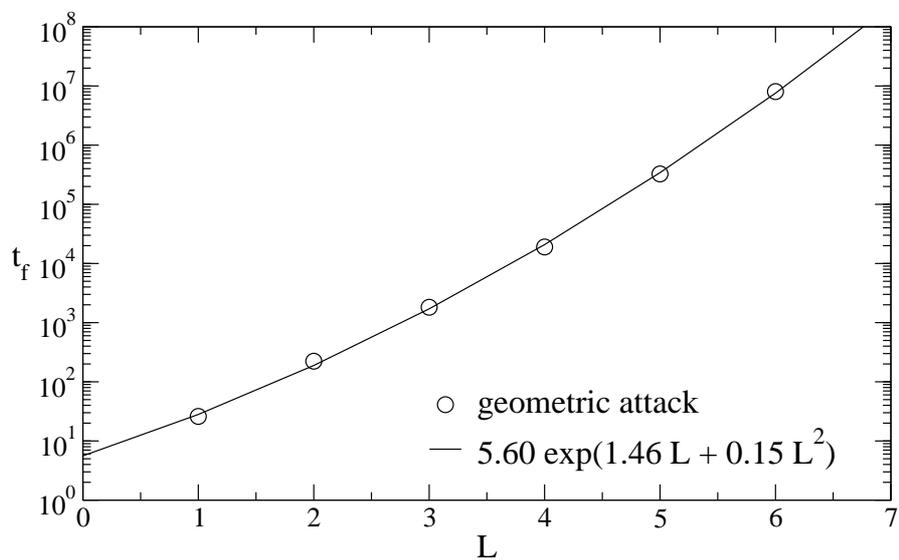}
  \caption{Time constant $t_\mathrm{f}$ for synchronization by
    fluctuations. Symbols denote results obtained in $1000$
    simulations of the geometric attack for $K=3$, $N=1000$, and
    random walk learning rule. The line shows a fit with
    (\ref{eq:tf}).}
  \label{fig:ftime}
\end{figure}

In the simplified model \cite{Ruttor:2007:DNC} with linear $\langle
\Delta \rho(\rho) \rangle$ the mean time needed to achieve full
synchronization starting at the fixed point is given by
\begin{equation}
  t_\mathrm{f} \approx \frac{1}{P(\rho=1)} = \sqrt{2 \pi}
  \sigma_\mathrm{f} e^{\frac{(1 - \rho_\mathrm{f})^2}{2
      \sigma_\mathrm{f}^2}}
\end{equation}
as long as the fluctuations are small. If $\sigma_\mathrm{f} \ll 1 -
\rho_\mathrm{f}$, the assumption is reasonable, that the distribution
of $\rho$ is not influenced by the presence of the absorbing state at
$\rho=1$. Hence one expects
\begin{equation}
  t_\mathrm{f} \propto e^{c L^2}
\end{equation}
for the scaling of the time constant, as $\sigma_\mathrm{f}$ changes
proportional to $L^{-1}$, while $\rho_\mathrm{f}$ stays nearly
constant. And figure~\ref{fig:ftime} shows that indeed $t_\mathrm{f}$
grows exponentially with increasing synaptic depth:
\begin{equation}
  \label{eq:tf}
  t_\mathrm{f} \propto e^{c_1 L + c_2 L^2} \,.
\end{equation}

Thus the partners A and B can control the complexity of attacks on the
neural key-exchange protocol by choosing $L$. Or if E's effort stays
constant, her success probability drops exponentially with increasing
synaptic depth. As shown in chapter~\ref{chap:security}, this effect
can be observed in the case of the geometric attack
\cite{Mislovaty:2002:SKE} and even for advanced methods
\cite{Ruttor:2005:NCQ, Ruttor:2006:GAN}.

\section{Synchronization time}
\label{sec:synctime}

As shown before the scaling of the average synchronization time
$\langle t_\mathrm{sync} \rangle$ with regard to the synaptic depth
$L$ depends on the function $\langle \Delta \rho(\rho) \rangle$ which
is different for bidirectional and unidirectional interaction.
However, one has to consider two other parameters. The probability of
repulsive steps $P_\mathrm{r}$ depends not only on the interaction,
but also on the number of hidden units. Therefore one can switch
between synchronization on average and synchronization by fluctuations
by changing $K$, which is the topic of section~\ref{sec:ksync}.
Additionally, the chosen learning rule influences the step sizes of
attractive and repulsive steps. Section~\ref{sec:rsync} shows that
this affects $\langle \Delta \rho(\rho) \rangle$ and consequently the
average synchronization time $\langle t_\mathrm{sync} \rangle$, too.

\subsection{Number of hidden units}
\label{sec:ksync}

As long as $K \leq 3$, A and B are able to synchronize on average. In
this case $\langle t_\mathrm{sync} \rangle$ increases proportional to
$L^2$. In contrast, E can only synchronize by fluctuations as soon as
$K > 1$, so that for her $\langle t_\mathrm{sync} \rangle$ grows
exponentially with the synaptic depth $L$. Consequently, A and B can
reach any desired level of security by choosing a suitable value for
$L$.

\begin{figure}
  \centering
  \includegraphics[scale=0.5]{pics/kfour.eps}
  \caption{Average change of the overlap for $L=10$, $N=1000$, random
    walk learning rule, and bidirectional synchronization. Symbols
    denote results obtained from $100$ simulations, while the lines
    have been calculated using (\ref{eq:average}).}
  \label{fig:kfour}
\end{figure}

\begin{figure}
  \centering
  \includegraphics[scale=0.5]{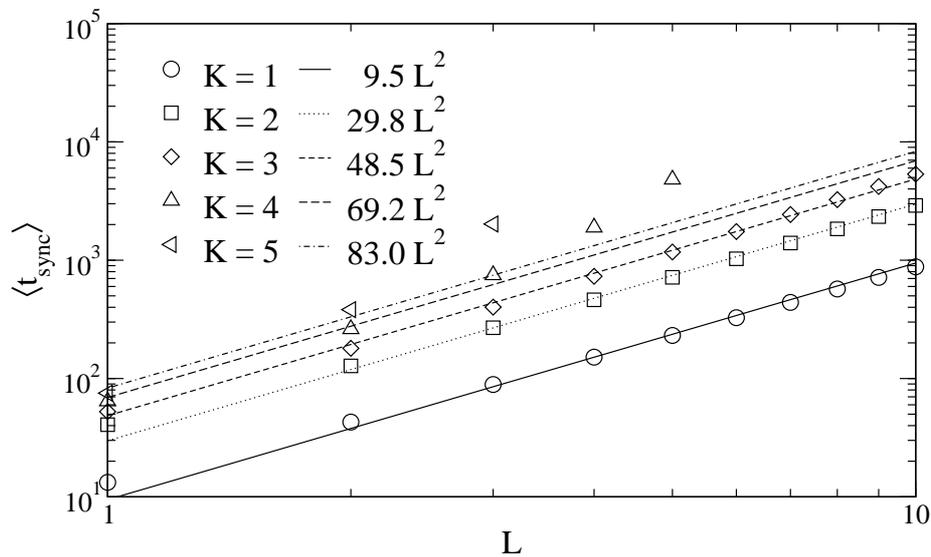}
  \caption{Synchronization time for bidirectional interaction,
    $N=1000$, and random walk learning rule. Symbols denote results
    averaged over $10\,000$ simulations and the lines represent fits of
    the model $\langle t_\mathrm{sync} \rangle \propto L^2$.}
  \label{fig:synctime}
\end{figure}

However, this is not true for $K > 3$. As shown in
figure~\ref{fig:kfour}, a fixed point at $\rho_\mathrm{f} < 1$ appears
in the case of bidirectional synchronization, too. Therefore
(\ref{eq:synctime}) is not valid any more and $\langle t_\mathrm{sync}
\rangle$ now increases exponentially with $L$. This is clearly visible
in figure~\ref{fig:synctime}. Consequently, Tree Parity Machines with
four and more hidden units cannot be used in the neural key-exchange
protocol, except if the synaptic depth is very small.

\begin{figure}
  \centering
  \includegraphics[scale=0.5]{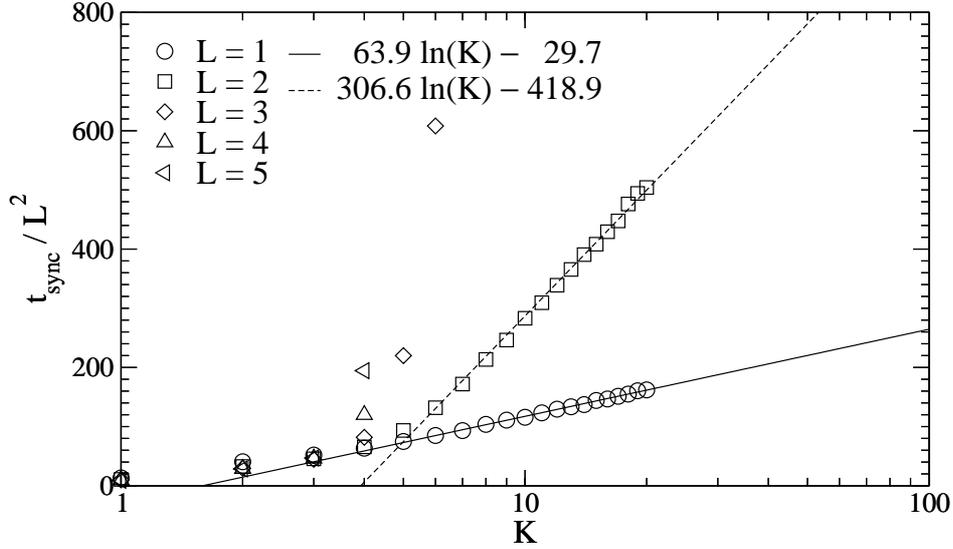}
  \caption{Synchronization time for bidirectional interaction,
    $N=1000$, and random walk learning rule, averaged over $1000$
    simulations.}
  \label{fig:ksync}
\end{figure}

Figure~\ref{fig:ksync} shows the transition between the two mechanisms
of synchronization clearly. As long as $K \leq 3$ the scaling law
$\langle t_\mathrm{sync} \rangle \propto L^2$ is valid, so that the
constant of proportionality $\langle t_\mathrm{sync} \rangle / L^2$ is
independent of the number of hidden units. Additionally, it increases
proportional to $\ln K N$, as the total number of weights in a Tree
Parity Machine is given by $K N$.

In contrast, $\langle t_\mathrm{sync} \rangle \propto L^2$ is not
valid for $K > 3$. In this case $\langle t_\mathrm{sync} \rangle /
L^2$ still increases proportional to $\ln K N$, but the steepness of
the curve depends on the synaptic depth, as the fluctuations of the
overlap decrease proportional to $L^{-1}$. Consequently, there are two
sets of parameters, which allow for synchronization using
bidirectional interaction in a reasonable number of steps: the
absorbing state $\rho = 1$ is reached on average for $K \leq 3$,
whereas large enough fluctuations drive the process of synchronization
in the case of $L \leq 3$ and $K \geq 4$. Otherwise, a huge number of
steps is needed to achieve full synchronization.

\subsection{Learning rules}
\label{sec:rsync}

Although the qualitative properties of neural synchronization are
independent of the chosen learning rule, one can observe quantitative
deviations for Hebbian and anti-Hebbian learning in terms of
finite-size effects. Of course, these disappear in the limit $L /
\sqrt{N} \rightarrow 0$.

\begin{figure}
  \centering
  \includegraphics[scale=0.5]{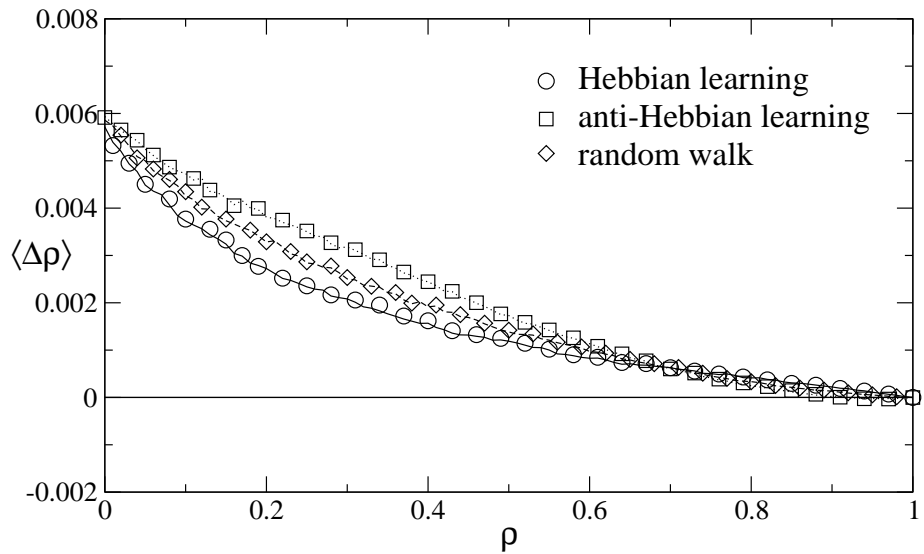}
  \caption{Average change of the overlap for $K=3$, $L=10$, $N=1000$,
    and bidirectional synchronization. Symbols denote results obtained
    from $100$ simulations, while the lines have been calculated using
    (\ref{eq:average}).}
  \label{fig:ravg}
\end{figure}

\begin{figure}
  \centering
  \includegraphics[scale=0.5]{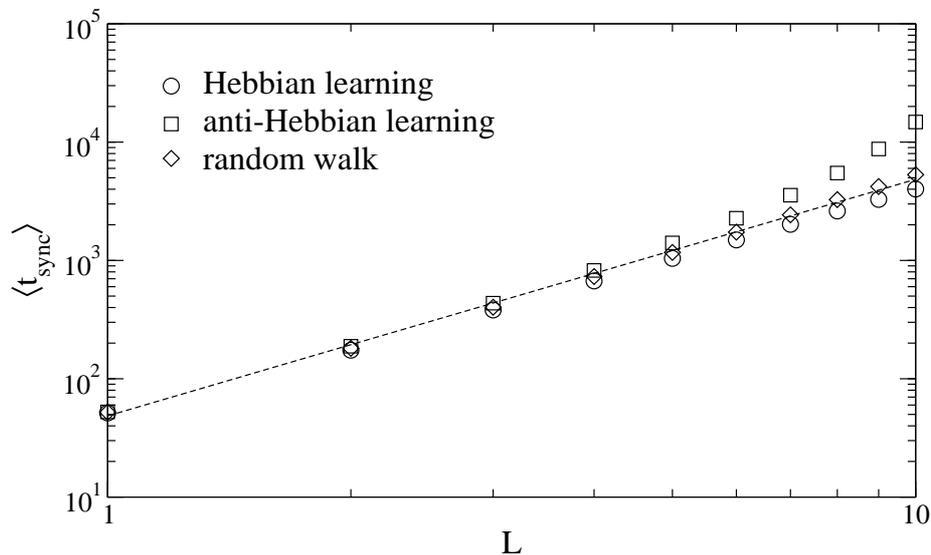}
  \caption{Synchronization time for bidirectional interaction and
    $K=3$. Symbols denote results averaged over $10\,000$ simulations
    and the line shows the corresponding fit from
    figure~\ref{fig:synctime} for the random walk learning rule.}
  \label{fig:rsync}
\end{figure}

As shown in section~\ref{sec:effect} Hebbian learning enhances the
effects of both repulsive and attractive steps. This results in a
decrease of $\langle \Delta \rho \rangle$ for small overlap, where a
lot of repulsive steps occur. But if A's and B's Tree Parity Machines
are nearly synchronized, attractive steps prevail, so that the average
change of the overlap is increased compared to the random walk
learning rule. This is clearly visible in figure~\ref{fig:ravg}. Of
course, anti-Hebbian learning reduces both step sizes and one observes
the opposite effect.

However, the average synchronization time $\langle t_\mathrm{sync}
\rangle$ is mainly influenced by the transition from $\rho \approx 1$
to $\rho = 1$, which is the slowest part of the synchronization
process. Therefore Hebbian learning decreases the average number of
steps needed to achieve full synchronization. This effect is clearly
visible in figure~\ref{fig:rsync}.

In contrast, anti-Hebbian learning increases $\langle t_\mathrm{sync}
\rangle$. Here finite-size effects cause problems for bidirectional
synchronization, because one can even observe $\langle \Delta \rho
\rangle < 0$ for $K=3$, if $L / \sqrt{N}$ is just sufficiently large.
Then the synchronization time increases faster than $L^2$.
Consequently, this learning rule is only usable in large systems,
where finite-size effects are small and the observed behavior is
similar to that of the random walk learning rule.

\chapter{Security of neural cryptography}
\label{chap:security}

The security of the neural key-exchange protocol is based on the
phenomenon analyzed in chapter~\ref{chap:neurosync}: two Tree Parity
Machines interacting with each other synchronize much faster than a
third neural network trained using their inputs and outputs as
examples. In fact, the effort of the partners grows only polynomially
with increasing synaptic depth, while the complexity of an attack
scales exponentially with $L$.

However, neural synchronization is a stochastic process driven by
random attractive and repulsive forces \cite{Ruttor:2004:NCF}.
Therefore A and B are not always faster than E, but there is a small
probability $P_E$ that an attacker is successful before the partners
have finished the key exchange. Because of the different dynamics
$P_E$ drops exponentially with increasing $L$, so that the system is
secure in the limit $L \rightarrow \infty$ \cite{Mislovaty:2002:SKE}.
And in practise, one can reach any desired level of security by just
increasing $L$, while the effort of generating a key only grows
moderately \cite{Ruttor:2006:GAN}.

Although this mechanism works perfectly, if the parameters of the
protocol are chosen correctly, other values can lead to an insecure
key exchange \cite{Shacham:2004:CAN}. Therefore it is necessary to
determine the scaling of $P_E$ for different configurations and all
known attack methods. By doing so, one can form an estimate regarding
the minimum synaptic depth needed for some practical applications,
too.

While $P_E$ directly shows whether neural cryptography is secure, it
does not reveal the cause of this observation. For that purpose, it is
useful to analyze the mutual information $I$ gained by partners and
attackers during the process of synchronization. Even though all
participants receive the same messages, A and B can select the most
useful ones for adjusting the weights. That is why they learn more
about each other than E, who is only listening. Consequently,
bidirectional interaction gives an advantage to the partners, which
cannot be exploited by a passive attacker.

Of course, E could try other methods instead of learning by listening.
Especially in the case of a brute-force attack, security depends on
the number of possible keys, which can be generated by the neural
key-exchange protocol. Therefore it is important to analyze the
scaling of this quantity, too.

\section{Success probability}
\label{sec:pe}

Attacks which are based on learning by listening have in common that
the opponent E tries to synchronize one or more Tree Parity Machines
with A's and B's neural networks. Of course, after the partners have
reached identical weight vectors, they stop the process of
synchronization, so that the number of available examples for the
attack is limited. Therefore E's online learning is only successful,
if she discovers the key before A and B finish the key exchange.

As synchronization of neural networks is a stochastic process, there
is a small probability that E synchronizes faster with A than B. In
actual fact, one could use this quantity directly to describe the
security of the key-exchange protocol. However, the partners may not
detect full synchronization immediately, so that E is even successful,
if she achieves her goal shortly afterwards. Therefore $P_E$ is
defined as the probability that the attacker knows 98 per cent of the
weights at synchronization time. Additionally, this definition reduces
fluctuations in the simulations, which are employed to determine $P_E$
\cite{Mislovaty:2002:SKE}.

\subsection{Attacks using a single neural network}

For both the simple attack and the geometric attack E only needs one
Tree Parity Machine. So the complexity of these methods is small. But
as already shown in section~\ref{sec:fsync} E can only synchronize by
fluctuations if $K > 1$, while the partners synchronize on average as
long as $K \leq 3$. That is why $t_\mathrm{sync}^E$ is usually much
larger than $t_\mathrm{sync}^B$ for $K = 2$ and $K = 3$. In fact, the
probability of $t_\mathrm{sync}^E \leq t_\mathrm{sync}^B$ in this case
is given by
\begin{equation}
  \label{eq:pint}
  P(t_\mathrm{sync}^E \leq t_\mathrm{sync}^B) = \int_{t=0}^\infty
  P_\mathrm{sync}^E(t) \, \frac{\mathrm{d}}{\mathrm{d}t}
  P_\mathrm{sync}^B(t) \, \mathrm{d}t
\end{equation}
under the assumption that the two synchronization times are
uncorrelated random variables. In this equation $P_\mathrm{sync}^B(t)$
and $P_\mathrm{sync}^E(t)$ are the cumulative probability
distributions of the synchronization time defined in (\ref{eq:async})
and (\ref{eq:fsync}), respectively.

In order to approximate this probability one has to look especially at
the fluctuations of the synchronization times $t_\mathrm{sync}^B$ and
$t_\mathrm{sync}^E$. The width of the Gumbel
distribution,
\begin{equation}
  \sqrt{\langle (t_\mathrm{sync}^B)^2 \rangle - \langle
    t_\mathrm{sync}^B \rangle^2} = \frac{\pi}{\sqrt{6}} \, t_\mathrm{b} \,,
\end{equation}
for A and B is much smaller than the standard deviation of the
exponential distribution,
\begin{equation}
  \sqrt{\langle (t_\mathrm{sync}^E)^2 \rangle - \langle
    t_\mathrm{sync}^E \rangle^2} = t_\mathrm{f} \,,
\end{equation}
for E because of $t_\mathrm{f} \gg t_\mathrm{b}$. Therefore one can
approximate $P_\mathrm{sync}^B(t)$ in integral (\ref{eq:pint}) by
$\Theta(t - \langle t_\mathrm{sync}^B \rangle)$, which leads to
\begin{equation}
  \label{eq:pfast}
  P(t_\mathrm{sync}^E \leq t_\mathrm{sync}^B) \approx 1 -
  \exp \left( -\frac{\langle t_\mathrm{sync}^B \rangle}{\langle
      t_\mathrm{sync}^E \rangle}
  \right) \,.
\end{equation}
Hence the success probability of an attack depends on the ratio of
both average synchronization times,
\begin{equation}
  \frac{\langle t_\mathrm{sync}^B \rangle}{\langle t_\mathrm{sync}^E
    \rangle} \propto \frac{L^2}{e^{c_1 L +c_2 L^2}} \,,
\end{equation}
which are functions of the synaptic depth $L$ according to
(\ref{eq:synctime}) and (\ref{eq:tf}). Consequently, $L$ is the most
important parameter for the security of the neural key-exchange
protocol.

In the case of $L \gg 1$ the ratio $\langle t_\mathrm{sync}^B \rangle
/ \langle t_\mathrm{sync}^E \rangle$ becomes very small, so that a
further approximation of (\ref{eq:pfast}) is possible. This yields the
result
\begin{equation}
  \label{eq:pasymp}
  P(t_\mathrm{sync}^E \leq t_\mathrm{sync}^B) \propto L^2 e^{- c_1 L}
  e^{- c_2 L^2} \,,
\end{equation}
which describes the asymptotic behavior of the success probability: if
A and B increase the synaptic depth of their Tree Parity Machines, the
success probability of an attack drops exponentially
\cite{Mislovaty:2002:SKE}. Thus the partners can achieve any desired
level of security by changing $L$.

\begin{figure}
  \centering
  \includegraphics[scale=0.5]{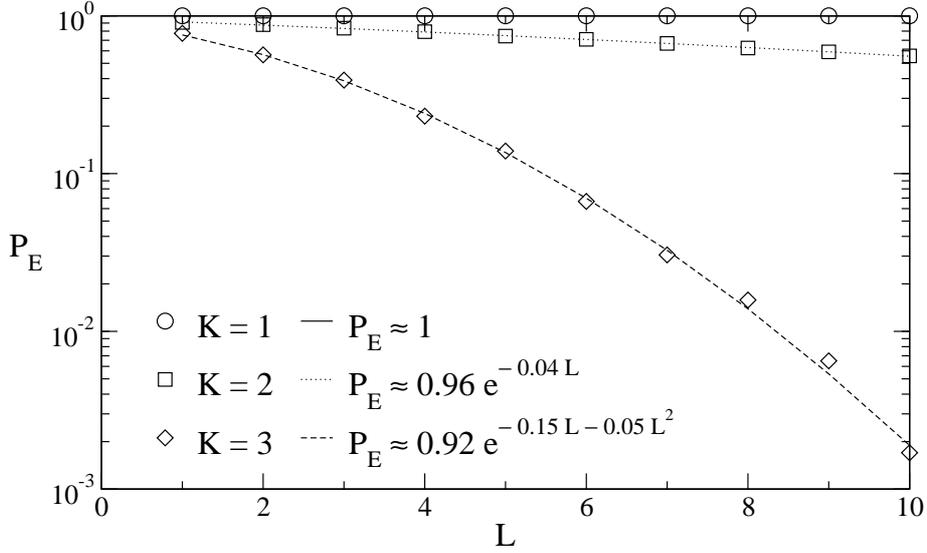}
  \caption{Success probability of the geometric attack as a function
    of $L$. Symbols denote results obtained in $10\,000$ simulations
    with $N=1000$ and random walk learning rule, while the lines
    represent fit results for model (\ref{eq:success}).}
  \label{fig:pgeo}
\end{figure}

Although $P_E$ is not exactly identical to $P(t_\mathrm{sync}^E \leq
t_\mathrm{sync}^B)$ because of its definition, it has the expected
scaling behavior,
\begin{equation}
  \label{eq:success}
  P_E \propto e^{-y_1 L - y_2 L^2} \,,
\end{equation}
which is clearly visible in figure~\ref{fig:pgeo}. However, the
coefficients $y_1$ and $y_2$ are different from $c_1$ and $c_2$ due to
interfering correlations between $t_\mathrm{sync}^B$ and
$t_\mathrm{sync}^E$, which have been neglected in the derivation of
$P(t_\mathrm{sync}^E \leq t_\mathrm{sync}^B)$.

Additionally, figure~\ref{fig:pgeo} shows that the success probability
of the geometric attack depends not only on the synaptic depth $L$,
but also on the number of hidden units $K$. This effect, which results
in different values of the coefficients, is caused by a limitation of
the algorithm: the output of at most one hidden unit is corrected in
each step. While this is sufficient to avoid all repulsive steps in
the case $K = 1$, there can be several hidden units with $\sigma_i^E
\not= \sigma_i^A$ for $K > 1$. And the probability for this event
grows with increasing $K$, so that more and more repulsive steps occur
in E's neural network.

\begin{figure}
  \centering
  \includegraphics[scale=0.5]{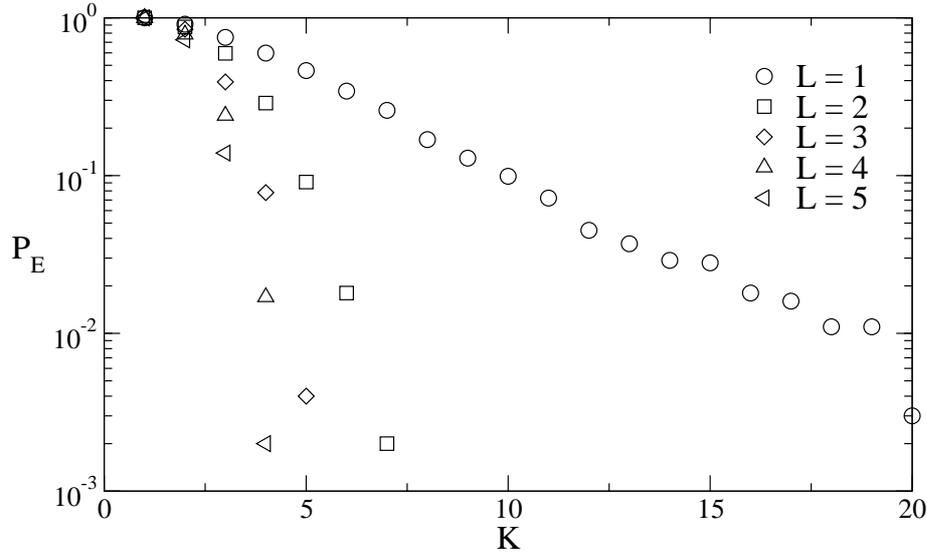}
  \caption{Success probability $P_E$ of the geometric attack as a
    function of $K$. Symbols denote results obtained in $1000$
    simulations using the random walk learning rule and $N=1000$.}
  \label{fig:kgeo}
\end{figure}

Consequently, A and B can achieve security against this attack not
only by increasing the synaptic depth $L$, but also by using a greater
number of hidden units $K$. Of course, for $K > 3$ large values of $L$
are not possible, as the process of synchronization is then driven by
fluctuations. Nevertheless, figure~\ref{fig:kgeo} shows that the
success probability $P_E$ for the geometric attack drops quickly with
increasing $K$ even in the case $L=1$.

As the geometric attack is an element of both advanced attacks,
majority attack and genetic attack, one can also defeat these methods
by increasing $K$. But then synchronization by mutual interaction and
learning by listening become more and more similar. Thus one has to
look at the success probability of the simple attack, too.

\begin{figure}
  \centering
  \includegraphics[scale=0.5]{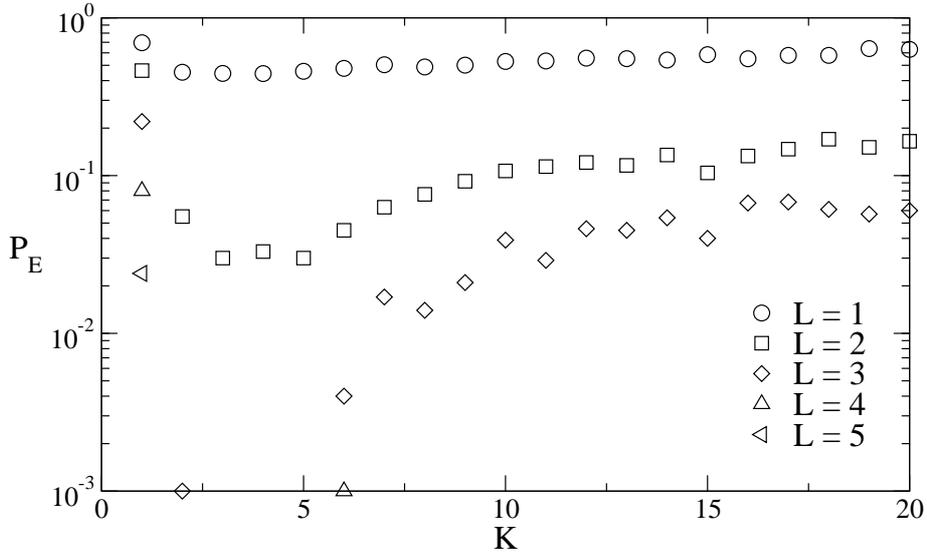}
  \caption{Success probability $P_E$ of the simple attack as a
    function of $K$. Symbols denote results obtained in $1000$
    simulations using the random walk learning rule and $N=1000$.}
  \label{fig:ksimple}
\end{figure}

As this method does not correct the outputs $\sigma_i^E$ of the hidden
units at all, the distance between the fixed point at $\rho_\mathrm{f}
< 1$ of the dynamics and the absorbing state at $\rho = 1$ is greater
than in the case of the geometric attack. That is why a simple
attacker needs larger fluctuations to synchronize and is less
successful than the more advanced attack as long as the number of
hidden units is small.

In principle, scaling law (\ref{eq:success}) is also valid for this
method. But one cannot find a single successful simple attack in
$1000$ simulations using the parameters $K=3$ and $L=3$
\cite{Kanter:2002:SEI}. This is clearly visible in
figure~\ref{fig:ksimple}. Consequently, the simple attack is not
sufficient to break the security of the neural key-exchange protocol
for $K \leq 3$.

But learning by listening without any correction works if the number
of hidden units is large. Here the probability of repulsive steps is
similar for both bidirectional and unidirectional interaction as shown
in section~\ref{sec:prob}. That is why $P_E$ approaches a non-zero
constant value in the limit $K \rightarrow \infty$.

These results show that $K=3$ is the optimal choice for the
cryptographic application of neural synchronization. $K=1$ and $K=2$
are too insecure in regard to the geometric attack. And for $K > 3$
the effort of A and B grows exponentially with increasing $L$, while
the simple attack is quite successful in the limit $K \rightarrow
\infty$. Consequently, one should only use Tree Parity Machines with
three hidden units for the neural key-exchange protocol.

\subsection{Genetic attack}
\label{sec:gensec}

In the case of the genetic attack E's success depends mainly on the
ability to determine the fitness of her neural networks. Of course,
the best quantity for this purpose would be the overlap $\rho^{AE}$
between an attacking network and A's Tree Parity Machine. However, it
is not available, as E only knows the weight vectors of her own
networks. Instead the attacker uses the frequency of the event $\tau^E
= \tau^A$ in recent steps, which gives a rough estimate of $\rho^{AE}$.

Therefore a selection step only works correctly, if there are clear
differences between attractive and repulsive effects. As the step
sizes $\langle \Delta \rho_\mathrm{a} \rangle$ and $\langle \Delta
\rho_\mathrm{r} \rangle$ decrease proportional to $L^{-2}$,
distinguishing both step types becomes more and more complicated for
E. Thus one expects a similar asymptotic behavior of $P_E$ in
the limit $L \rightarrow \infty$ as observed before.

\begin{figure}
  \centering
  \includegraphics[scale=0.5]{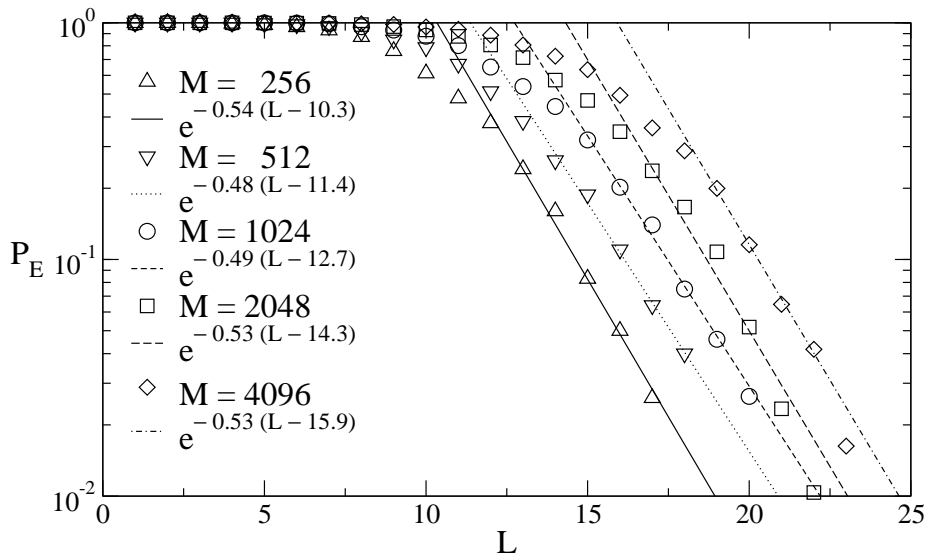}
  \caption{Success probability of the genetic attack. Symbols denote
    results obtained in $1000$ simulations with $K=3$, $N=1000$, and
    random walk learning rule. The lines show fit results using
    (\ref{eq:pe}) as a model.}
  \label{fig:genetic}
\end{figure}

Figure~\ref{fig:genetic} shows that this is indeed the case. The
success probability drops exponentially with increasing synaptic depth
$L$,
\begin{equation}
  \label{eq:pe}
  P_E \sim e^{-y (L - L_0)} \,,
\end{equation}
as long as $L > L_0$ \cite{Ruttor:2006:GAN}. But for $L < L_0$ E is
nearly always successful. Consequently, A and B have to use Tree
Parity Machines with large synaptic depth in order to secure the
key-exchange protocol against this attack.

\begin{figure}
  \centering
  \includegraphics[scale=0.5]{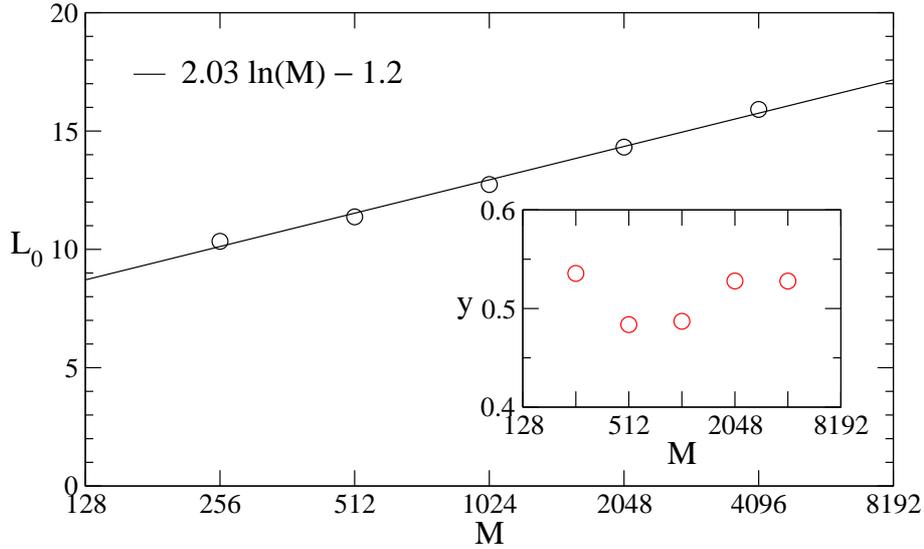}
  \caption{Coefficients $L_0$ and $y$ for the genetic attack as a
    function of the number of attackers $M$. Symbols denote the
    results obtained in figure \ref{fig:genetic} for $K=3$, $N=1000$,
    and random walk learning rule.}
  \label{fig:genpar}
\end{figure}

In contrast to the geometric method, E is able to improve her success
probability by increasing the maximum number of networks used for the
genetic attack. As shown in figure~\ref{fig:genpar} this changes
$L_0$, but the coefficient $y$ remains approximately constant.
However, it is a logarithmic effect:
\begin{equation}
  \label{eq:genpar}
  L_0(M) = L_0(1) + L_E \ln M \,.
\end{equation}
That is why the attacker has to increase the number of her Tree Parity
Machines exponentially,
\begin{equation}
  M \propto e^{L / L_E}
\end{equation}
in order to compensate a change of $L$ and maintain a constant success
probability $P_E$. But the effort needed to generate a key only
increases proportional to $L^2$. Consequently, the neural key-exchange
protocol is secure against the genetic attack in the limit $L
\rightarrow \infty$.

\subsection{Majority attack}

An opponent who has an ensemble of $M$ Tree Parity Machines, can also
use the majority attack to improve $P_E$. This method does neither
generate variants nor select the fittest networks, so that their
number remains constant throughout the process of synchronization.
Instead the majority decision of the $M$ Tree Parity Machines
determines the internal representation $(\sigma_1^E, \dots,
\sigma_K^E)$ used for the purpose of learning. Therefore this
algorithm implements, in fact, the optimal Bayes learning rule
\cite{Engel:2001:SML, Urbanczik:2000:OLE, Watkin:1993:OLN}.

In order to describe the state of the ensemble one can use two order
parameters. First, the mean value of the overlap between corresponding
hidden units in A's and E's neural networks,
\begin{equation}
  \label{eq:rae}
  \rho_i^{AE} = \frac{1}{M} \sum_{m=1}^M \frac{\mathbf{w}_i^A \cdot
    \mathbf{w}_i^{E,m}}{||\mathbf{w}_i^A|| \, ||\mathbf{w}_i^{E,m}||}
  \,,
\end{equation}
indicates the level of synchronization. Here the index $m$ denotes the
$m$-th attacking network. Similar to other attacks, E starts without
knowledge in regard to A, so that $\rho^{AE}=0$ at the beginning. And
finally she is successful, if $\rho^{AE}=1$ is reached.

Second, the average overlap between two attacking networks,
\begin{equation}
  \label{eq:ree}
  \rho_i^{EE} = \frac{1}{M(M - 1)} \sum_{m=1}^M \sum_{n \not= m}
  \frac{\mathbf{w}_i^{E,m} \cdot
    \mathbf{w}_i^{E,n}}{||\mathbf{w}_i^{E,m}|| \,
    ||\mathbf{w}_i^{E,n}||} \,,
\end{equation}
describes the correlations in E's ensemble. At the beginning
of the synchronization $\rho^{EE}=0$, because all weights are
initialized randomly and uncorrelated. But as soon as $\rho^{EE}=1$ is
reached, E's networks are identical, so that the performance of the
majority attack is reduced to that of the geometric method.

In the large $M$ limit, the majority vote of the ensemble is identical
to the output values $\sigma_i$ of a single neural network, which is
located at the center of mass \cite{Urbanczik:2000:OLE}. Its weight
vectors are given by the normalized average over all of E's Tree
Parity Machines:
\begin{equation}
  \mathbf{w}_i^{E,\mathrm{cm}} = \frac{1}{M} \sum_{m=1}^M
  \frac{\mathbf{w}_i^{E,m}}{||\mathbf{w}_i^{E,m}||} \,.
\end{equation}
The normalization of $\mathbf{w}_i^{E,m}$ corresponds to the fact that
each member has exactly one vote. Using (\ref{eq:rae}) and
(\ref{eq:ree}) one can calculate the overlap between the center of
mass and A's tree Parity Machine:
\begin{equation}
  \rho_i^\mathrm{cm} = \frac{\mathbf{w}_i^A \cdot
    \mathbf{w}_i^{E,\mathrm{cm}}}{||\mathbf{w}_i^A|| \,
    ||\mathbf{w}_i^{E,\mathrm{cm}}||} =
  \frac{\rho_i^{AE}}{\sqrt{\rho_i^{EE} + \frac{1}{M} \left( 1 -
      \rho_i^{EE} \right) }} \,.
\end{equation}
Consequently, the effective overlap between A and E is given by
\begin{equation}
  \rho_i^\mathrm{cm} \sim \frac{\rho_i^{AE}}{\sqrt{\rho_i^{EE}}}
\end{equation}
in the limit $M \rightarrow \infty$. This result is important for the
dynamics of the synchronization process between A and E, because
$\rho_i^\mathrm{cm}$ replaces $\rho_i^{AE}$ in the calculation of the
transition probabilities $P_\mathrm{a}(\rho)$ and
$P_\mathrm{r}(\rho)$, whenever the majority vote is used to adjust the
weights. But the step sizes $\langle \Delta \rho_\mathrm{a}(\rho)
\rangle$ and $\langle \Delta \rho_\mathrm{r}(\rho) \rangle$ are not
affected by this modification of the algorithm. Therefore the average
change of the overlap between A and E is given by
\begin{equation}
  \langle \Delta \rho_i^{AE} \rangle =
  P_\mathrm{a}(\rho_i^\mathrm{cm}) \langle \Delta
  \rho_\mathrm{a}(\rho_i^{AE}) \rangle +
  P_\mathrm{r}(\rho_i^\mathrm{cm}) \langle \Delta
  \rho_\mathrm{r}(\rho_i^{AE}) \rangle \,,
\end{equation}
if the majority vote is used for updating the weight vectors. Although
this equation is strictly correct only in the limit $M \rightarrow
\infty$, $M=100$ is already sufficient for the majority attack
\cite{Shacham:2004:CAN}.

However, as all attacking networks learn the same internal
representation, the internal overlap $\rho_i^{EE}$ is increased by the
resulting attractive effect:
\begin{equation}
  \langle \Delta \rho_i^{EE} \rangle = \frac{1}{2} \, \langle \Delta
  \rho_\mathrm{a}(\rho_i^{EE}) \rangle \,.
\end{equation}
Hence $\rho_i^{EE}$ grows faster than $\rho_i^{AE}$ in these steps, so
that the advantage of the majority vote decreases whenever it is used
\cite{Kang:1995:LPP}.

\begin{figure}
  \centering
  \includegraphics[scale=0.5]{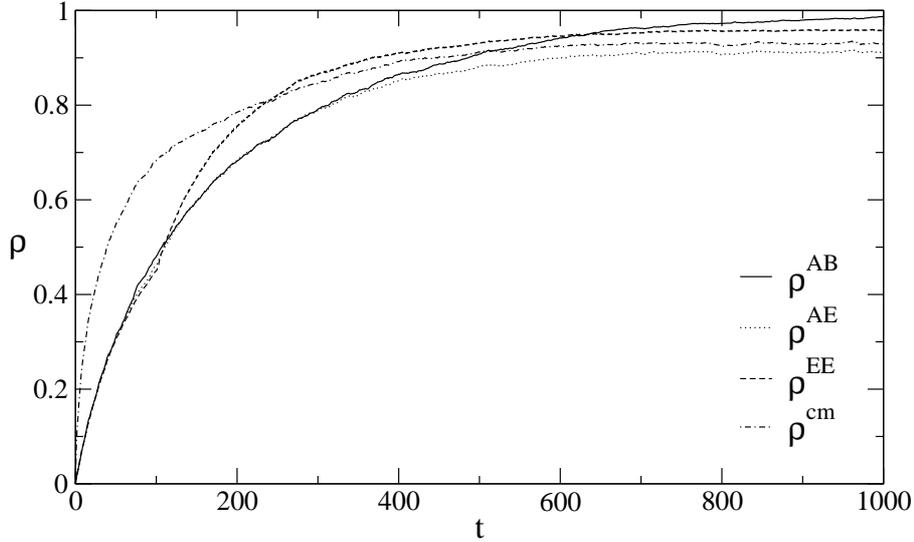}
  \caption{Process of synchronization in the case of a majority attack
    with $M=100$ attacking networks, averaged over $1000$ simulations
    for $K=3$, $L=5$, $N=1000$, and random walk learning rule.}
  \label{fig:overlap}
\end{figure}

This is clearly visible in figure~\ref{fig:overlap}. In the first
$100$ steps the attacker only uses the geometric attack. Here
$\rho^{EE} \approx \rho^{AE}$, which can also be observed for an
ensemble of perceptrons learning according to the Bayes rule
\cite{Engel:2001:SML}. At $t = 100$, using the majority vote gives E a
huge advantage compared to the geometric attack, because
$\rho^\mathrm{cm} \approx \sqrt{\rho^{AE}} > \rho^{AE}$, so that the
probability of repulsive steps is reduced. Therefore the attacker is
able to maintain $\rho^{AE} \approx \rho^{AB}$ for some time. Later
$\rho^{EE}$ increases and this benefit vanishes.

\begin{figure}
  \centering
  \includegraphics[scale=0.5]{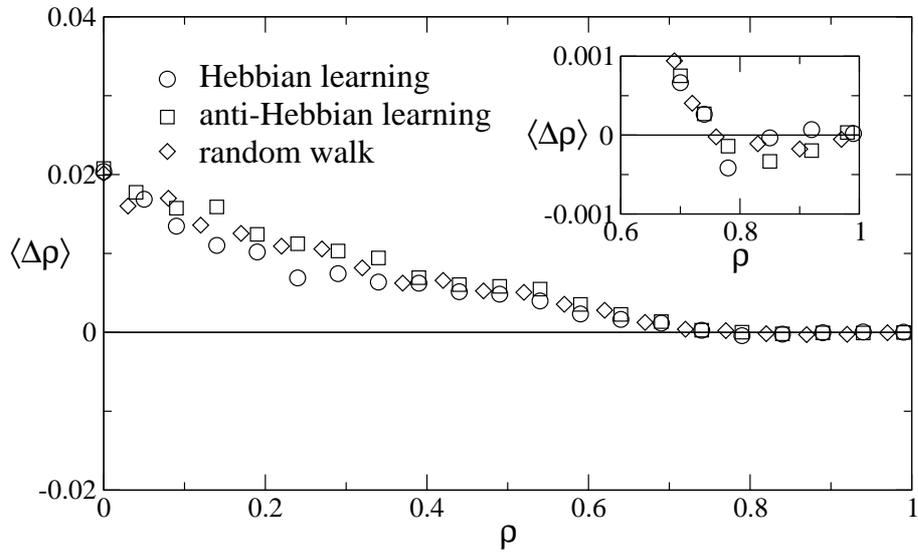}
  \caption{Average change of the overlap for the majority attack with
    $M=100$ attacking networks. Symbols denote results obtained in
    $200$ simulations using $K=3$, $L=5$, and $N=1000$.}
  \label{fig:maverage}
\end{figure}

However, the attacker is unable to reach full synchronization on
average. As shown in figure~\ref{fig:maverage}, there is still a fixed
point at $\rho_\mathrm{f} < 1$ in the case of the majority attack,
although its distance to the absorbing state is smaller than for the
geometric attack. Consequently, one expects a higher success
probability $P_E$, but similar scaling behavior.

\begin{figure}
  \centering
  \includegraphics[scale=0.5]{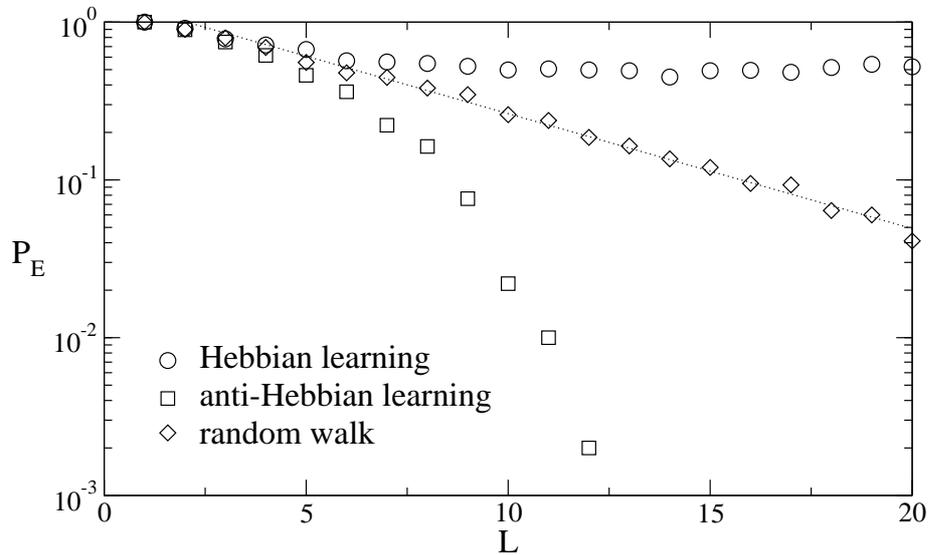}
  \caption{Success probability $P_E$ of the majority attack with
    $M=100$ attacking networks for $K=3$ and $N=1000$. Symbols denote
    results obtained in $10\,000$ simulations and the line shows the
    corresponding fit from figure~\ref{fig:rsuccess}.}
  \label{fig:majority}
\end{figure}

Figure~\ref{fig:majority} shows that this is indeed the case for the
random walk learning rule and for anti-Hebbian learning. But if A and
B use the Hebbian learning rule instead, $P_E$ reaches a constant
non-zero value in the limit $L \rightarrow \infty$
\cite{Shacham:2004:CAN}. Apparently, the change of the weight
distribution caused by Hebbian learning is enough to break the
security of the neural key-exchange protocol. Consequently, A and B
cannot use this learning rule for cryptographic purposes.

While anti-Hebbian learning is secure against the majority attack, a
lot of finite size effects occur in smaller systems, which do not
fulfill the condition $L \ll \sqrt{N}$. In this case $\langle
t_\mathrm{sync} \rangle$ increases faster than $L^2$ as shown in
section \ref{sec:synctime}. Fortunately, A and B can avoid this
problem by just using the random walk learning rule.

\subsection{Comparison of the attacks}
\label{sec:seccomp}

As E knows the parameters of the neural key-exchange protocol, she is
able to select the best method for an attack. Consequently, one has to
compare the available attacks in order to determine the maximum of
$P_E$.

\begin{figure}
  \centering
  \includegraphics[scale=0.5]{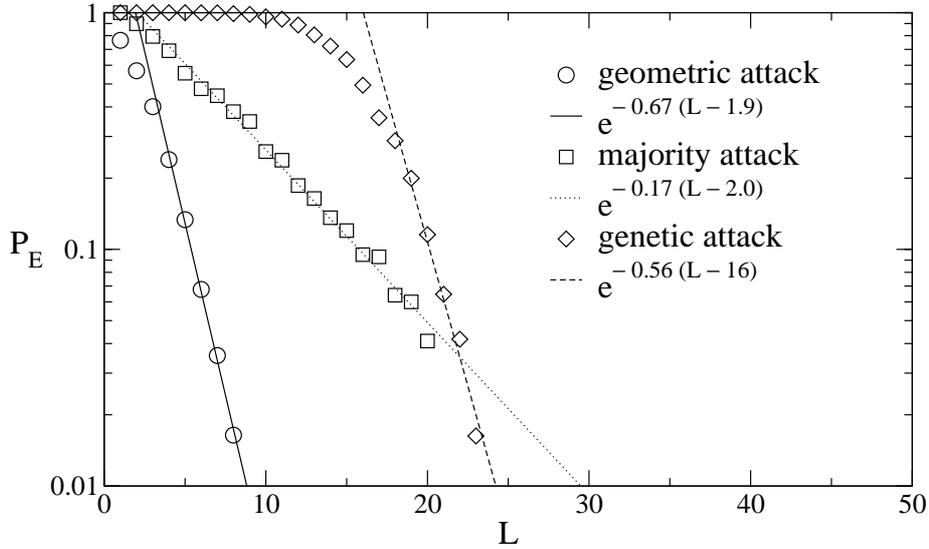}
  \caption{Success probability of different attacks as a function of
    the synaptic depth $L$. Symbols denote results obtained in $1000$
    simulations using the random walk learning rule, $K=3$, and
    $N=1000$, while the lines show fit results using model
    (\ref{eq:pe}). The number of attacking networks is $M=4096$ for
    the genetic attack and $M=100$ for the majority attack.}
  \label{fig:rsuccess}
\end{figure}

Figure~\ref{fig:rsuccess} shows the result. Here (\ref{eq:pe}) has
been used as fit model even for the geometric attack, which is a
special case of both advanced attacks for $M=1$. Of course, by doing
so the curvature visible in figure~\ref{fig:pgeo} is neglected, so
that extrapolating $P_E$ for $L \rightarrow \infty$ overestimates the
efficiency of this method.

All three attacks have similar scaling behavior, but the coefficients
$L_0$ and $y$ obtained by fitting with (\ref{eq:pe}) depend
on the chosen method. The geometric attack is the simplest method
considered in figure~\ref{fig:rsuccess}. Therefore its success
probability is lower than for the more advanced methods. As the
exponent $y$ is large, A and B can easily secure the key-exchange
protocol against this method by just increasing $L$.

In the case of the majority attack, $P_E$ is higher, because the
cooperation between the attacking networks reduces the coefficient
$y$. A and B have to compensate this by further stepping up $L$. In
contrast, the genetic attack merely increases $L_0$, but $y$ does not
change significantly compared to the geometric attack. Therefore the
genetic attack is only better if $L$ is not too large. Otherwise E
gains most by using the majority attack \cite{Ruttor:2006:GAN}.

While A and B can reach any desired level of security by increasing
the synaptic depth, this is difficult to realize in practise.
Extrapolation of (\ref{eq:pe}) shows clearly that $P_E \approx
10^{-4}$ is achieved for $K=3$, $L=57$, $N=1000$, and random walk
learning rule. But the average synchronization time $\langle
t_\mathrm{sync} \rangle \approx 1.6 \cdot 10^5$ is quite large in this
case. Consequently, it is reasonable to develop an improved neural
key-exchange protocol \cite{Mislovaty:2003:PCC, Ruttor:2004:NCF,
  Ruttor:2005:NCQ}, which is the topic of chapter \ref{chap:queries}.

\section{Security by interaction}
\label{sec:secint}

The main difference between the partners and the attacker in neural
cryptography is that A and B are able to influence each other by
communicating their output bits $\tau^A$ and $\tau^B$, while E can
only listen to these messages. Of course, A and B use their advantage
to select suitable input vectors for adjusting the weights. As shown
in chapter~\ref{chap:dynamics} this finally leads to different
synchronization times for partners and attackers.

However, there are more effects, which show that the two-way
communication between A and B makes attacking the neural key-exchange
protocol more difficult than simple learning of examples. These
confirm that the security of neural cryptography is based on the
bidirectional interaction of the partners.

\subsection{Version space}

The time series of pairs $(\tau^A, \tau^B)$ of output bits produced by
two interacting Tree Parity Machines depends not only on the sequence
of input vectors $\mathbf{x}_i(t)$, but also on the initial weight
vectors $\mathbf{w}_i^{A/B}(0)$ of both neural networks. Of course, E
can reproduce the time series $\tau^B(t)$ exactly, if she uses a third
Tree Parity Machine with $\mathbf{w}_i^E(0) = \mathbf{w}_i^B(0)$,
because the learning rules are deterministic. But choosing other
initial values for the weights in E's neural network may lead to the
same sequence of output bits. The set of initial configurations with
this property is called \emph{version space} \cite{Engel:2001:SML}.
Its size $n_\mathrm{vs}$ is a monotonically decreasing function of the
length $t$ of the sequence, because each new element $\mathbf{x}_i(t)$
imposes an additional condition on $\mathbf{w}_i^E(0)$. Of course, it
does not matter whether E uses the simple attack or the geometric
attack, as both algorithms are identical under the condition
$\tau^E(t) = \tau^B(t)$.

\begin{figure}
  \centering
  \includegraphics[scale=0.5]{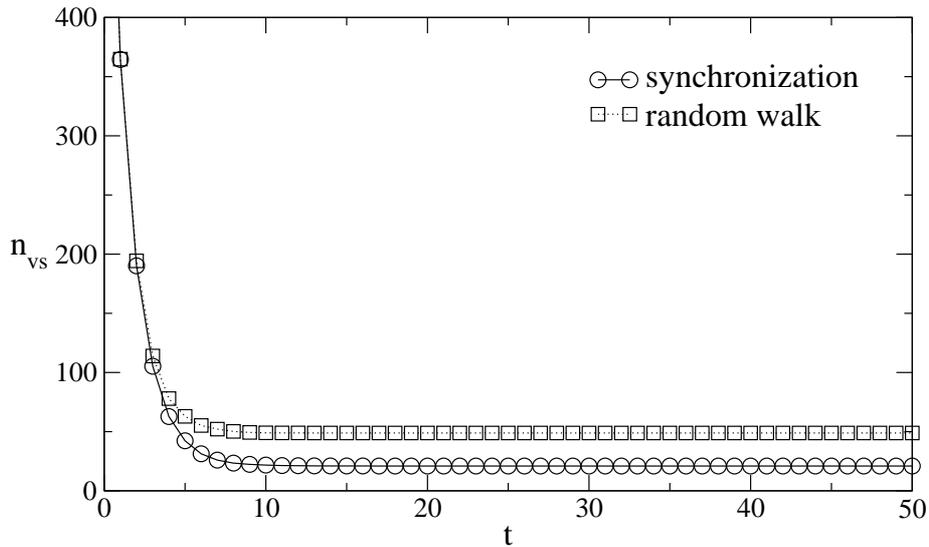}
  \caption{Version space of interacting Tree Parity Machines with
    $K=3$, $L=1$, $N=2$, and random walk learning rule, averaged over
    $1000$ simulations.}
  \label{fig:vspace}
\end{figure}

Figure~\ref{fig:vspace} shows that the size of the version space
shrinks by a factor $1/2$ in each step at the beginning of the time
series. Here the possible configurations of the weights are still
uniformly distributed, so that each output $\tau^B$ gives one bit of
information about the initial conditions.

But later neural networks which have started with similar weight
vectors synchronize. That is why the configurations are no longer
uniformly distributed and the shrinking of the $n_\mathrm{vs}$ becomes
slower and slower. Finally, all Tree Parity Machines in the version
space have synchronized with each other. From then on $n_\mathrm{vs}$
stays constant.

However, the size of the version space in the limit $t \rightarrow
\infty$ depends on the properties of the time series. If A and B start
fully synchronized, they do not need to influence each other and all
input vectors in the sequence are used to update the weights. In this
case E has to learn randomly chosen examples of a time-dependent rule
\cite{Metzler:2000:INN}. In contrast, if A and B start with
uncorrelated weight vectors, they select a part of the input sequence
for adjusting their weights. For them this interaction increases the
speed of the synchronization process, because only suitable input
vectors remain. But it also decreases $n_\mathrm{vs}(t \rightarrow
\infty)$, so that imitating B is more difficult for E.

Consequently, the two-way communication between the partners gives
them an advantage, which cannot be exploited by a passive attacker.
Therefore bidirectional interaction is important for the security of
the neural key-exchange protocol.

\subsection{Mutual information}

Instead of using the overlap $\rho$ as order parameter, which is
closely related to the dynamics of the neural networks and the theory
of learning, one can look at the process of synchronization from the
point of view of information theory, too. For this purpose, the mutual
information $I^{AB}(t)$ defined in (\ref{eq:minfo}) describes A's and
B's knowledge about each other. Similarly $I^{AE}(t)$ measures, how
much information E has gained in regard to A at time $t$ by listening
to the communication of the partners.

All participants start with zero knowledge about each other, so that
$I^{AB} = 0$ and $I^{AE} = 0$ at the beginning of the key exchange. In
each step there are several sources of information. The input vectors
$\mathbf{x}_i(t)$ determine, in which directions the weights are
moved, if the learning rule is applied.  And, together with the
outputs $\tau^A(t)$ and $\tau^B(t)$, they form an example, which gives
some information about the current weight vectors in A's and B's Tree
Parity Machines.  Although all participants have access to
$\mathbf{x}_i(t)$, $\tau^A(t)$, and $\tau^B(t)$, the increase of the
mutual information $I$ depends on the algorithm used to adjust the
weights.

\begin{figure}
  \centering
  \includegraphics[scale=0.5]{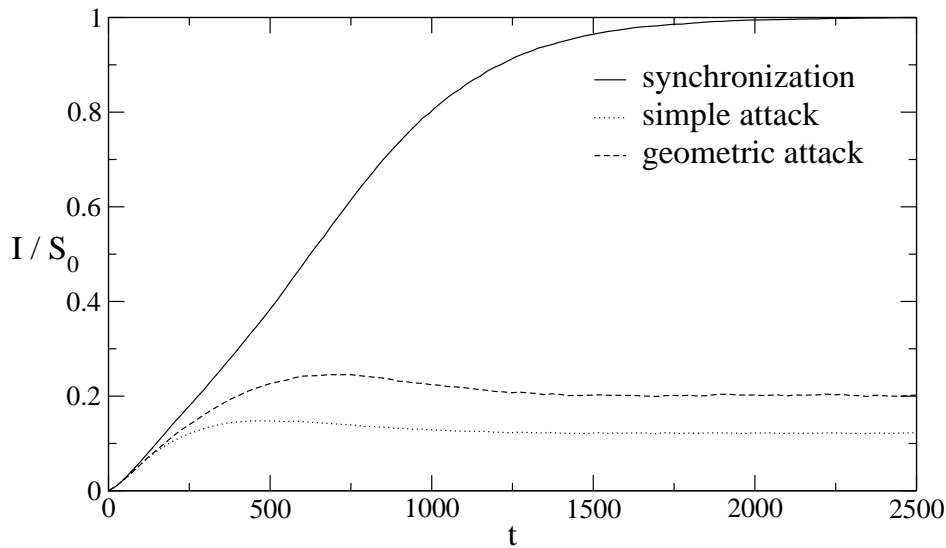}
  \caption{Mutual information between partners and attackers. Symbols
    denote simulation results for $K=3$, $L=5$, $N=1000$, and random
    walk learning rule, while the lines show the results of
    corresponding iterative calculations for $N \rightarrow \infty$.}
  \label{fig:minfo}
\end{figure}

This is clearly visible in figure~\ref{fig:minfo}. While the partners
reach full synchronization with $I^{AB} = S_0$ quickly, the attacker
is much slower. And E performs better if she uses the geometric method
instead of the simple attack. Of course, these observations correspond
to those presented in chapter~\ref{chap:dynamics}.

While the differences between E's attack methods are caused by the
learning algorithms, which transform the available information into
more or less knowledge about A's and B's weights, this is not the only
reason for $I^{AE}(t) < I^{AB}(t)$. In order to synchronize the
partners have to agree on some weight vectors $\mathbf{w}_i$, which
are, in fact, functions of the sequence $\mathbf{x}_i(t)$ and the
initial conditions $\mathbf{w}_i^{A/B}(0)$. So they already have some
information, which they share during the process of synchronization.
Therefore the partners gain more mutual information $I$ from each
message than an attacker, who has no prior knowledge about the outcome
of the key exchange. Consequently, it is the bidirectional interaction
which gives A and B an advantage over E.

\section{Number of keys}
\label{sec:keys}

Although all attacks on the neural key-exchange protocol known up to
now are based on the training of Tree Parity Machines with examples
generated by the partners A and B, this is not a necessary condition.
Instead of that the opponent could try a brute-force attack. Of
course, a success of this method is nearly impossible without
additional information, as there are $(2 L + 1)^{K N}$ different
configurations for the weights of a Tree Parity Machine. However, E
could use some clever algorithm to determine which keys are generated
with high probability for a given input sequence. Trying out these
would be a feasible task as long as there are not too many.
Consequently, a large number of keys is important for the security of
neural cryptography, especially against brute-force attacks.

\subsection{Synchronization without interaction}

If the weights are chosen randomly, there are $(2 L + 1)^{2 K N}$
possible configurations for a pair of Tree Parity Machines. But the
neural key-exchange protocol can generate at most $(2 L + 1)^{K N}$
different keys. Consequently, sets of different initial conditions
exist, which lead to identical results. That is why synchronization
even occurs without any interaction between neural networks besides a
common sequence of input vectors.

In order to analyze this effect the following system consisting of two
pairs of Tree Parity Machines is used:
\begin{eqnarray}
  \mathbf{w}_i^{A+} &=& g(\mathbf{w}_i^A + f(\sigma_i^A, \tau^A,
  \tau^B) \mathbf{x}_i) \,, \\
  \mathbf{w}_i^{B+} &=& g(\mathbf{w}_i^B + f(\sigma_i^B, \tau^A,
  \tau^B) \mathbf{x}_i) \,, \\
  \mathbf{w}_i^{C+} &=& g(\mathbf{w}_i^C + f(\sigma_i^C, \tau^C,
  \tau^D) \mathbf{x}_i) \,, \\
  \mathbf{w}_i^{D+} &=& g(\mathbf{w}_i^D + f(\sigma_i^D, \tau^C,
  \tau^D) \mathbf{x}_i) \,.
\end{eqnarray}
All four neural networks receive the same sequence of input vectors
$\mathbf{x}_i$, but both pairs communicate their output bits only
internally. Thus A and B as well as C and D synchronize using one of
the available learning rules, while correlations caused by common
inputs are visible in the overlap $\rho_i^{AC}$. Because of the
symmetry in this system, $\rho_i^{AD}$, $\rho_i^{BC}$, and
$\rho_i^{BD}$ have the same properties as this quantity, so that it is
sufficient to look at $\rho_i^{AC}$ only.

Of course, synchronization of networks which do not interact with each
other, is much more difficult and takes a longer time than performing
the normal key-exchange protocol. Thus full internal synchronization
of the pairs usually happens well before A's and C's weight vectors
become identical, so that $\rho_i^{AB} = 1$ and $\rho_i^{CD} = 1$ are
assumed for the calculation of $\langle \Delta
\rho_i^{AC}(\rho_i^{AC}) \rangle$.

As before, both attractive and repulsive steps are possible. In the
case of identical overlap between corresponding hidden units, random
walk learning rule, and $K > 1$, the probability of these step types
is given by:
\begin{eqnarray}
  P_\mathrm{a} &=& \frac{1}{2} \sum_{i=0}^{(K-1)/2} {K - 1 \choose 2
    i} (1 - \epsilon)^{K - 2 i} \, \epsilon^{2 i} \nonumber \\
  &+& \frac{1}{2} \sum_{i=0}^{(K-1)/2} {K - 1 \choose 2 i} (1 -
  \epsilon)^{K - 2 i - 1} \epsilon^{2 i + 1} \,, \\
  P_\mathrm{r} &=& \sum_{i=1}^{K/2} {K - 1 \choose 2 i - 1} (1 -
  \epsilon)^{K - 2 i} \epsilon^{2 i} \nonumber \\
  &+& \sum_{i=1}^{K/2} {K - 1 \choose 2 i - 1} (1 - \epsilon)^{K -
    2 i + 1} \, \epsilon^{2 i - 1} \,.
\end{eqnarray}
Here $\epsilon$ denotes the generalization error defined in equation
(\ref{eq:generr}) in regard to $\rho^{AC}$. For $K=1$ only attractive
steps occur, so that $P_\mathrm{a}=1$, which is similar to the
geometric attack. But in the case of $K=3$, one finds
\begin{eqnarray}
  P_\mathrm{a} &=& \frac{1}{2} \left[ 1 - 2 (1 - \epsilon) \epsilon
  \right] \,, \\
  P_\mathrm{r} &=& 2 (1 - \epsilon) \epsilon \,.
\end{eqnarray}
As long as $\rho^{AC}>0$ the probability of repulsive steps is higher
than $P_\mathrm{r}^E=\epsilon$ for the simple attack. Consequently,
one expects that the dynamics of $\rho^{AC}$ has a fixed point at
$\rho_\mathrm{f}^{AC} < \rho_\mathrm{f}^E < 1$ and synchronization is
only possible by fluctuations.

\begin{figure}
  \centering
  \includegraphics[scale=0.5]{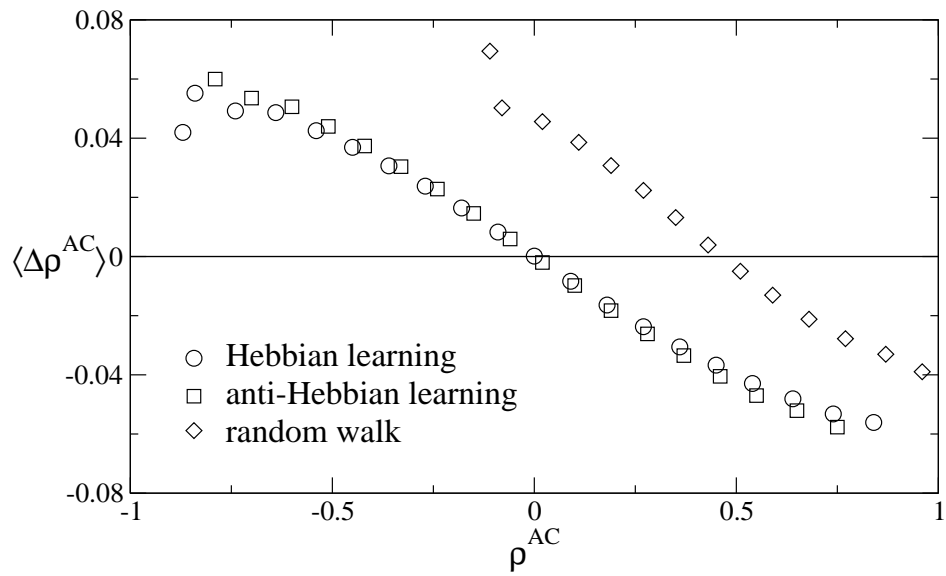}
  \caption{Average change of the overlap between A and C for $K=3$,
    $L=3$, and $N=1000$, obtained in $100$ simulations with $100$
    pairs of neural networks.}
  \label{fig:nkstep}
\end{figure}

Figure~\ref{fig:nkstep} shows that this is indeed the case. As more
repulsive steps occur, the probability for full synchronization here
is much smaller than for a successful simple attack. In fact, large
enough fluctuations which lead from $\rho^{AC}=0$ to $\rho^{AC}=1$
without interaction only occur in small systems. But the common input
sequence causes correlations between $\mathbf{w}_i^A$ and
$\mathbf{w}_i^C$ even for $L \gg 1$ and $N \gg 1$.

\begin{figure}
  \centering
  \includegraphics[scale=0.5]{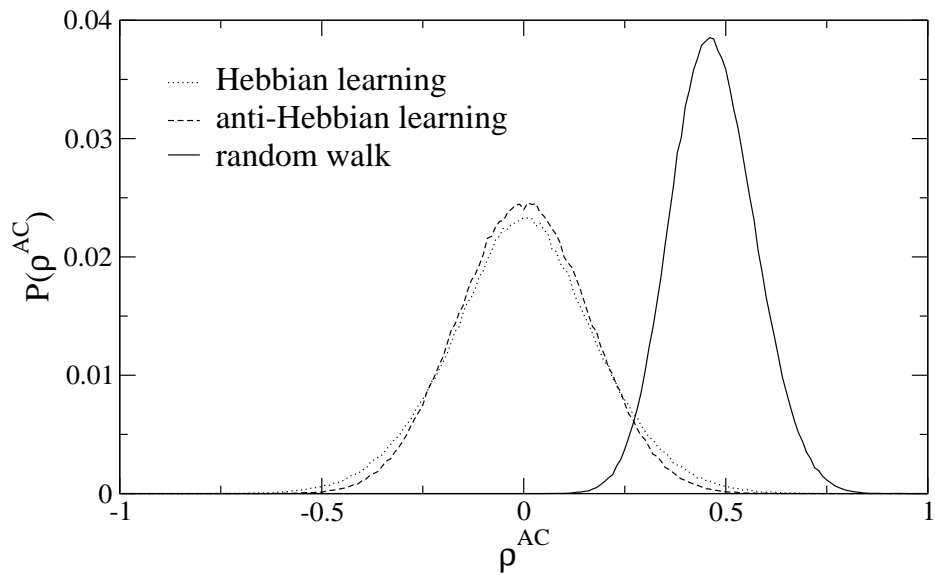}
  \caption{Distribution of the overlap $\rho^{AC}$ after $1000$ steps
    for $K=3$, $L=3$, and $N=100$, obtained in $100$ simulations with
    $100$ pairs of Tree Parity Machines.}
  \label{fig:nkdist}
\end{figure}

This is clearly visible in figure~\ref{fig:nkdist}. However, if
Hebbian or anti-Hebbian learning is used instead of the random walk
learning rule, one observes a somewhat different behavior: the fixed
point of the dynamics for $K=3$ is located at $\rho_\mathrm{f}=0$.
According to these two learning rules the weights of corresponding
hidden units move in opposite directions, if both $\tau^A \not=
\tau^C$ and $\sigma_i^A \not= \sigma_i^C$. The average step size of
such an \emph{inverse attractive} step is given by
\begin{equation}
  \langle \Delta \rho_\mathrm{i}(\rho) \rangle = - \langle \Delta
  \rho_\mathrm{a}(-\rho) \rangle  \,.
\end{equation}
While $P_\mathrm{r}$ is independent of the learning rule, one finds
\begin{eqnarray}
  P_\mathrm{a} &=& \frac{1}{2} \sum_{i=0}^{(K-1)/2} {K - 1 \choose 2
    i} (1 - \epsilon)^{K - 2 i} \, \epsilon^{2 i} \,, \\
  P_\mathrm{i} &=& \frac{1}{2} \sum_{i=0}^{(K-1)/2} {K - 1 \choose 2
    i} (1 - \epsilon)^{K - 2 i - 1} \epsilon^{2 i + 1}
\end{eqnarray}
for Hebbian or anti-Hebbian learning and $K>1$. If $K$ is odd, the
effects of all types of steps cancel out exactly at $\rho=0$, because
$\langle \Delta \rho_\mathrm{r}(0) \rangle = 0$ and \linebreak
$P_\mathrm{i}(0) = P_\mathrm{a}(0)$. Otherwise, the two transition
probabilities, $P_\mathrm{a}(0)$ for attractive steps and
$P_\mathrm{i}(0)$ for inverse attractive steps, are only approximately
equal. Thus one observes $\rho_\mathrm{f} \approx 0$ independent of
$K$, so that the correlations between A and C are smaller in this case
than for the random walk learning rule.

But if the initial overlap between A and C is already large, it is
very likely that both pairs of Tree Parity Machines generate the same
key regardless of the learning rule. Consequently, the number of keys
$n_\mathrm{key}$ is smaller than the number of weight configurations
$n_\mathrm{conf} = (2 L + 1)^{K N}$ of a Tree Parity Machine.

\subsection{Effective key length}

In order to further analyze the correlations between A's and C's
neural networks the entropy 
\begin{equation}
  S^{AC} = \sum_{i=1}^K S_i^{AC}
\end{equation}
of their weight distribution is used. Here $S_i^{AC}$ is the entropy
of a single hidden unit defined in (\ref{eq:entropy}). Of course, one
can assume here that the weights stay uniformly distributed during the
process of synchronization, either because the system size is large
($N \gg 1$) or the random walk learning rule is used. Therefore the
entropy of a single network is given by $S_0 = K N \ln (2 L + 1)$.

Consequently, $S^{AC} - S_0$ is the part of the total entropy, which
describes the correlations caused by using a common input sequence. It
is proportional to the \emph{effective length} of the generated
cryptographic key,
\begin{equation}
  l_\mathrm{key} = \frac{S^{AC} - S_0}{\ln 2} \,,
\end{equation}
which would be the average number of bits needed to represent it using
both an optimal encoding without redundancy and the input sequence as
additional knowledge. If the possible results of the neural
key-exchange protocol are uniformly distributed, each one can be
represented by a number consisting of $l_\mathrm{key}$ bits. In this
case
\begin{equation}
  \label{eq:keys}
  n_\mathrm{key} = 2^{l_\mathrm{key}} = e^{S^{AC} - S_0}
\end{equation}
describes exactly the number of keys which can be generated using
different initial weights for the same input sequence. Otherwise, the
real number is larger, because mainly configurations, which occur with
high probability, are relevant for the calculation of $S^{AC}$.
However, an attacker is only interested in those prevalent keys.
Therefore $n_\mathrm{key}$ as defined in equation (\ref{eq:keys}) is,
in fact, a lower bound for the number of cryptographic relevant
configurations.

\begin{figure}
  \centering
  \includegraphics[scale=0.5]{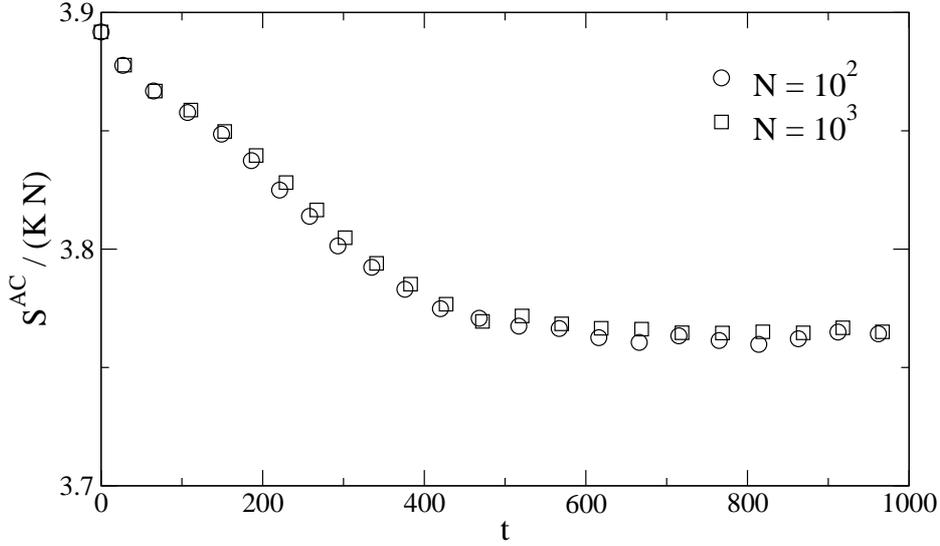}
  \caption{Entropy per weight for A and C for $K=3$, $L=3$, and random
    walk learning rule, obtained in $100$ simulations with $100$ pairs
    of neural networks.}
  \label{fig:nkentropy}
\end{figure}

\begin{figure}
  \centering
  \includegraphics[scale=0.5]{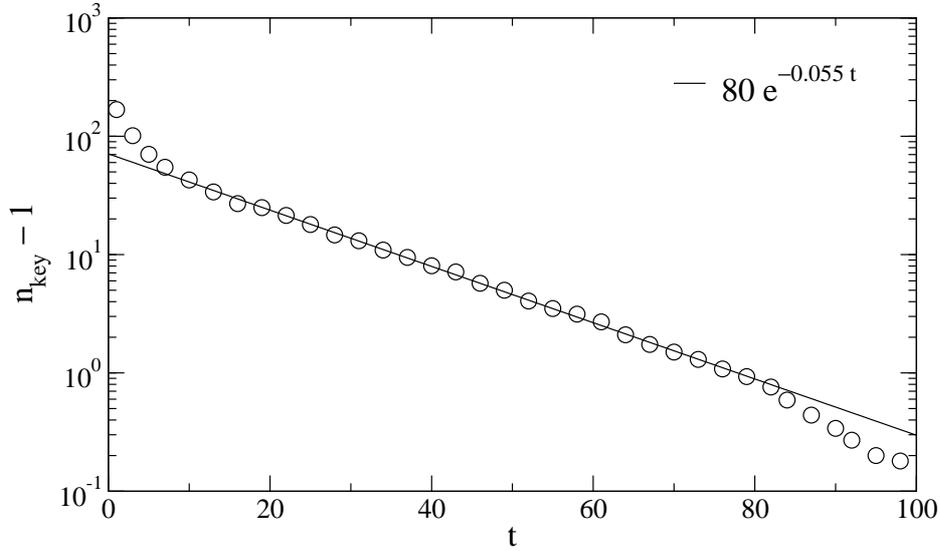}
  \caption{Number of keys for $K=3$, $L=1$, $N=2$, and random walk
    learning rule, obtained by exhaustive search and averaged over
    $100$ random input sequences.}
  \label{fig:keys}
\end{figure}

Figure~\ref{fig:nkentropy} shows the time evolution of the entropy.
First $S^{AC}$ shrinks linearly with increasing $t$, as the overlap
$\rho$ between A and C grows, while it approaches the stationary
state. This behavior is consistent with an exponentially decreasing
number of keys, which can be directly observed in very small systems
as shown in figure~\ref{fig:keys}. Of course, after the system has
reached the fixed point, the entropy stays constant. This minimum
value of the entropy is then used to determine the effective number
$n_\mathrm{key}$ of keys according to (\ref{eq:keys}).

It is clearly visible that there are two scaling relations for
$S^{AC}$:
\begin{itemize}
\item Entropy is an extensive quantity. Thus both $S^{AC}$ and $S_0$
  are proportional to the number of weights $N$. Consequently, the
  number of keys, which can be generated by the neural key-exchange
  protocol for a given input sequence, grows exponentially with
  increasing system size $N$.
\item The relevant time scale for all processes related to the
  synchronization of Tree Parity Machines is defined by the step sizes
  of attractive and repulsive steps which are asymptotically
  proportional to $L^{-2}$. Therefore the time needed to reach the
  fixed point $\rho_\mathrm{f}^{AC}$ is proportional to $L^2$, similar
  to $\langle t_\mathrm{sync} \rangle$. In fact, it is even of the
  same order as the average synchronization time.
\end{itemize}

\begin{figure}
  \centering
  \includegraphics[scale=0.5]{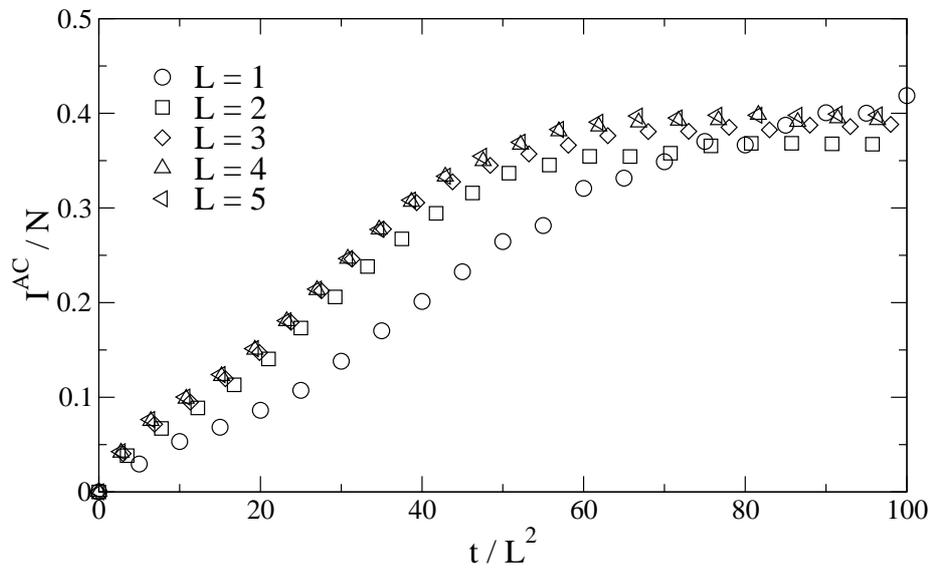}
  \caption{Mutual information between A and C for $K=3$, $N=1000$, and
    random walk learning rule, obtained in $1000$ simulations with $10$
    pairs of neural networks.}
  \label{fig:nkinfo}
\end{figure}

Instead of using the entropy directly, it is better to look at the
mutual information $I^{AC} = 2 S_0 - S^{AC}$ shared by A and C, which
comes from the common input sequence and is visible in the
correlations of the weight vectors. Using (\ref{eq:s0}) and
(\ref{eq:keys}) leads to \cite{Ruttor:2007:DNC}
\begin{equation}
  I^{AC} = - \ln \left( \frac{n_\mathrm{key}}{n_\mathrm{conf}} \right)
  \,.
\end{equation}
Therefore the effective number of keys is given by
\begin{equation}
  n_\mathrm{key} = n_\mathrm{conf} \, e^{- I^{AC}} = (2 L + 1)^{K N}
  e^{- I^{AC}} \,.
\end{equation}
As shown in figure~\ref{fig:nkinfo} the mutual information $I^{AC}$ at
the end of the synchronization process becomes asymptotically
independent of the synaptic depth in the limit $L \rightarrow \infty$.
Consequently, the ratio $n_\mathrm{key} / n_\mathrm{conf}$ is constant
except for finite-size effects occurring in small systems.

\begin{figure}
  \centering
  \includegraphics[scale=0.5]{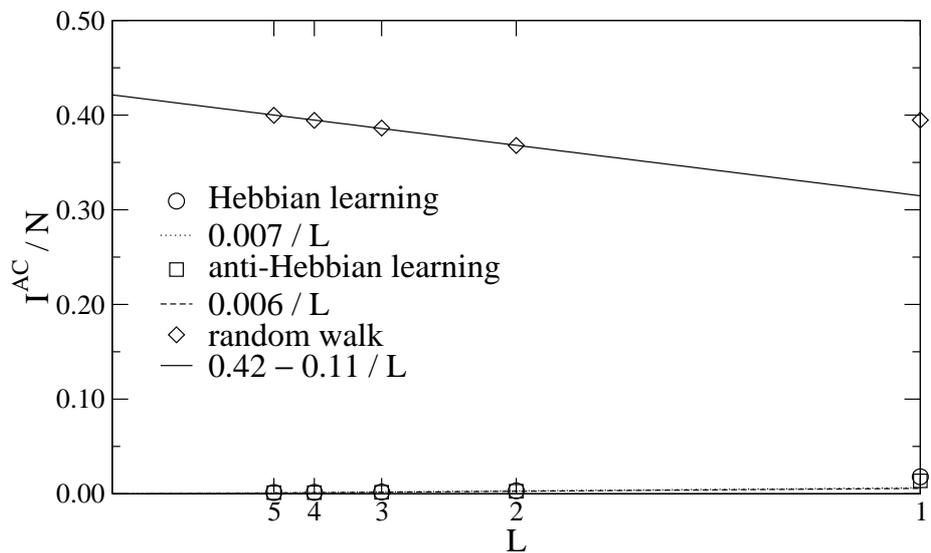}
  \caption{Extrapolation of $I^{AC}$ for $K=3$, $L \rightarrow \infty$,
    and $N=1000$. Symbols denote the average value of $I^{AC}(t)$ in
    the range $80 \, L^2 \leq t \leq 100 \, L^2$, which has been
    obtained in $1000$ simulations with $10$ pairs of neural
    networks.}
  \label{fig:nkextra}
\end{figure}

The amount of correlations between A and C depends on the distribution
of the overlap $\rho^{AC}$ in the steady state, which can be described
by its average value $\rho_\mathrm{f}$ and its standard deviation
$\sigma_\mathrm{f}$. As before, $\sigma_\mathrm{f}$ decreases
proportional to $L^{-1}$ due to diminishing fluctuations in the limit
$L \rightarrow \infty$, while $\rho_\mathrm{f}$ stays nearly constant.
Hence $I^{AC}$ consists of two parts, one independent of $L$ and one
proportional to $L^{-1}$ as shown in figure \ref{fig:nkextra}. 

In the case of the random walk learning rule the mutual information
increases with $L$, because fluctuations are reduced which just
disturb the correlations created by the common sequence of input
vectors. Extrapolating $I^{AC}$ yields the result
\cite{Ruttor:2007:DNC}
\begin{equation}
  n_\mathrm{key} \approx \left[ 0.66 (2 L + 1)^3 \right]^N \,,
\end{equation}
which is valid for $K=3$ and $1 \ll L \ll \sqrt{N}$. Consequently,
$n_\mathrm{key}$ grows exponentially with $N$, so that there are
always enough possible keys in larger systems to prevent successful
brute-force attacks on the neural key-exchange protocol.

Using Hebbian or anti-Hebbian learning, however, improves the
situation further. Because of $\rho_\mathrm{f}=0$ one finds
$n_\mathrm{key} \rightarrow n_\mathrm{conf}$ in the limit $L
\rightarrow \infty$. Therefore the input sequence does not restrict
the set of possible keys in very large systems using $K=3$, $1 \ll L
\ll \sqrt{N}$, and one of these two learning rules.

\section{Secret inputs}

The results of section \ref{sec:keys} indicate that the input vectors
are an important source of information for the attacker. Thus keeping
$\mathbf{x}_i$ at least partially secret should improve the security of
the neural key-exchange protocol.

\subsection{Feedback mechanism}

\begin{figure}[h]
  \centering
  \includegraphics{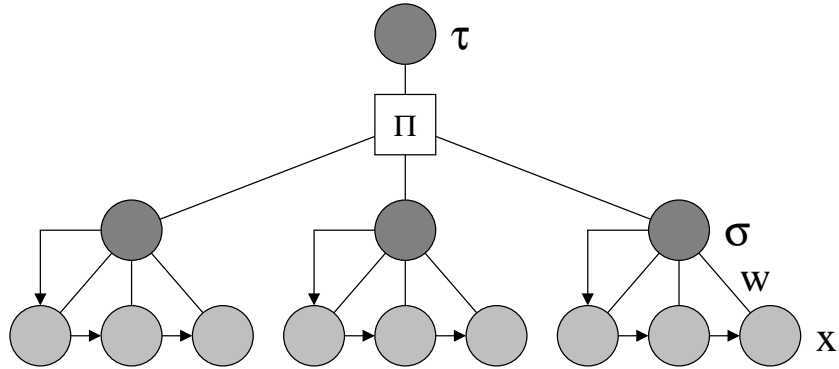}
  \caption{A Tree Parity Machine with $K=3$, $N=3$, and feedback.}
  \label{fig:ctpm}
\end{figure}

In order to reduce the amount of input vectors transmitted over the
public channel, the partners have to use an alternative source for the
input bits.  For that purpose they can modify the structure of the
Tree Parity Machines, which is shown in figure \ref{fig:ctpm}
\cite{Ruttor:2004:NCF}.

Here the generation of the input values is different. Of course, A and
B still start with a set of $K$ randomly chosen public inputs
$\mathbf{x}_i$. But in the following time steps each input vector is
shifted,
\begin{equation}
  x_{i,j}^{A/B+} = x_{i,j-1}^{A/B} \quad \mbox{for $j>1$,}
\end{equation}
and the output bit $\sigma_i$ of the corresponding hidden unit is used
as the new first component,
\begin{equation}
  x_{i,1}^{A/B+} = \sigma_i^{A/B} \,.
\end{equation}
This feedback mechanism \cite{Eisenstein:1995:GPT} replaces the public
sequence of random input vectors. Additionally, the anti-Hebbian
learning rule (\ref{eq:hebb_minus}) is used to update the weights. By
doing so one avoids the generation of trivial keys, which would be the
result of the other learning rules \cite{Ruttor:2004:NCF}. Thus the
hidden units of both Tree Parity Machines work as \emph{confused bit
  generators} \cite{Metzler:2001:GUT}.

\begin{figure}
  \centering
  \includegraphics[scale=0.5]{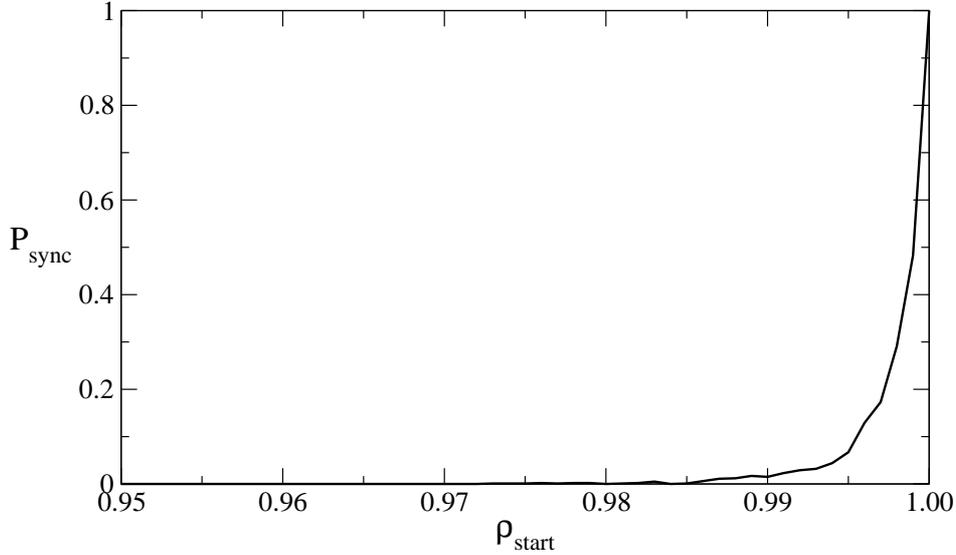}
  \caption{Probability of synchronization for two Tree Parity Machines
    with feedback as a function of the initial overlap
    $\rho_\mathrm{start}$. Symbols denote results obtained in $1000$
    simulations with $K=3$, $L=3$, and $N=100$.}
  \label{fig:fbsync}
\end{figure}

However, synchronization is not possible without further information,
as the bit sequence produced by such a neural network is unpredictable
\cite{Eisenstein:1995:GPT, Bialek:2001:PCL, Zhu:1998:APS} for another
one of the same type \cite{Metzler:2001:GUT, Ruttor:2004:NCF}. This is
clearly visible in figure~\ref{fig:fbsync}. The reason is that the
input vectors of A's and B's Tree Parity Machines become more and more
different, because each occurrence of $\sigma_i^A \not= \sigma_i^B$
reduces the number of identical input bits by one for the next $N$
steps. Of course, the partners disagree on the outputs $\sigma_i$
quite often at the beginning of the synchronization process, so that
they soon have completely uncorrelated input vectors and mutual
learning is no longer possible.

\subsection{Synchronization with feedback}

As the feedback mechanism destroys the common information about the
inputs of the Tree Parity Machines, an additional mechanism is
necessary for synchronization, which compensates this detrimental
effect sufficiently. For that purpose A and B occasionally reset the
input vectors of their Tree Parity Machines if too many steps with
$\tau^A \not= \tau^B$ occur.

In fact, the following algorithm is used \cite{Ruttor:2004:NCF}:
\begin{itemize}
\item If $\tau^A = \tau^B$, the weights are updated according to the
  anti-Hebbian learning rule (\ref{eq:hebb_minus}) and the feedback
  mechanism is used to generate the next input.
\item If the output bits disagree, $\tau^A \not= \tau^B$, the input
  vectors are shifted, too, but all pairs of input bits
  $x_{i,1}^{A/B}$ are set to common public random values.
\item After $R$ steps with different output, $\tau^A \not= \tau^B$,
  all inputs are reinitialized using a set of $K$ randomly chosen
  public input vectors.
\end{itemize}
Of course, setting $R=0$ leads to synchronization without feedback,
while no synchronization is possible in the limit $R \rightarrow
\infty$.

\begin{figure}
  \centering
  \includegraphics[scale=0.5]{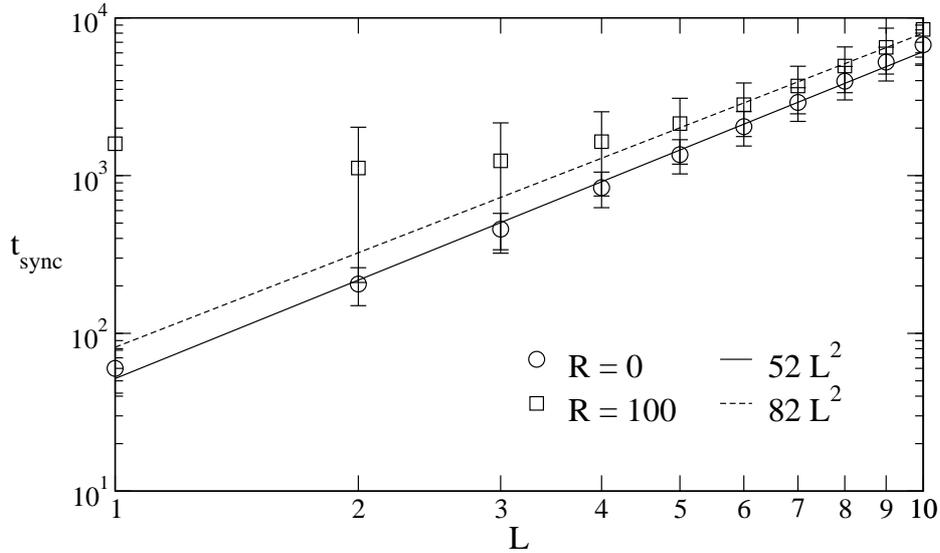}
  \caption{Average synchronization time and its standard deviation for
    neural cryptography with feedback, obtained in $10\,000$ simulations
    with $K=3$ and $N=10\,000$.}
  \label{fig:fbtime}
\end{figure}

Figure \ref{fig:fbtime} shows that using the feedback mechanism
increases the average number of steps needed to achieve full
synchronization. While there are strong finite size-effects, the
scaling relation $\langle t_\mathrm{sync} \rangle \propto L^2$ is
still valid for $1 \ll L \ll \sqrt{N}$. Only the constant of
proportionality is larger than before \cite{Ruttor:2004:NCF}.

\begin{figure}
  \centering
  \includegraphics[scale=0.5]{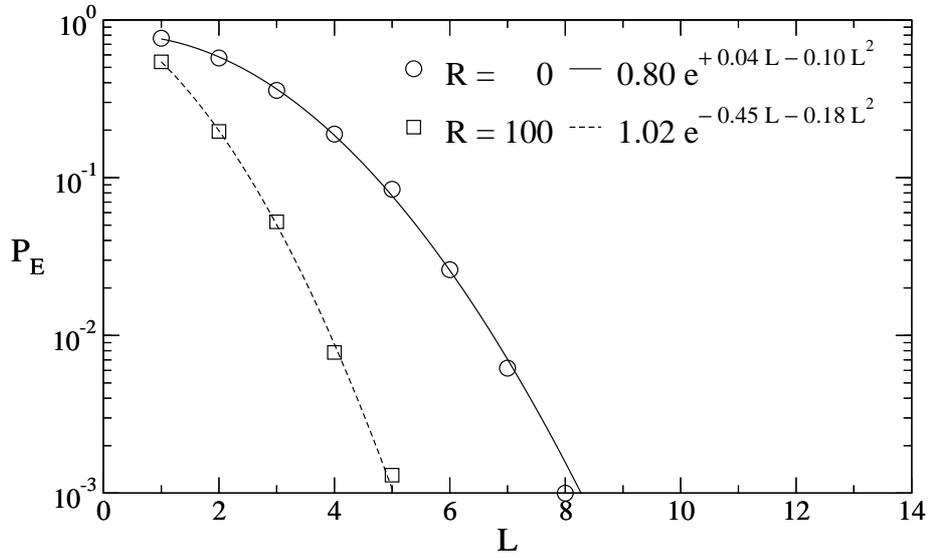}
  \caption{Success probability $P_E$ of the geometric attack as a
    function of the synaptic depth $L$. Symbols denote results
    averaged over $10\,000$ simulations for $K=3$ and $N=1000$.}
  \label{fig:fbsuccess}
\end{figure}

\begin{figure}
  \centering
  \includegraphics[scale=0.5]{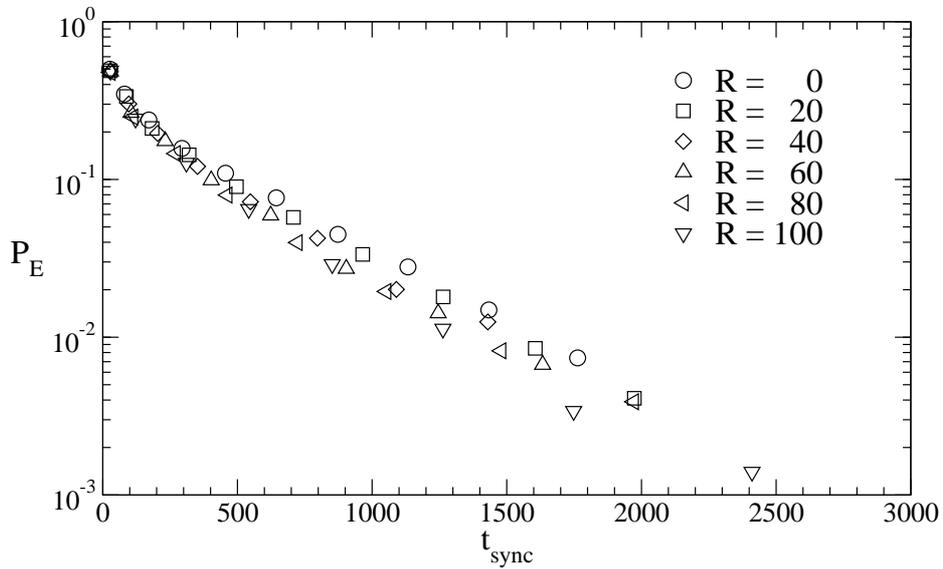}
  \caption{Success probability $P_E$ of the geometric attack as a
    function of the average synchronization time $\langle
    t_\mathrm{sync} \rangle$. Symbols denote results of $10\,000$
    iterative calculations for $K=3$ and $N \rightarrow \infty$. Here
    successful synchronization has been defined as $\rho > 0.9$
    \cite{Ruttor:2004:NCF}.}
  \label{fig:fbresult}
\end{figure}

As shown in figure \ref{fig:fbsuccess} a similar result can be
observed in regard to the success probability of the geometric attack.
As before, $P_E$ drops exponentially with increasing synaptic depth, so
that A and B can achieve any desired level of security by changing
$L$. But with feedback smaller values of $L$ are sufficient, because
the factors $y_1$ and $y_2$ in the scaling law (\ref{eq:success}) are
larger. Therefore using the feedback mechanism improves the security
of neural cryptography by keeping input values partially secret.

However, A and B usually want to keep their effort constant. Then one
has to look at the function $P_E(\langle t_\mathrm{sync} \rangle)$
instead of $P_E(L)$, which is plotted in figure \ref{fig:fbresult} for
several values of the feedback parameter $R$. It is clearly visible
that $P_E(\langle t_\mathrm{sync} \rangle)$ does not depend much on
$R$. Consequently, using feedback only yields a small improvement of
security unless the partners accept an increase of the average
synchronization time \cite{Ruttor:2004:NCF}.

\subsection{Key exchange with authentication}

Synchronization of Tree Parity Machines by mutual learning only works
if they receive a common sequence of input vectors. This effect can be
used to implement an authentication mechanism for the neural
key-exchange protocol \cite{Volkmer:2004:ATP, Volkmer:2006:EAA}.

For that purpose each partner uses a separate, but identical
pseudo-random number generator. As these devices are initialized with
a secret seed state shared by A and B, they produce exactly the same
sequence of bits, which is then used to generate the input vectors
$\mathbf{x}_i$ needed during the synchronization process. By doing so
A and B can synchronize their neural networks without transmitting
input values over the public channel.

Of course, an attacker does not know the secret seed state. Therefore
E is unable to synchronize due to the lack of information about the
input vectors. Even an active man-in-the-middle attack does not help
in this situation, although it is always successful for public inputs.

Consequently, reaching full synchronization proves that both
participants know the secret seed state. Thus A and B can authenticate
each other by performing this variant of the neural key exchange. As
one cannot derive the secret from the public output bits, it is a
zero-knowledge protocol \cite{Volkmer:2004:ATP}.

\chapter{Key exchange with queries}
\label{chap:queries}

The process of neural synchronization is driven by the sequence of
input vectors, which are really used to adjust the weights of the Tree
Parity Machines according to the learning rule. As these are selected
by the partners participating in the key exchange, A and B have an
important advantage over E, who can only listen to their
communication. Up to now the partners just avoid repulsive steps by
skipping some of the randomly generated input vectors.

However, they can use their advantage in a better way. For this
purpose the random inputs are replaced by queries
\cite{Kinzel:1990:ING}, which A and B choose alternately according to
their own weight vectors. In fact, the partners ask each other
questions and learn only the answers, on which they reach an
agreement.

Of course, the properties of the synchronization process now depend
not only on the synaptic depth $L$ of the Tree Parity Machines, but
also on the chosen queries. Thus there is an additional parameter $H$,
which fixes the absolute value $|h_i|$ of the local fields in the
neural network generating the current query. As the prediction error
of a hidden unit is a function of both the overlap $\rho_i$ and the
local field $h_i$, the partners modify the probability of repulsive
steps $P_\mathrm{r}(\rho)$ if they change $H$. By doing so A and B are
able to adjust the difficulty of neural synchronization and learning
\cite{Ruttor:2005:NCQ}.

In order to achieve a secure key exchange with queries the partners
have to choose the parameter $H$ in such a way that they synchronize
quickly, while an attacker is not successful. Fortunately, this is
possible for all known attacks \cite{Ruttor:2006:GAN}. Then one finds
the same scaling laws again, which have been observed in the case of
synchronization with random inputs. But because of the new parameter
$H$ one can reach a higher level of security for the neural
key-exchange protocol without increasing the average synchronization
time \cite{Ruttor:2005:NCQ, Ruttor:2006:GAN}.

However, queries make additional information available to the
attacker, as E now knows the absolute value of the local fields in
either A's or B's hidden units. In principle, this information might
be used in specially adapted methods. But knowing $H$ does not help E
in the case of the geometric attack and its variants, so that using
queries does not introduce obvious security risks.

\section{Queries}

In the neural key-exchange protocol as proposed in
\cite{Kanter:2002:SEI} the input vectors $\mathbf{x}_i$ are generated
randomly and independent of the current weight vectors
$\mathbf{w}_i^{A/B}$ of A's and B's Tree Parity Machines. Of course,
by interacting with each other the partners are able to select which
inputs they want to use for the movements of the weights. But they use
their influence on the process of synchronization only for skipping
steps with $\tau^A \not= \tau^B$ in order to avoid repulsive effects.
Although this algorithm for choosing the relevant inputs is sufficient
to achieve a more or less secure key-exchange protocol, A and B could
improve it by taking more information into account.

In contrast, E uses the local field $h_i^E$ of the hidden units in her
Tree Parity Machines in order to correct their output bits
$\sigma_i^E$ if necessary. While this algorithm, which is part of all
known attack methods except the simple attack, is not suitable for A
and B, they could still use the information contained in $h_i^{A/B}$.
Then the probability for $\sigma_i^A \not= \sigma_i^B$ or $\sigma_i^E
\not= \sigma_i^{A/B}$ is no longer given by the generalization error
(\ref{eq:generr}), but by the prediction error (\ref{eq:perr}) of the
perceptron \cite{Ein-Dor:1999:CPN}.

Consequently, the partners are able to distinguish input vectors
$\mathbf{x}_i$ which are likely to cause either attractive or
repulsive steps if they look at the local field. In fact, A's and B's
situation is quite similar to E's in the case of the geometric attack.
A low value of $|h_i^{A/B}|$ indicates a high probability for
$\sigma_i^A \not= \sigma_i^B$. These input vectors may slow down the
process of synchronization due to repulsive effects, so that it is
reasonable to omit them. And a high value of $|h_i^{A/B}|$ indicates
that $\sigma_i^E = \sigma_i^{A/B}$ is very likely, which would help E.
Therefore A and B could try to select only input vectors
$\mathbf{x}_i$ with $|h_i| \approx H$ for the application of the
learning rule, whereas the parameter $H$ has to be chosen carefully in
order to improve the security of the neural key-exchange protocol.

While it is indeed possible to use the random sequence of input
vectors and just skip unsuitable ones with $|h_i| \not\approx H$, this
approach does not work well. If the range of acceptable local fields
is small, then a lot of steps do not change the weights and $\langle
t_\mathrm{sync} \rangle$ increases. But otherwise only small effects
can be observed, because most input vectors with $\tau^A = \tau^B$ are
accepted as before.

That is why the random inputs are replaced by queries
\cite{Kinzel:1990:ING}, so that the partners ask each other questions,
which depend on their own weight vectors $\mathbf{w}_i^{A/B}$. In odd
(even) steps A (B) generates $K$ input vectors $\mathbf{x}_i$ with
$h_i^A \approx \pm H$ ($h_i^B \approx \pm H$) using the algorithm
presented in appendix \ref{chap:qgen}. By doing so it is not necessary
to skip steps in order to achieve the desired result: the absolute
value of the local field $h_i$ is approximately given by the parameter
$H$, while its sign $\sigma_i$ is chosen randomly
\cite{Ruttor:2005:NCQ}.

\begin{figure}
  \centering
  \includegraphics[scale=0.5]{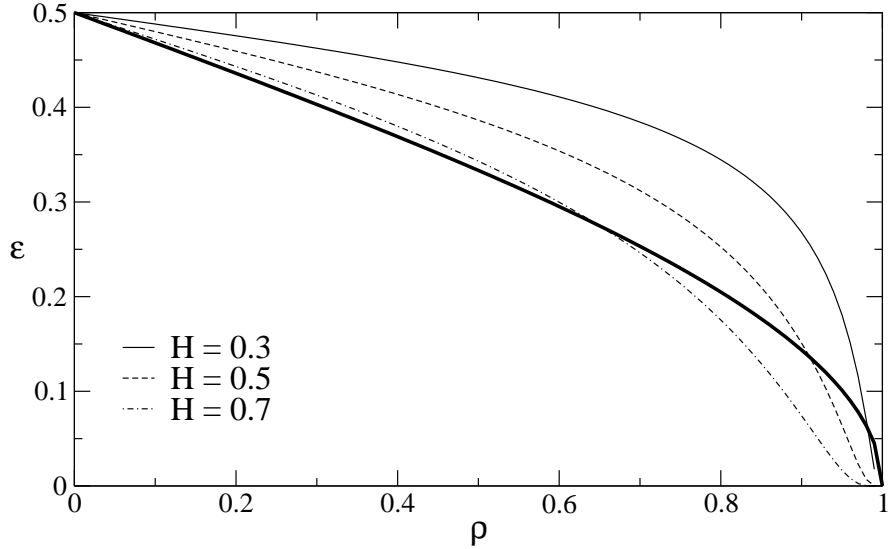}
  \caption{Probability of disagreeing hidden units in the case of
    queries with different parameter $H$ and $Q=1$. The thick line
    shows $P(\sigma_i^A \not= \sigma_i^B)$ for random inputs.}
  \label{fig:qerror}
\end{figure}

As shown in figure \ref{fig:qerror} using queries affects the
probability that two corresponding hidden units disagree on their
output $\sigma_i$. Compared to the case of a random input sequence,
this event occurs more frequently for small overlap, but less for
nearly synchronized neural networks. Hence queries are especially a
problem for the attacker. As learning is slower than synchronization,
$\rho_i^\mathrm{AE}$ is typically smaller than $\rho_i^\mathrm{AB}$.
In this situation queries increase the probability of repulsive steps
for the attacker, while the partners are able to regulate this effect
by choosing $H$ in a way that it does not interfere much with their
process of synchronization. Consequently, using queries gives A and B
more control over the difficulty of both synchronization and learning.

\section{Synchronization time}
\label{sec:qtime}

Because queries change the relation between the overlap $\rho_i^{AB}$
and the probability of repulsive steps $P_\mathrm{r}^B(\rho_i^{AB})$,
using them affects the number of steps needed to reach full
synchronization. But the neural key-exchange protocol is only useful
in practice, if the synchronization time $t_\mathrm{sync}$ is not too
large. Otherwise, there is no advantage compared to classical
algorithms based on number theory. This condition, of course,
restricts the usable range of the new parameter $H$.

\begin{figure}
  \centering
  \includegraphics[scale=0.5]{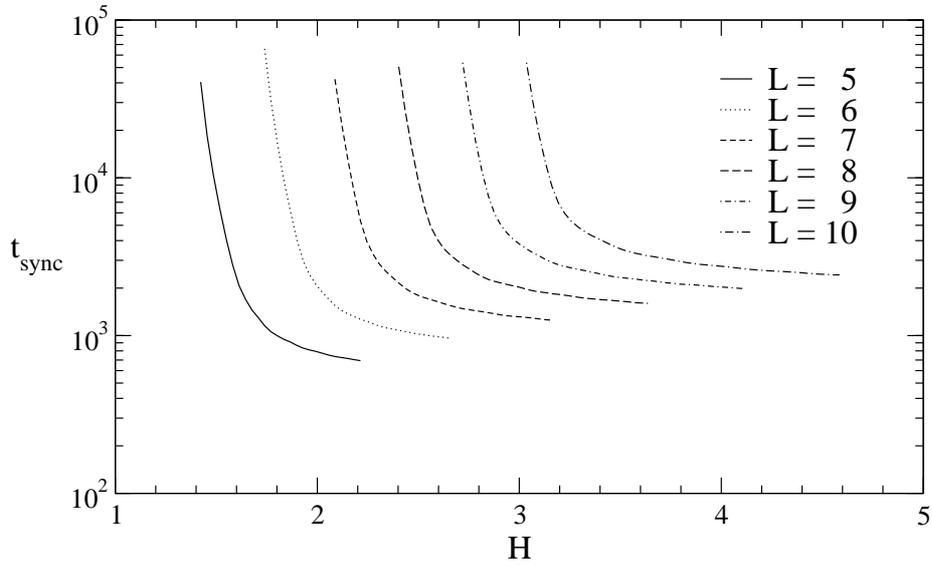}
  \caption{Synchronization time of two Tree Parity Machines with
    $K=3$, $N=1000$, and random walk learning rule, averaged over
    $10\,000$ simulations.}
  \label{fig:qtime}
\end{figure}

As shown in \ref{fig:qtime}, $\langle t_\mathrm{sync} \rangle$
diverges for $H \rightarrow 0$. In this limit the prediction error
$\epsilon^\mathrm{p}$ reaches $1/2$ independent of the overlap, so
that the effect of the repulsive steps inhibits synchronization. But
as long as $H$ is chosen large enough, it does not have much influence
on the effort of generating a key \cite{Ruttor:2005:NCQ}.

\begin{figure}
  \centering
  \includegraphics[scale=0.5]{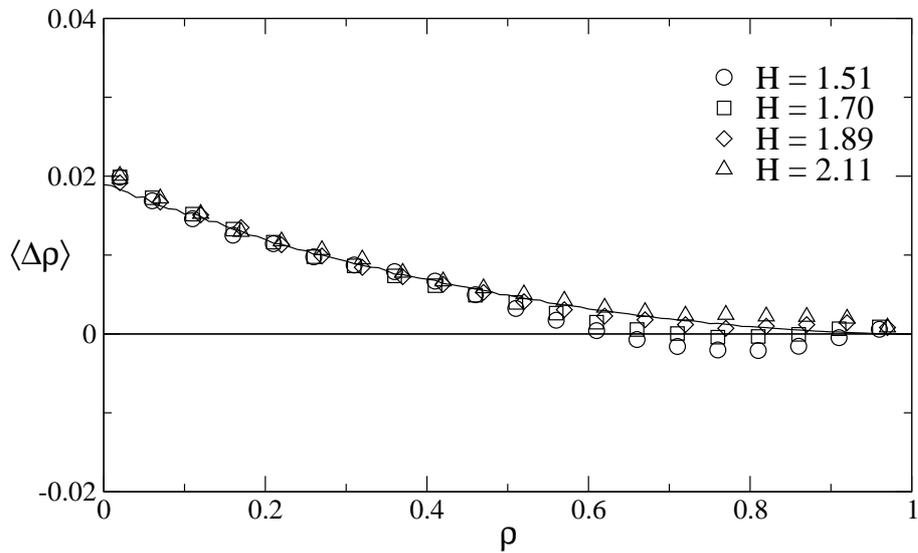}
  \caption{Average change of the overlap for synchronization with
    queries using $K=3$, $L=5$, $N=1000$, and the random walk learning
    rule. Symbols denote results obtained in $10\,000$ simulations,
    while the line shows $\langle \Delta \rho \rangle$ for
    synchronization with random inputs.}
  \label{fig:qaverage}
\end{figure}

In fact, A and B can switch the mechanism of synchronization by
modifying $H$. This is clearly visible in figure \ref{fig:qaverage}.
If the absolute value of the local fields is so large that $\langle
\Delta \rho \rangle > 0$ for all $\rho < 1$, synchronization on
average happens, which is similar to the normal key-exchange protocol
using a random sequence of input vectors. But decreasing $H$ below a
certain value $H_\mathrm{f}$ creates a new fixed point of the dynamics
at $\rho_\mathrm{f} < 1$. In this case synchronization is only
possible by fluctuations. As the gap with $\langle \Delta \rho \rangle
< 0$ grows with decreasing $H < H_\mathrm{f}$, one observes a steep
increase of the average synchronization time $\langle t_\mathrm{sync}
\rangle$. If A and B use the random walk learning rule together with
$K=3$, $L=5$, and $N=1000$, one finds $H_\mathrm{f} \approx 1.76$ in
this case.

Additionally, figure~\ref{fig:qtime} shows a dependency on the
synaptic depth $L$ of $\langle t_\mathrm{sync} \rangle$, which is
caused by two effects \cite{Ruttor:2005:NCQ}:
\begin{itemize}
\item The speed of synchronization is proportional to the step sizes
  $\langle \Delta \rho_\mathrm{a} \rangle$ for attractive and $\langle
  \Delta \rho_\mathrm{r} \rangle$ for repulsive steps. As shown in
  section~\ref{sec:steps}, these quantities decrease proportional to
  $L^{-2}$. Therefore the average synchronization time increases
  proportional to the square of the synaptic depth as long as $H >
  H_\mathrm{f}$:
  \begin{equation}
    \label{eq:qtscale}
    \langle t_\mathrm{sync} \rangle \propto L^2 \,.
  \end{equation}
  This causes the vertical shift of the curves in
  figure~\ref{fig:qtime}.
\item If queries are used, the probabilities $P_\mathrm{a}$ for
  attractive and $P_\mathrm{r}$ for repulsive steps depend not only on
  the overlap $\rho_i$, but also on quantity $H / \sqrt{Q_i}$
  according to (\ref{eq:perr}). In the case of the random walk
  learning rule the weights stay uniformly distributed, so that the
  length of the weight vectors grows proportional to $L$ as shown in
  section \ref{sec:wdist}. That is why one has to increase $H$
  proportional to the synaptic depth,
  \begin{equation}
    \label{eq:hcscale}
    H = \alpha L \,,
  \end{equation}
  in order to achieve identical transition probabilities and
  consequently the same average synchronization time. This explains
  the horizontal shift of the curves in figure \ref{fig:qtime}.
\end{itemize}
Using both scaling laws (\ref{eq:qtscale}) and (\ref{eq:hcscale}) one
can rescale $\langle t_\mathrm{sync} \rangle$ in order to obtain
functions $f_L(\alpha)$ which are nearly independent of the synaptic
depth except for finite-size effects \cite{Ruttor:2005:NCQ}:
\begin{equation}
  \langle t_\mathrm{sync} \rangle = L^2 f_L \! \left( \frac{H}{L}
  \right) \,.
\end{equation}

\begin{figure}
  \centering
  \includegraphics[scale=0.5]{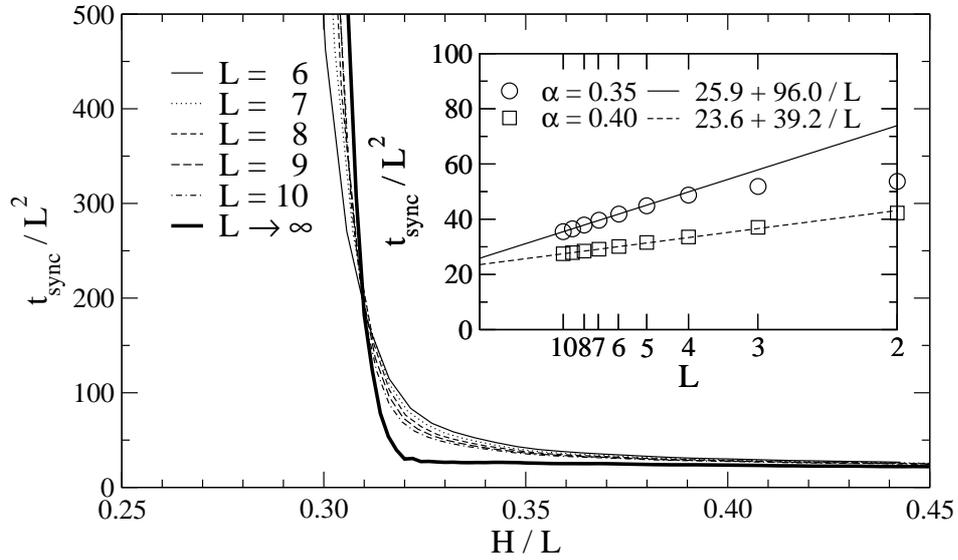}
  \caption{Scaling behavior of the synchronization time. The thick
    curve denotes the universal function $f(\alpha)$ defined in
    (\ref{eq:qf}). It has been obtained by finite-size scaling, which
    is shown in the inset.}
  \label{fig:qtscale}
\end{figure}

Figure \ref{fig:qtscale} shows these functions for different
values of $L$. It is clearly visible that $f_L(\alpha)$ converges to a
universal scaling function $f(\alpha)$ in the limit $L \rightarrow
\infty$:
\begin{equation}
  \label{eq:qf}
  f(\alpha) = \lim_{L \rightarrow \infty} f_L(\alpha) \,.
\end{equation}
Additionally, the finite-size effects have a similar behavior in
regard to $L$ as the fluctuations of the overlap $\rho_i$, which have
been analyzed in section \ref{sec:fsync}. That is why the distance
$|f_L(\alpha) - f(\alpha)|$ shrinks proportional to $L^{-1}$.
Therefore the universal function $f(\alpha)$ can be determined by
finite-size scaling, which is shown in figure \ref{fig:qtscale}, too.

\begin{figure}
  \centering
  \includegraphics[scale=0.5]{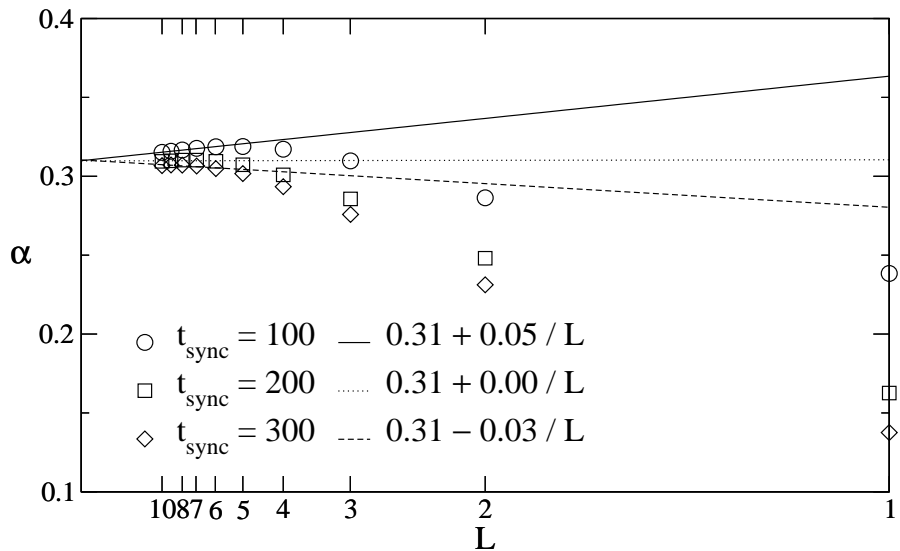}
  \caption{Extrapolation of the inverse function $f_L^{-1}$ to $L
    \rightarrow \infty$. Symbols denote the values extracted from
    figure \ref{fig:qtscale} for different average synchronization
    times.}
  \label{fig:qtinv}
\end{figure}

This function diverges for $\alpha < \alpha_\mathrm{c}$. The critical
value $\alpha_\mathrm{c} = H_\mathrm{c} / L$ can be estimated by
extrapolating the inverse function $f_L^{-1}$, which is shown in
\ref{fig:qtinv}. By doing so one finds $\alpha_\mathrm{c} \approx
0.31$ for $K=3$ and $N=1000$, if A and B use the random walk learning
rule \cite{Ruttor:2006:GAN}. Consequently, synchronization is only
achievable for $H > \alpha_\mathrm{c} L$ in the limit $L \rightarrow
\infty$. However, in the case of finite synaptic depth synchronization
is even possible slightly below $H_\mathrm{c}$ due to fluctuations
\cite{Ruttor:2005:NCQ}.

\begin{figure}
  \centering
  \includegraphics[scale=0.5]{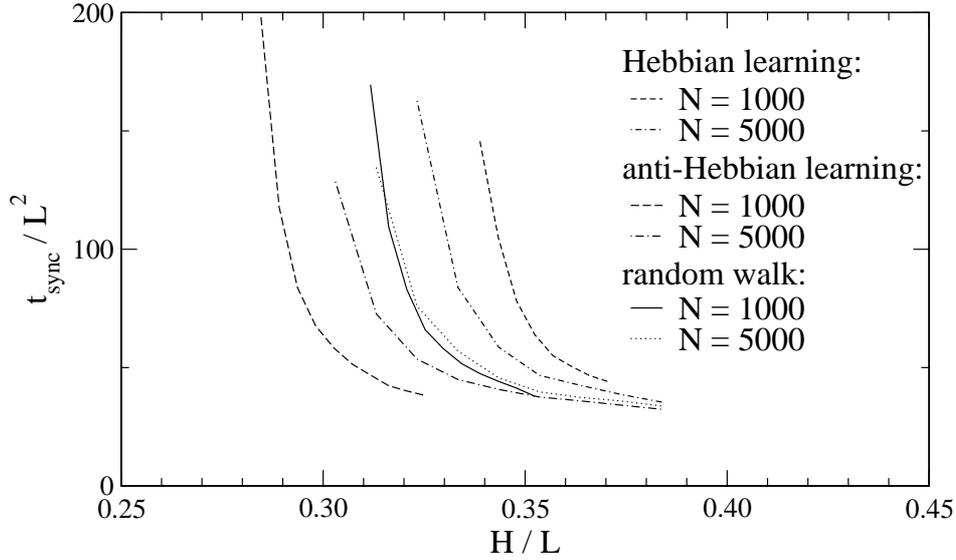}
  \caption{Synchronization time for neural cryptography with queries.
    These results have been obtained in $100$ simulations with $K=3$
    and $L=7$.}
  \label{fig:qtrule}
\end{figure}

Although the weights do not stay uniformly distributed in the case of
Hebbian and anti-Hebbian learning, one observes qualitatively the same
behavior of $\langle t_\mathrm{sync} \rangle$ as a function of the
parameters $H$ and $L$. This is clearly visible in figure
\ref{fig:qtrule}. As the length of the weight vectors is changed by
these learning rules, the critical local field $H_\mathrm{c} =
\alpha_\mathrm{c} L$ for synchronization is different. In the case of
$K=3$ and $N=1000$, one finds $\alpha_\mathrm{c} \approx 0.36$ for
Hebbian learning \cite{Ruttor:2005:NCQ} and $\alpha_\mathrm{c} \approx
0.25$ for anti-Hebbian learning. But in the limit $N \rightarrow
\infty$ the behavior of both learning rules converges to that of the
random walk learning rule \cite{Ruttor:2006:GAN}, which is also
visible in figure~\ref{fig:qtrule}.

\section{Security against known attacks}

Because of the cryptographic application of neural synchronization it
is important that the key-exchange protocol using queries is not only
efficient, but also secure against the attacks known up to now.
Therefore it is necessary to determine how different absolute values
of the local field influence the security of the system. Of course,
the results impose further restrictions upon the usable range of the
parameter $H$.

\subsection{Dynamics of the overlap}

Replacing random inputs with queries gives A and B an additional
advantage over E. Now they can choose a suitable value of the new
parameter $H$, which influences the probability of repulsive steps as
shown in figure \ref{fig:qerror} (on page~\pageref{fig:qerror}). By
doing so the partners are able to modify the dynamics of the
synchronization process, not only for themselves, but also for an
attacker. And because $\langle \Delta \rho^{AB}(\rho) \rangle$ is
greater than $\langle \Delta \rho^{AE}(\rho) \rangle$, A and B can
generate queries in such a way that a fixed point of the dynamics at
$\rho_\mathrm{f} < 1$ only exists for E. Then the neural key-exchange
protocol is secure in principle, because $\langle t_\mathrm{sync}^E
\rangle$ grows exponentially with increasing synaptic depth while
$\langle t_\mathrm{sync}^B \rangle \propto L^2$.

\begin{figure}
  \centering
  \includegraphics[scale=0.5]{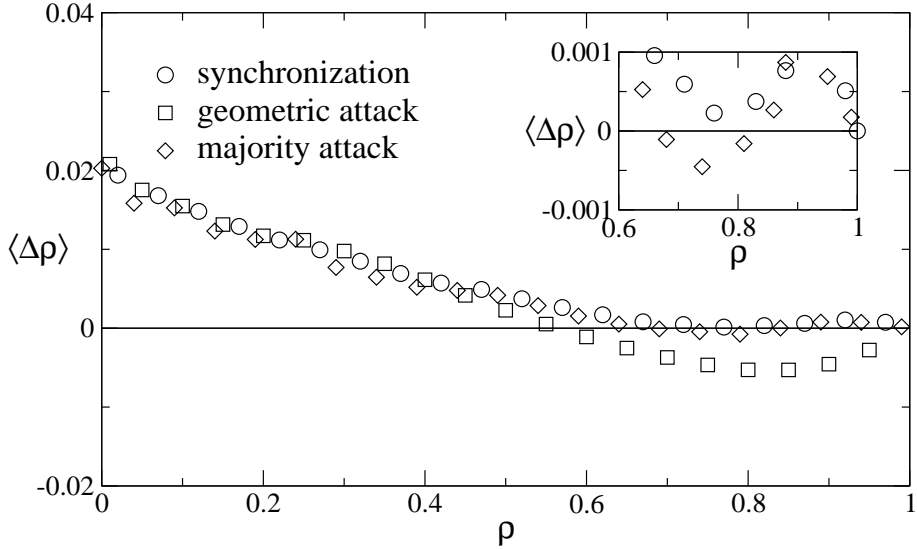}
  \caption{Average change of the overlap for $K=3$, $L=5$, $N=1000$,
    $H=1.77$, and $M=100$. Symbols denote results obtained in $200$
    simulations using the random walk learning rule.}
  \label{fig:qgeo}
\end{figure}

Figure \ref{fig:qgeo} shows that this is indeed possible. Here A and B
have chosen $H \approx H_\mathrm{f}$, so that they just synchronize on
average. In contrast, E can reach the absorbing state at $\rho=1$ only
by fluctuations, as there is a fixed point of the dynamics at
$\rho_\mathrm{f} < 1$ for both the geometric attack and the majority
attack. In principle, this situation is similar to that observed in
the case of random inputs. However, the gap between the fixed point
and the absorbing state is larger, so that the success probability of
both attacks is decreased. This is clearly visible by comparing figure
\ref{fig:qaverage} with figure \ref{fig:average} (on
page~\pageref{fig:average}) and figure \ref{fig:maverage} (on
page~\pageref{fig:maverage}).

\subsection{Success probability}

In practice it is necessary to look at the success probability $P_E$
of the known attacks in order to determine the level of security
provided by neural cryptography with queries.

\begin{figure}
  \centering
  \includegraphics[scale=0.5]{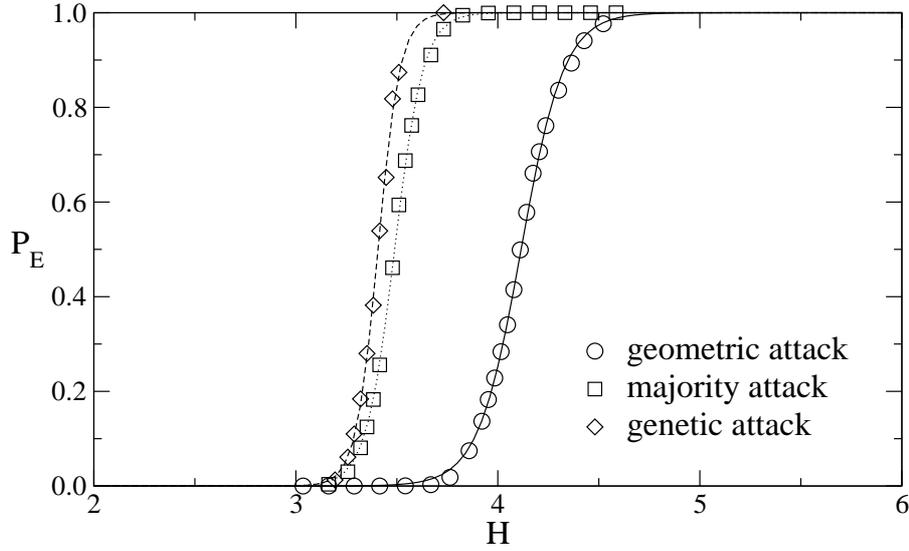}
  \caption{Success probability $P_E$ as a function of $H$. Symbols
    denote the results obtained in $1000$ simulations using $K=3$,
    $L=10$, $N=1000$, and the random walk learning rule, while the
    lines show fit results for model (\ref{eq:fermi}). The number of
    attacking networks is $M=4096$ for the genetic attack and $M=100$
    for the majority attack.}
  \label{fig:qsuccess}
\end{figure}

As shown in figure \ref{fig:qsuccess}, E is nearly always successful
in the case of large $H$, because she is able to synchronize on
average similar to A and B. But if $H$ is small, the attacker can
reach full synchronization only by fluctuations, so that $P_E$ drops
to zero. In fact, one can use a Fermi-Dirac distribution
\begin{equation}
  \label{eq:fermi}
  P_\mathrm{E} = \frac{1}{1 + \exp(-\beta (H - \mu))}
\end{equation}
as a suitable fitting function in order to describe $P_E$ as a
function of $H$. This model is suitable for both the majority attack
\cite{Ruttor:2005:NCQ} and the genetic attack \cite{Ruttor:2006:GAN}.
Of course, one can also use it to describe $P_E(H)$ of the geometric
attack, which is the special case $M=1$ of the more advanced attacks.
Comparing these curves in figure \ref{fig:qsuccess} reveals directly
that the genetic attack is the best choice for the attacker in this
case.

\begin{figure}
  \centering
  \includegraphics[scale=0.5]{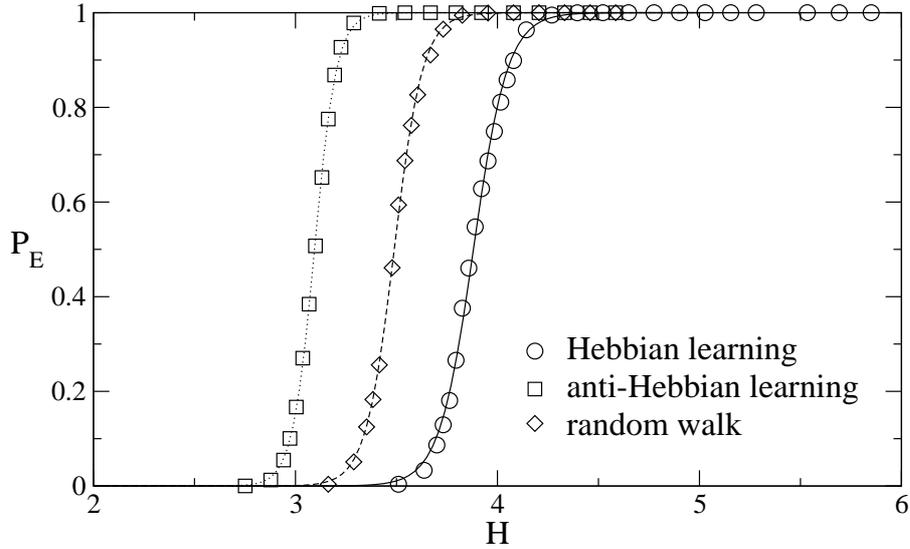}
  \caption{Success probability $P_E$ of the majority attack for $K=3$,
    $L=10$, $N=1000$, and $M=100$. Symbols denote results obtained in
    $10\,000$ simulations, while lines represents fits with model
    (\ref{eq:fermi}).}
  \label{fig:qsrule}
\end{figure}

Additionally, one observes a similar behavior for all three learning
rules. This is clearly visible in figure \ref{fig:qsrule}. Only the
fit parameters are different due to the changed length of the weight
vectors. Hebbian learning increases $Q_i$, so that an higher value of
$H$ is needed in order to achieve the same value of the success
probability. In contrast, the anti-Hebbian learning rule decreases
$Q_i$, so that one observes a similar behavior with a lower value of
$H$. Consequently, equation (\ref{eq:fermi}) is a universal model,
which describes the success probability $P_E$ as a function of the
absolute local field $H$ for all known attacks.

However, it is not sufficient to know the fit parameters $\mu$ and
$\beta$ for only one value of the synaptic depth. In order to estimate
the security of the neural key-exchange protocol with queries, one has
to look at the scaling behavior of these quantities in regard to $L$.

\begin{figure}
  \centering
  \includegraphics[scale=0.5]{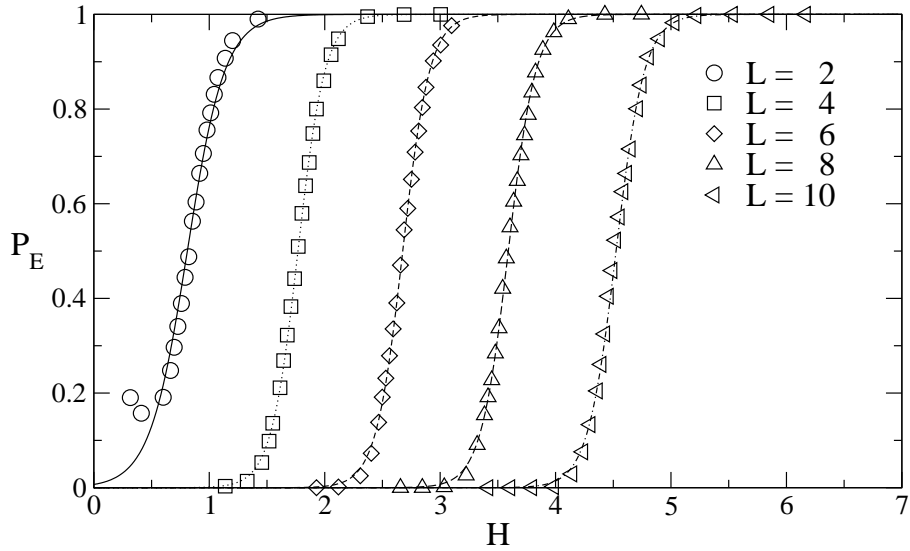}
  \caption{Success probability of the geometric attack for $K=3$,
    $N=1000$, and the Hebbian learning rule. Symbols denote results
    obtained in $10\,000$ simulations, while lines show fits with
    model (\ref{eq:fermi}).}
  \label{fig:gsuccess}
\end{figure}

\begin{figure}
  \centering
  \includegraphics[scale=0.5]{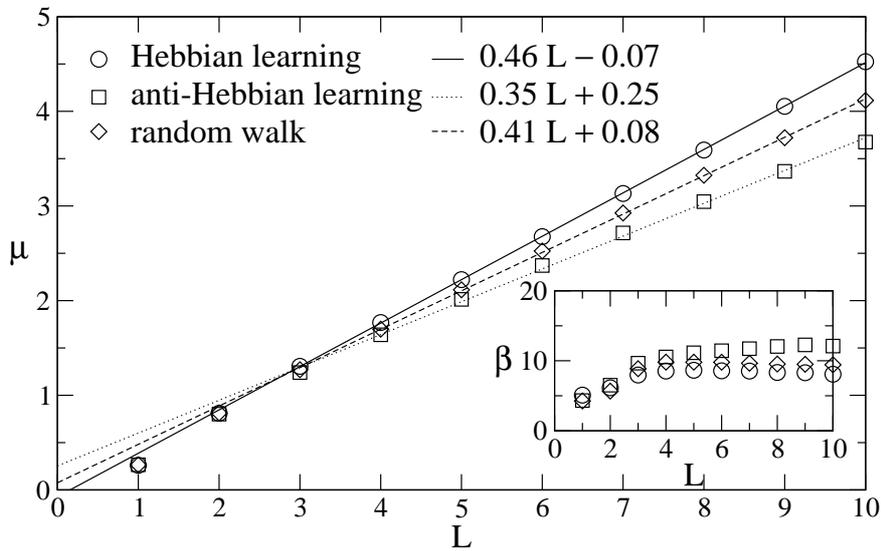}
  \caption{Parameters $\mu$ and $\beta$ as a function of the synaptic
    depth $L$ for the geometric attack. Symbols denote the results of
    fits using model (\ref{eq:fermi}), based on $10\,000$ simulations
    with $K=3$ and $N=1000$.}
  \label{fig:gparam}
\end{figure}

\begin{figure}
  \centering
  \includegraphics[scale=0.5]{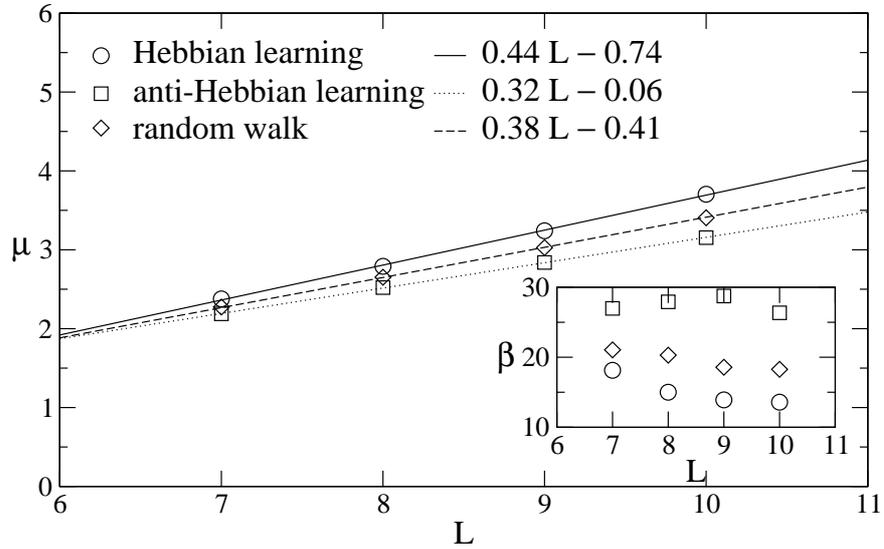}
  \caption{Parameter $\mu$ and $\beta$ as a function of $L$ for the
    genetic attack with $K=3$, $N=1000$, and $M=4096$. The symbols
    represent results from $1000$ simulations and the lines show a fit
    using the model given in (\ref{eq:linear}).}
  \label{fig:qparam}
\end{figure}

\begin{figure}
  \centering
  \includegraphics[scale=0.5]{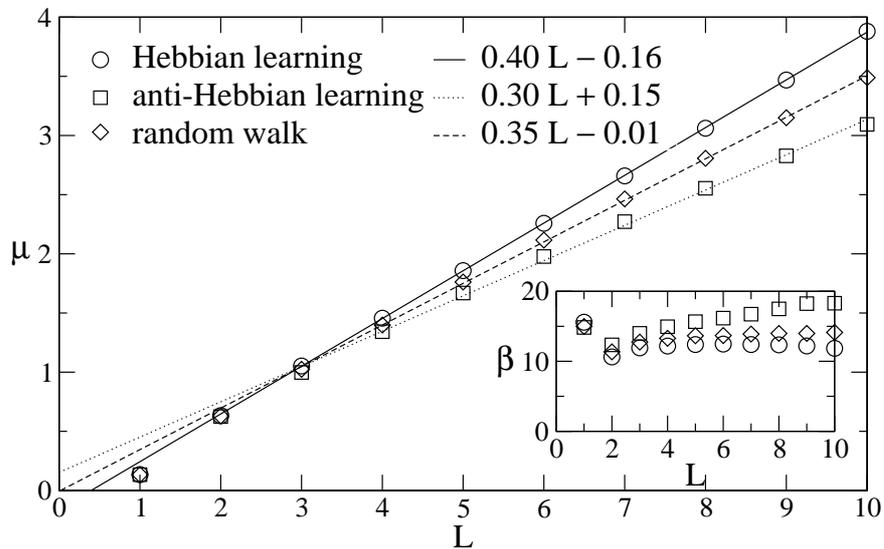}
  \caption{Parameters $\mu$ and $\beta$ as a function of the synaptic
    depth $L$ for the majority attack. Symbols denote the results of
    fits using model (\ref{eq:fermi}), based on $10\,000$ simulations
    with $K=3$, $N=1000$, and $M=100$.}
  \label{fig:mparam}
\end{figure}

Figure \ref{fig:gsuccess} shows that increasing the synaptic depth
does not change the shape of $P_E(H)$ much, so that the steepness
$\beta$ is nearly constant for $L > 3$. But there is a horizontal
shift of the curves due to the growing length of the weight vectors.
In fact, the position $\mu$ of the smooth step increases linearly with
the synaptic depth $L$,
\begin{equation}
  \label{eq:linear}
  \mu = \alpha_\mathrm{s} L + \delta \,,
\end{equation}
which is shown in figure \ref{fig:gparam}. As before, the method
chosen by E does not matter, because equation (\ref{eq:linear}) is
valid in all cases \cite{Ruttor:2005:NCQ, Ruttor:2006:GAN}. Only the
parameters $\alpha_\mathrm{s}$ and $\delta$ depend on the learning
rule and the attack. This is clearly visible in figure
\ref{fig:qparam} and in figure \ref{fig:mparam}.

Combining (\ref{eq:fermi}) and (\ref{eq:linear}) yields
\begin{equation}
  \label{eq:qsuccess}
  P_\mathrm{E} = \frac{1}{1 + \exp(\beta \, \delta) \exp(\beta \,
    (\alpha_\mathrm{s} - \alpha) L)}
\end{equation}
for the success probability of any known attack. As long as A and B
choose $\alpha = H / L$ according to the condition $\alpha <
\alpha_\mathrm{s}$, $P_E$ vanishes for $L \rightarrow \infty$. In this
case its asymptotic behavior is given by
\begin{equation}
  P_\mathrm{E} \sim \mathrm{e}^{-\beta \, \delta} \,
  \mathrm{e}^{-\beta \, (\alpha_\mathrm{s} - \alpha) L} \,,
\end{equation}
which is consistent with the observation
\begin{equation}
  \label{eq:scale}
  P_\mathrm{E} \sim \mathrm{e}^{-y (L - L_0)}
\end{equation}
found for neural cryptography with random inputs
\cite{Mislovaty:2002:SKE}. Comparing the coefficients in both
equations reveals
\begin{eqnarray}
  y   &=& \beta \, (\alpha_\mathrm{s} - \alpha) \,, \\
  L_0 &=& - \delta / ( \alpha_\mathrm{s} - \alpha) \,.
\end{eqnarray}
Thus replacing random inputs with queries gives A and B direct
influence on the scaling of $P_E$ in regard to $L$, as they can change
$y$ by modifying $\alpha$. Finally, the results indicate that there
are two conditions for a fast and secure key-exchange protocol based
on neural synchronization with queries:
\begin{itemize}
\item As shown in section \ref{sec:qtime} the average synchronization
  time $\langle t_\mathrm{sync} \rangle$ diverges in the limit $L
  \rightarrow \infty$, if $H$ is too small. Therefore A and B have to
  choose this parameter according to $H > \alpha_\mathrm{c} L$.
\item And if $H$ is too large, the key-exchange becomes insecure,
  because $P_E = 1$ is reached in the limit $L \rightarrow \infty$.
  So the partners have to fulfill the condition $H < \alpha_\mathrm{s}
  L$ for all known attacks.
\end{itemize}
Fortunately, A and B can always choose a fixed $\alpha = H / L$
according to
\begin{equation}
  \alpha_\mathrm{c} < \alpha < \alpha_\mathrm{s} \,,
\end{equation}
as there is no known attack with $\alpha_\mathrm{s} \leq
\alpha_\mathrm{c}$. Then $\langle t_\mathrm{sync} \rangle$ grows
proportional to $L^2$, but $P_E$ drops exponentially with increasing
synaptic depth. Consequently, A and B can reach any desired level of
security by just changing $L$ \cite{Ruttor:2006:GAN}.

\subsection{Optimal local field}

For practical aspects of security, however, it is important to look at
the relation between the average synchronization time and the success
probability, as a too complex key-exchange protocol is nearly as
unusable as an insecure one. That is why A and B want to minimize
$P_E$ for a given value of $\langle t_\mathrm{sync} \rangle$ by
choosing $L$ and $H$ appropriately. These optimum values can be
determined by analyzing the function $P_E(\langle t_\mathrm{sync}
\rangle)$.

\begin{figure}
  \centering
  \includegraphics[scale=0.5]{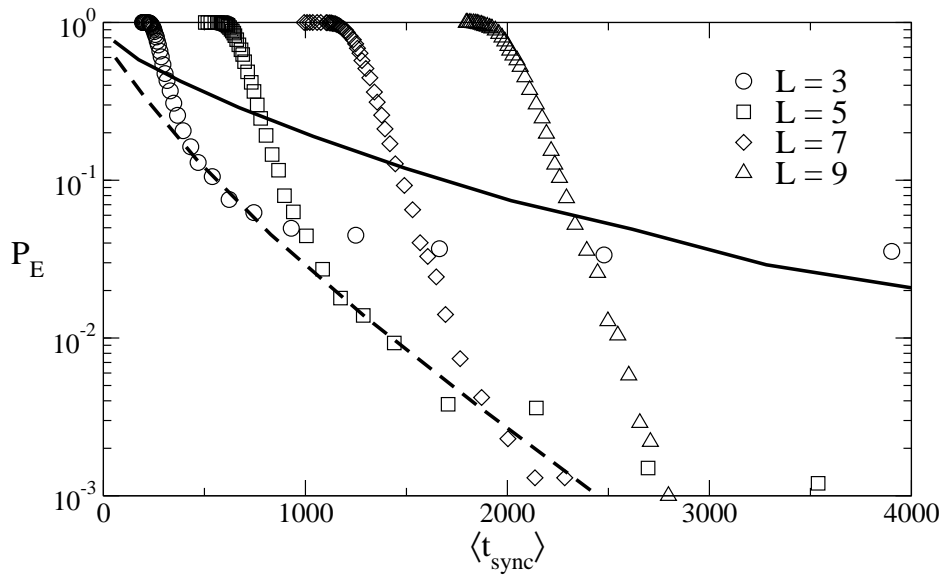}
  \caption{Success probability of the geometric attack as a function
    of $\langle t_\mathrm{sync} \rangle$. Symbols denote results
    obtained in $10\,000$ simulations using the Hebbian learning rule,
    $K=3$, and $N=1000$. The solid curve represents $P_E$ in the case
    of random inputs and the dashed line marks $H = 0.36 \, L$.}
  \label{fig:gresult}
\end{figure}

Figure \ref{fig:gresult} shows the result for the geometric attack.
The optimum value of $H$ lies on the envelope of all functions
$P_\mathrm{E}(\langle t_\mathrm{sync} \rangle)$. This curve is
approximately given by $H = \alpha_\mathrm{c} L$, as this choice
maximizes $\alpha_\mathrm{s} - \alpha$, while synchronization is still
possible \cite{Ruttor:2005:NCQ}. It is also clearly visible that
queries improve the security of the neural key-exchange protocol
greatly for a given average synchronization time.

\begin{figure}
  \centering
  \includegraphics[scale=0.5]{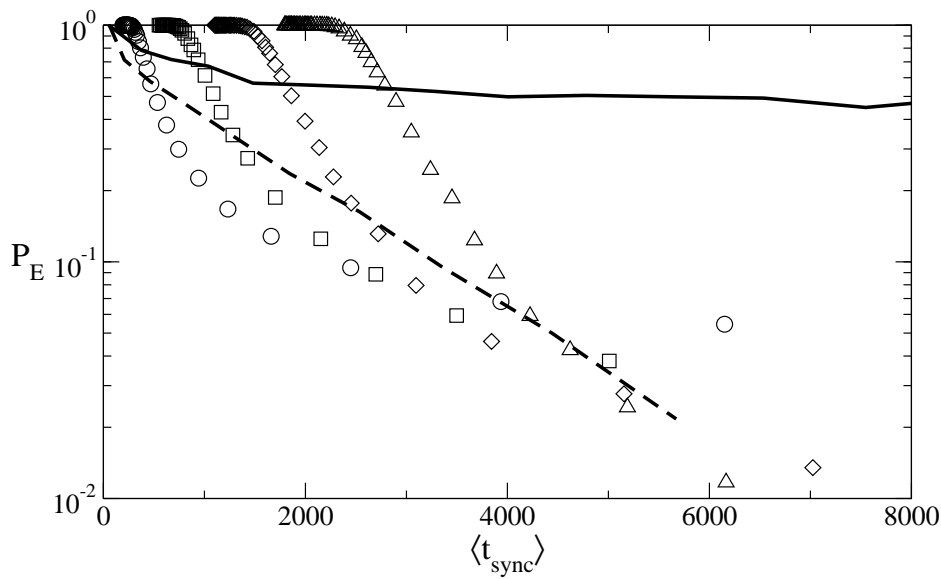}
  \caption{Success probability of the majority attack as a function of
    $\langle t_\mathrm{sync} \rangle$. Symbols denote results obtained
    in $10\,000$ simulations using the Hebbian learning rule, $K=3$,
    $M=100$, and $N=1000$. The solid curve represents $P_E$ in the
    case of random inputs and the dashed line marks $H = 0.36 \, L$.}
  \label{fig:mresult}
\end{figure}

A similar result is obtained for the majority attack. Here figure
\ref{fig:mresult} shows that the partners can even do better by using
queries with $H < \alpha_\mathrm{c} L$ as long as $L$ is not too
large. This effect is based on fluctuations which enable
synchronization, but vanish in the limit $L \rightarrow \infty$. Thus
the optimum value of $H$ is still given by $H \approx
\alpha_\mathrm{c} L$ if $L \gg 1$. Additionally, figure
\ref{fig:mresult} indicates that A and B can even employ the Hebbian
learning rule for neural cryptography with queries, which led to an
insecure key-exchange protocol in the case of random inputs
\cite{Shacham:2004:CAN,Ruttor:2005:NCQ}.

\subsection{Genetic attack}

\begin{figure}
  \centering
  \includegraphics[scale=0.5]{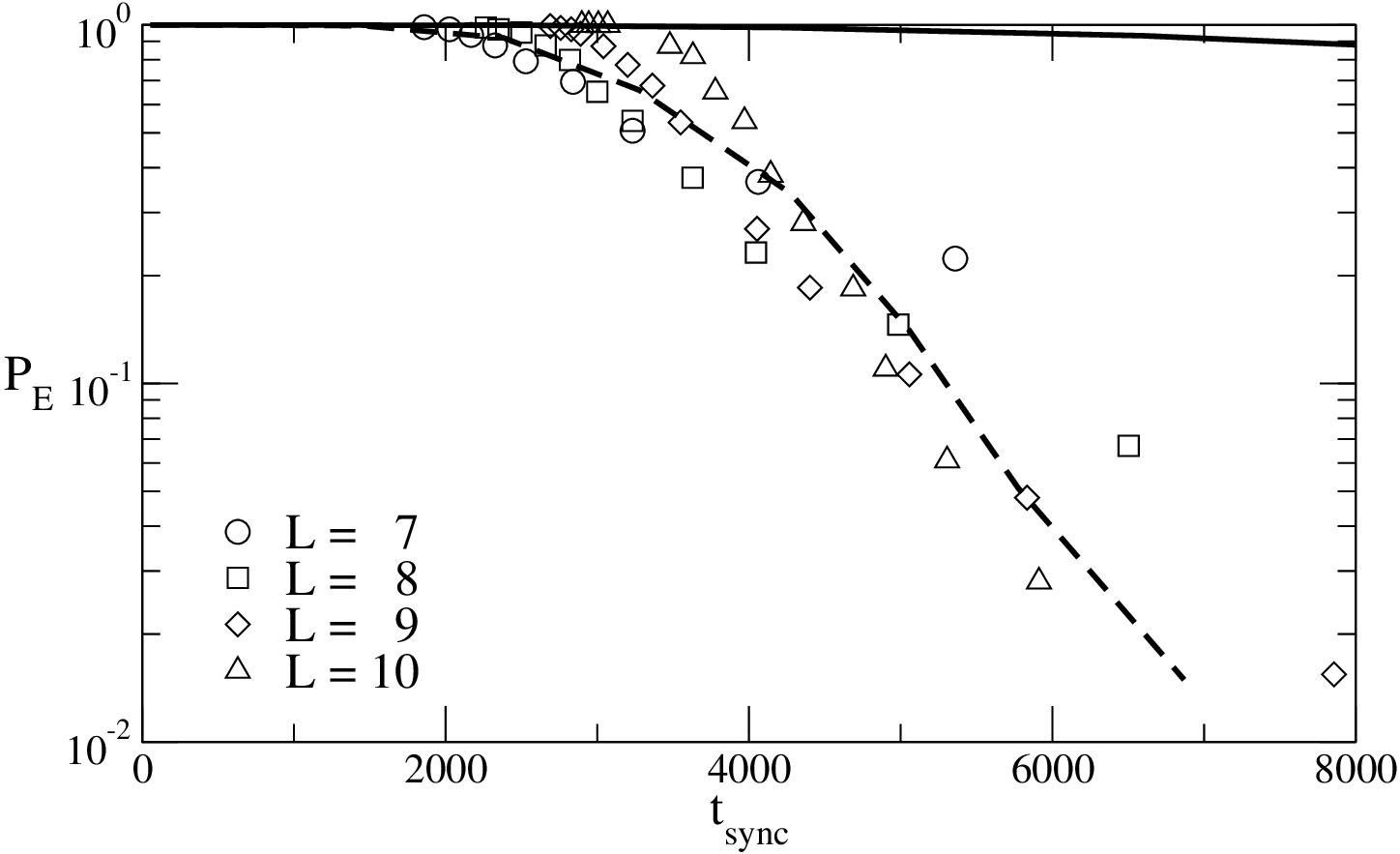}
  \caption{Success probability of the genetic attack as a function of
    $\langle t_\mathrm{sync} \rangle$. Symbols denote results obtained
    in $1000$ simulations using the random walk learning rule, $K=3$,
    $M=4096$, and $N=1000$. The solid curve represents $P_E$ in the
    case of random inputs and the dashed line marks $H = 0.32 \, L$.}
  \label{fig:result}
\end{figure}

Compared to the other methods the genetic attack is in a certain way
different. First, it is especially successful, if $L$ is small. That
is why A and B have to use Tree Parity Machines with large synaptic
depth $L$ regardless of the parameter $H$. Of course, this sets a
lower limit for the effort of performing the neural key-exchange
protocol as shown in figure \ref{fig:result}.

Second, the genetic attack is a rather complicated algorithm with a
lot of parameters. Of course, E tries to optimize them in order to
adapt to special situations. Here the number $M$ of attacking networks
is clearly the most important parameter, because it limits the number
of mutation steps $t_\mathrm{s}$ which can occur between two selection
steps:
\begin{equation}
  t_\mathrm{s} \leq \frac{1}{K - 1} \frac{\ln M}{\ln 2} \,.
\end{equation}
Thus E can test different variants of the internal representation
$(\sigma_1, \dots \sigma_K)$ for at most $t_\mathrm{s}$ steps, before
she has to select the fittest Tree Parity Machines. And more time
results in better decisions eventually. Therefore one expects that E
can improve $P_E$ by increasing $M$ similar to the effect observed for
random inputs in section \ref{sec:gensec}.

\begin{figure}
  \centering
  \includegraphics[scale=0.5]{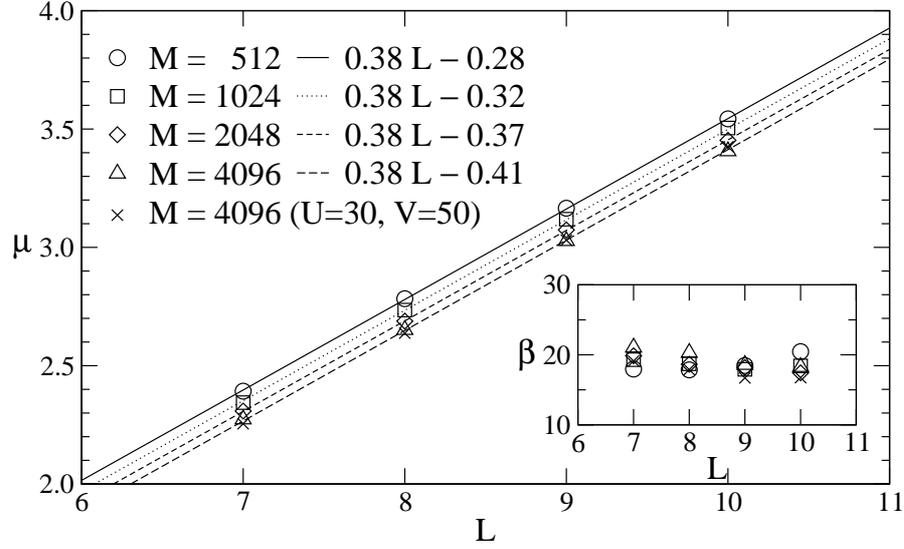}
  \caption{Parameter $\mu$ and $\beta$ as a function of $L$ for the
    genetic attack with $K=3$, $N=1000$, and the random walk learning
    rule. Symbols denote results of fitting simulation data with
    (\ref{eq:fermi}) and the lines were calculated using the model
    given in (\ref{eq:linear}).}
  \label{fig:qgenetic}
\end{figure}

\begin{figure}
  \centering
  \includegraphics[scale=0.5]{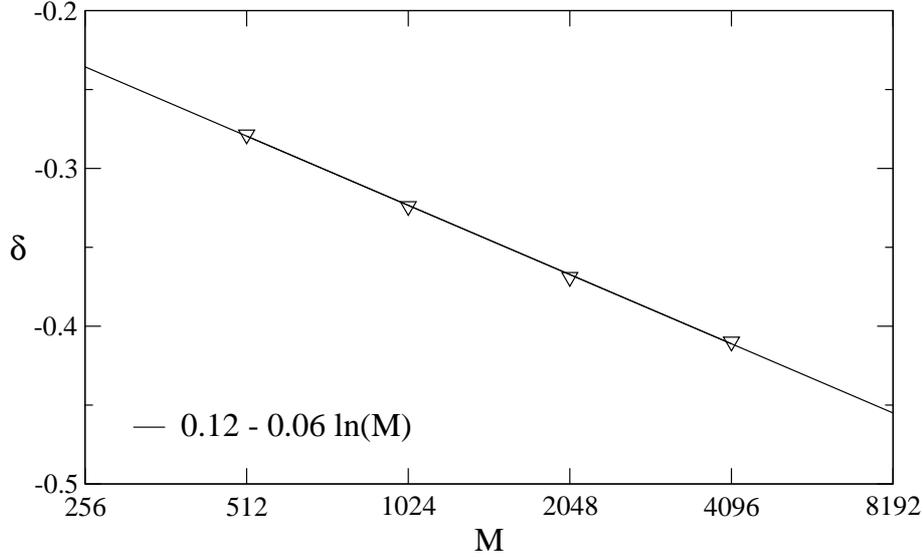}
  \caption{Offset $\delta$ as a function of the number of attackers
    $M$, for the genetic attack with $K=3$, $N=1000$, and the random
    walk learning rule. Symbols and the line were obtained by a fit
    with (\ref{eq:qgenpar}).}
  \label{fig:qgenpar}
\end{figure}

Figure \ref{fig:qgenetic} shows that this is indeed the case. While
$\alpha_\mathrm{s}$ stays constant, the offset $\delta$ decreases with
increasing $M$. As before, it is a logarithmic effect,
\begin{equation}
  \label{eq:qgenpar}
  \delta(M) = \delta(1) - \delta_E \ln M \,,
\end{equation}
which is clearly visible in figure \ref{fig:qgenpar}. Therefore E
gains a certain horizontal shift $\delta_E \ln 2$ of the smooth step
function $P_E(H)$ by doubling the effort used for the genetic attack
\cite{Ruttor:2006:GAN}. Combining (\ref{eq:genpar}) and
(\ref{eq:qsuccess}) yields
\begin{equation}
  P_E = \frac{1}{1 + \exp(\beta (\delta(1) - \delta_E \ln M))
    \exp(\beta (\alpha_s - \alpha) L)}
\end{equation}
for the success probability of this method. Then the asymptotic
behavior for $L \gg 1$ is given by
\begin{equation}
  \label{eq:scale1}
  P_E \sim e^{-\beta (\delta(1) - \delta_E \ln M)} e^{-\beta
    (\alpha_s - \alpha) L}
\end{equation}
as long as $\alpha < \alpha_\mathrm{s}$. Similar to neural
cryptography with random inputs E has to increase the number of
attacking networks exponentially,
\begin{equation}
  \label{eq:qscale}
  M \propto e^{[(\alpha_s - \alpha) / \delta_E] L} \,,
\end{equation}
in order to maintain a constant success probability $P_E$, if A and B
change the synaptic depth $L$. But, due to limited computer power,
this is often not feasible.

\begin{figure}
  \centering
  \includegraphics[scale=0.5]{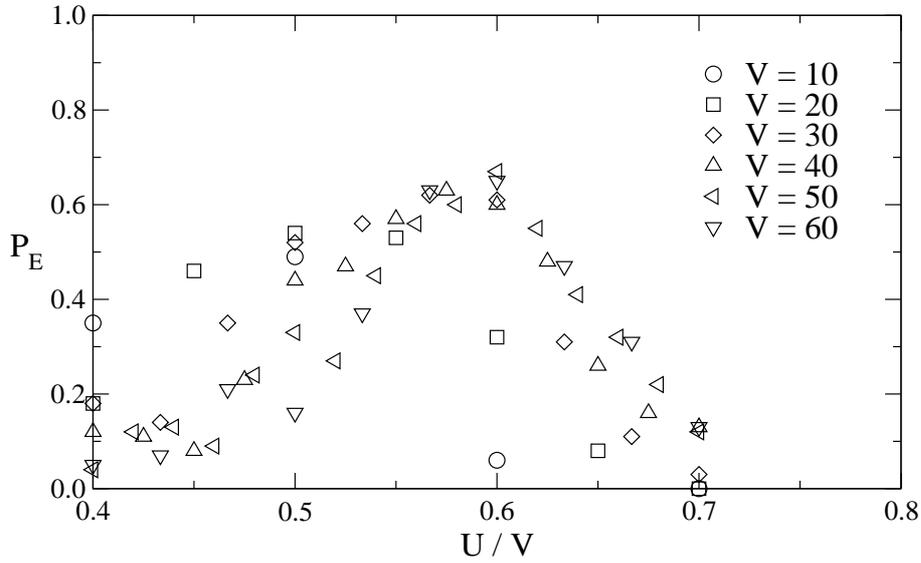}
  \caption{Success probability of the genetic attack in the case of
    $K=3$, $L=7$, $N=1000$, $M=4096$, $H=2.28$, and random walk
    learning rule. These results were obtained by averaging over $100$
    simulations.}
  \label{fig:gwh}
\end{figure}

However, the attacker could also try to improve $P_E$ by changing the
other parameters $U$ and $V$ of the genetic attack. Instead of the
default values $U=10$, $V=20$ E could use $U=30$, $V=50$, which
maximize $P_E$ without greatly changing the complexity of the attack
\cite{Ruttor:2006:GAN}. But this optimal choice, which is clearly
visible in figure \ref{fig:gwh}, does not help much as shown in figure
\ref{fig:qgenetic}. Only $\beta$ is lower for the optimized attack,
while $\alpha_\mathrm{s}$ remains nearly the same. Therefore the
attacker gains little, as the scaling relation (\ref{eq:qscale}) is
not affected. Consequently, the neural key-exchange protocol with
queries is even secure against an optimized variant of the genetic
attack in the limit $L \rightarrow \infty$.

\subsection{Comparison of the attacks}

Of course, the opponent E always employs the best method, which is
available to her in regard to computing power and other resources.
Therefore it is necessary to compare all known attack methods in order
to estimate the level of security achieved by a certain set of
parameters.

\begin{figure}
  \centering
  \includegraphics[scale=0.5]{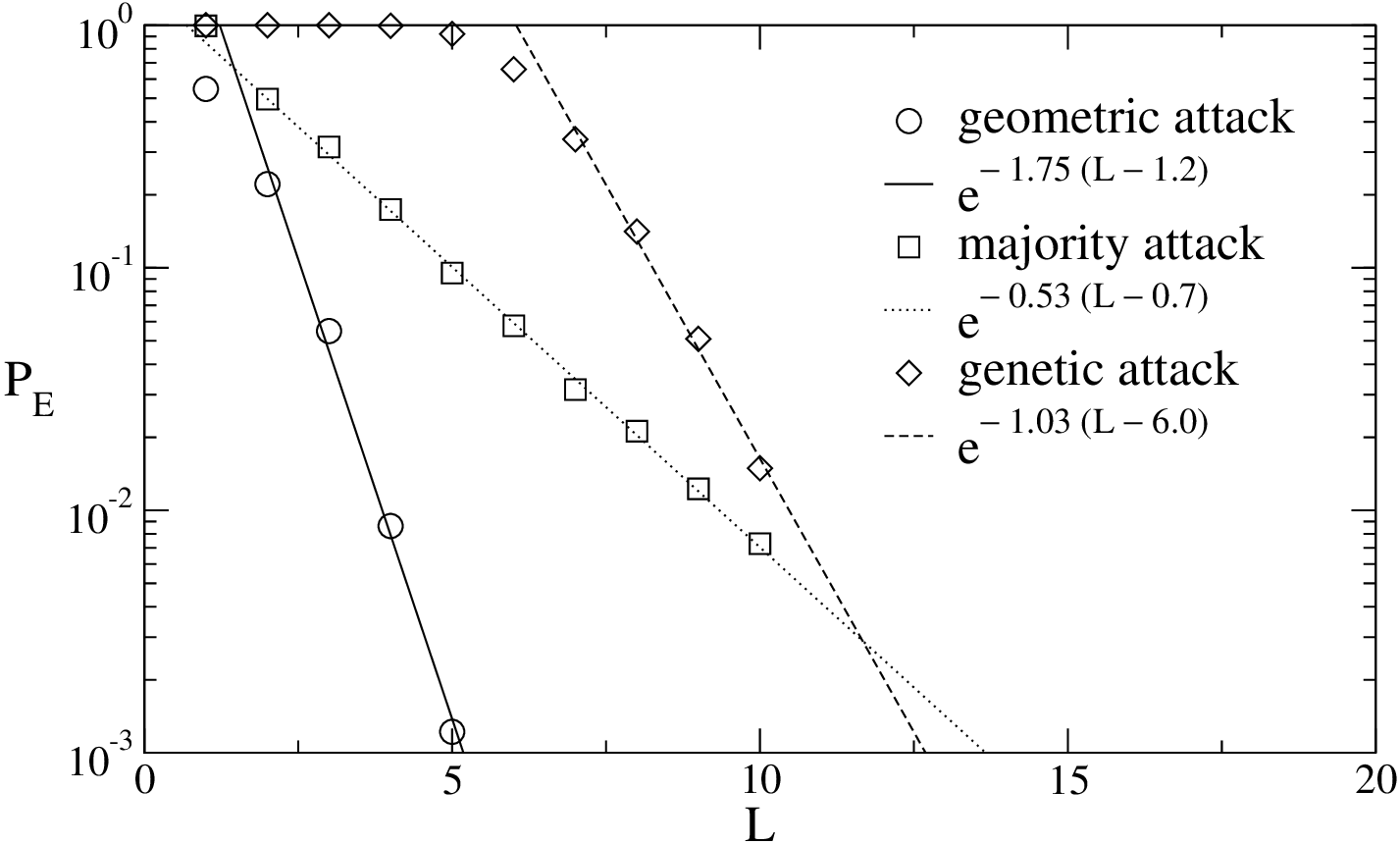}
  \caption{Success probability of different attacks as a function of
    the synaptic depth $L$. Symbols denote results obtained in $1000$
    simulations using the random walk learning rule, $K=3$, $H = 0.32
    L$, and $N=1000$, while the lines show fit results for model
    (\ref{eq:scale}). Here E has used $M=4096$ networks for the
    genetic attack and $M=100$ for the majority attack.}
  \label{fig:queries}
\end{figure}

The result for neural cryptography with queries is shown in figure
\ref{fig:queries}. It is qualitatively similar to that observed in
section \ref{sec:seccomp} in the case of synchronization with random
inputs. As the majority attack has the minimum value of
$\alpha_\mathrm{s}$, it is usually the best method for the attacker.
Only if A and B use Tree Parity Machines with small synaptic depth,
the genetic attack is better.

However, comparing figure \ref{fig:queries} with figure
\ref{fig:rsuccess} (on page \pageref{fig:rsuccess}) reveals, that
there are quite large quantitative differences, as replacing random
inputs with queries greatly improves the security of the neural
key-exchange protocol. Extrapolation of (\ref{eq:scale}) shows that
$P_E \approx 10^{-4}$ is achieved for $K=3$, $L=18$, $N=1000$, $H =
5.76$, and random walk learning rule. This is much easier to realize
than $L=57$, which would be necessary in order to reach the same level
of security in the case of random inputs.

\section{Possible security risks}

Although using queries improves the security of the neural
key-exchange protocol against known attacks, there is a risk that a
clever attacker may improve the success probability $P_E$ by using
additional information revealed through the algorithm generating the
input vectors. Two obvious approaches are analyzed here. First, E
could use her knowledge about the absolute local field $H$ to improve
the geometric correction of the internal representation $(\sigma_1^E,
\dots, \sigma_K^E)$. Second, each input vector $\mathbf{x}_i$ is
somewhat correlated to the corresponding weight vector $\mathbf{w}_i$
of the generating network. This information could be used for a new
attack method. 

\subsection{Known local field}

If the partners use queries, the absolute value of the local field in
either A's or B's hidden units is given by $H$. And E knows the local
fields $h_i^E$ in her own Tree Parity Machine. In this situation the
probability of $\sigma_i^E \not= \sigma_i^A$ is no longer given by
(\ref{eq:generr}) or (\ref{eq:perr}), if it is A's turn to generate
the input vectors. Instead, one finds
\begin{equation}
  P(\sigma_i^E \not= \sigma_i^A) = \left[ 1 + \exp \left( \frac{2
        \rho_i^{AE}}{1 - (\rho_i^{AE})^2} \frac{H}{\sqrt{Q_i^A}}
      \frac{|h_i^E|}{\sqrt{Q_i^E}} \right) \right]^{-1} \,.
\end{equation}
Although one might assume that this probability is minimal for
$|h_i^E| \approx H$, it is not the case. In contrast, $P(\sigma_i^E
\not= \sigma_i^A)$ reaches its maximum at $|h_i^E|=0$ and is a
strictly decreasing function of $|h_i^E|$ as before.

\begin{figure}
  \centering
  \includegraphics[scale=0.5]{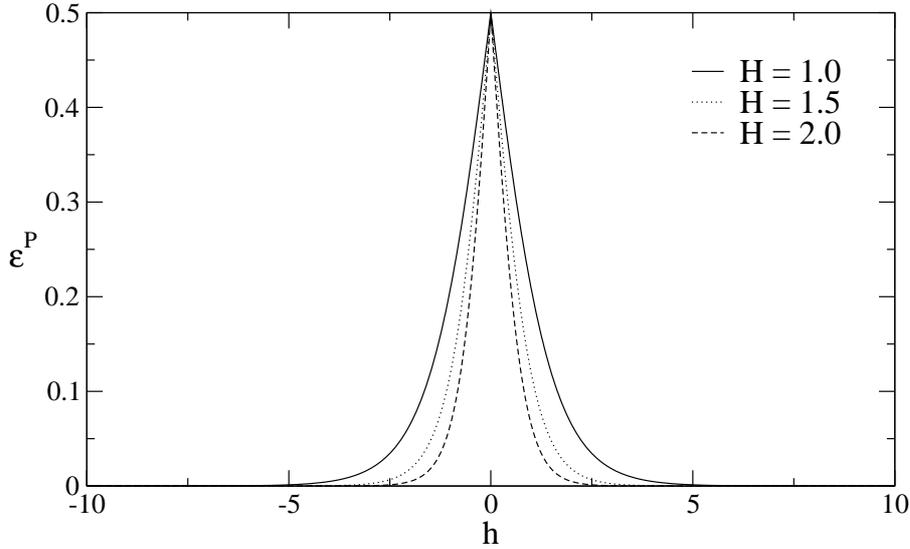}
  \caption{Prediction error $\epsilon_i^\mathrm{p}$ as a function of
    the local field $h_i^E$ for $Q_i^A=1$, $Q_i^E=1$, and $\rho=0.5$.}
  \label{fig:pquery}
\end{figure}

This is clearly visible in figure \ref{fig:pquery}. As there is no
qualitative difference compared to synchronization with random inputs,
it is not possible to improve the geometric attack by using $H$ as
additional information. Instead, it is still optimal for E to flip the
output of that hidden unit, which has the minimum absolute value of
the local field.

\subsection{Information about weight vectors}

While $H$ cannot be used directly in the geometric attack, queries
give E additional information about the weight vectors in A's and B's
Tree Parity Machines. But fortunately the absolute local field $H$
used for synchronization with queries is lower than the average value
\begin{equation}
  \langle |h_k| \rangle = \sqrt{2 Q_i / \pi} \approx 0.8 \sqrt{Q_i}
\end{equation}
observed for random inputs. Hence the overlap
\begin{equation}
  \rho_{i,\mathrm{in}} = \frac{\mathbf{w}_i \cdot
    \mathbf{x}_i}{\sqrt{\mathbf{w}_i
      \cdot \mathbf{w}_i} \sqrt{\mathbf{x}_i \cdot \mathbf{x}_i}} =
  \frac{1}{\sqrt{N}} \frac{h_i}{\sqrt{Q_i}}
\end{equation}
between input vector and weight vector is very small and converges to
zero in the limit $N \rightarrow \infty$, although $H > 0$.
Consequently, $\mathbf{x}_i$ and $\mathbf{w}_i$ are nearly
perpendicular to each other, so that the information revealed by
queries is minimized \cite{Ruttor:2005:NCQ}.

In fact, for a given value of $H$ the number of weight vectors, which
are consistent with a given query, is still exponentially large. As an
example, there are $2.8 \times 10^{129}$ possible weight vectors for a
query with $H=10$, $L=10$, and $N=100$ \cite{Ruttor:2005:NCQ}.
Consequently, E cannot benefit from the information contained in the
input vectors generated by A and B.

\chapter{Conclusions and outlook}
\label{chap:conclusion}

In this thesis the synchronization of neural networks by learning from
each other has been analyzed and discussed. At a glance this effect
looks like an extension of online learning to a series of examples
generated by a time dependent rule. However, it turns out that neural
synchronization is a more complicated dynamical process, so that new
phenomena occur.

This is especially true for Tree Parity Machines, whereas
synchronization is driven by stochastic attractive and repulsive
forces. Because this process does not have a self-averaging order
parameter, one has to take the whole distribution of the weights into
account instead of using just the average value of the order parameter
to determine the dynamics of the system. This can be done using direct
simulations of the variables $w_{i,j}$ for finite $N$ or a iterative
calculation of their probability distribution in the limit $N
\rightarrow \infty$.

While one can use different learning rules both for bidirectional
synchronization and unidirectional learning, they show similar
behavior and converge to the random walk learning rule in the limit $N
\rightarrow \infty$. So the deviations caused by Hebbian and
anti-Hebbian learning are, in fact, finite-size effects, which become
only relevant for $L \gg O(\sqrt{N})$.

In contrast, numerical simulations as well as iterative calculations
show a phenomenon, which is significant even in very large systems: In
the case of Tree Parity Machines learning by listening is much slower
than mutual synchronization. This effect is caused by different
possibilities of interaction. Two neural networks, which can influence
each other, are able to omit steps, if they caused a repulsive effect.
This is an advantage compared to a third Tree Parity Machine, which is
trained using the examples produced by the other two and cannot select
the most suitable input vectors for learning. Consequently, if
interaction is only possible in one direction, the frequency of
repulsive steps is higher than in the case of bidirectional
communication.

Although the overlap $\rho$ is not a self-averaging quantity, one can
describe neural synchronization as a random walk in $\rho$-space. Here
the average step sizes $\langle \Delta \rho_ \mathrm{a} \rangle$ and
$\langle \Delta \rho_ \mathrm{r} \rangle$ are the same for
synchronization and learning. But the transition probabilities
$P_\mathrm{a}(\rho)$ and $P_\mathrm{r}(\rho)$ depend on the type of
interaction. As a result one can observe qualitative differences
regarding the dynamics of the overlap. In the case of $K=3$ and
bidirectional interaction the average change of the overlap $\langle
\Delta \rho \rangle$ is strictly positive, so that synchronization by
mutual learning happens on average. But for $K>3$ or unidirectional
interaction the higher probability of repulsive steps causes a fixed
point of the dynamics at $\rho_\mathrm{f} < 1$. Then reaching the
absorbing state at $\rho=1$ is only possible by means of fluctuations.

While both mechanisms lead to full synchronization eventually, one
observes two different distributions of the synchronization time
depending on the function $\langle \Delta \rho(\rho) \rangle$. In the
case of synchronization on average, it is a Gumbel distribution,
because one has to wait until the last weight has synchronized.
Analytical calculations for systems without repulsive steps yield the
result $\langle t_\mathrm{sync} \rangle \propto L^2 \ln N$. And a few
repulsive steps do not change this scaling behavior, but simply
increase the constant of proportionality.

In contrast, if synchronization is only possible by means of
fluctuations, there is a constant probability per step to get over the
gap with $\langle \Delta \rho(\rho) \rangle < 0$ between the fixed
point and the absorbing state. Of course, this yields an exponential
distribution of the synchronization time. However, the fluctuations of
the overlap in the steady state decrease proportional to $L^{-1}$. As
they are essential for reaching $\rho=1$ in this case, the
synchronization time grows exponentially with increasing synaptic
depth of the Tree Parity Machines.

Without this difference a secure key-exchange protocol based on the
synchronization of neural networks would be impossible. But as A's and
B's Tree Parity Machines indeed synchronize faster than E's neural
networks, the partners can use the synchronized weight vectors as a
secret session key. Of course, there is a small probability $P_E$ that
E is successful before A and B have finished their key exchange due to
the stochastic nature of the synchronization process.  But fortunately
$P_E$ drops exponentially with increasing $L$ for nearly all
combinations of learning rules and attack methods. Thus A and B can
achieve any level of security by just increasing the synaptic depth
$L$.

Additionally, there are other observations which indicate that
bidirectional interaction is an advantage for A and B compared to a
passive attacker E. For a time series generated by two Tree Parity
Machines the version space of compatible initial conditions is larger,
if both are already synchronized at the beginning, than if the neural
networks start unsynchronized. So it is harder for an attacker to
imitate B because of the interaction between the partners. And, of
course, the attack methods are unable to extract all the information
which is necessary to achieve full synchronization. This effect is
mainly caused by the fact, that A and B can choose the most useful
input vectors from the random sequence, but E does not have this
ability \cite{Maurer:1993:SKA}.

Thus the partners can improve the security of neural cryptography
further, if they use a more advanced algorithm to select the input
vectors. This approach eventually leads to synchronization with
queries. In this variant of the key-exchange protocol A and B ask each
other questions, which depend on the weights in their own networks. In
doing so they are able to choose the absolute value of the local field
in the Tree Parity Machine generating the current query. Of course,
this affects both synchronization and attacks. However, E is at a
disadvantage compared to A and B, because she needs a higher absolute
value of the local field than the partners in order to synchronize on
average. Therefore it is possible to adjust the new parameter $H$ in
such a way, that A and B synchronize fast, but E is not successful
regardless of the attack method.

However, the algorithm generating the input vectors does not matter
for the opponent. E has no influence on it and the relative efficiency
of the attacks stays the same, whether a random input sequence or
queries are used. In both cases the majority attack is the best method
as long as the synaptic depth is large. Only if $L$ is small, the
genetic attack is better. Of course, both advanced attacks are always
more successful than the geometric attack. And the simple attack is
only useful for $K \gg 3$.

In any case, the effort of the partners grows only polynomially, while
the success probability of an attack drops exponentially, if the
synaptic depth increases. Similar scaling laws can be found, if one
looks at other cryptographic systems. Only the parameter is different.
While the security of conventional cryptography
\cite{Beutelspacher:2002:K, Stinson:1995:CTP} depends on the length of
the key, the synaptic depth of the Tree Parity Machines plays the same
role in the case of neural cryptography \cite{Ruttor:2006:GAN}.

Brute-force attacks are not very successful, either. Here the number
of keys grows exponentially with the system size $N$, while the
synchronization time is only proportional to $\log N$. Thus A and B
can use large systems without much additional effort in order to
prevent successful guessing of the generated key.

Consequently, the neural key-exchange protocol is secure against all
attacks known up to now. However, there is always the risk that one
might find a clever attack, which breaks the security of neural
cryptography completely, because it is hardly ever possible to prove
the security of such an algorithm \cite{Stinson:1995:CTP}.

However, the neural key-exchange protocol is different from
conventional cryptographic algorithms in one aspect. Here effects in a
physical system, namely attractive and repulsive stochastic forces,
are used instead of those found in number theory. In fact, the trap
door function is realized by a dynamics, which is different for
partners and attackers based on their possibilities of interaction
with the other participants. Of course, neural networks are not the
only type of systems with these properties. Any other system showing
similar effects can be used for such a cryptographic application, too.

Interesting systems include chaotic maps and coupled lasers
\cite{Gross:2005:FPC, Klein:2005:PCC, Klein:2006:PCC, Klein:2006:SIS}.
In both cases one observes that synchronization is achieved faster for
bidirectional than for unidirectional coupling. As this underlying
effect is very similar, one can use nearly the same cryptographic
protocol by just substituting the neural networks. Of course, this
applies to the attack methods, too. For example, the algorithms of the
majority attack and the genetic attack are so general, that they are
also useful methods for attacks on key-exchange protocols using
chaotic maps. In contrast, the geometric correction algorithm is
rather specific for neural networks, so that it has to be replaced by
appropriate methods.

Consequently, the neural key-exchange protocol is only the first
element of a class of new cryptographic algorithms. Of course, all
these proposals have to be analyzed in regard to efficiency and
security. For that purpose the methods used in this thesis can
probably act as a guidance. Especially synchronization by fluctuations
and synchronization on average are rather abstract concepts, so that
one should be able to observe them in a lot of systems.

Another interesting direction is the implementation of the neural
key-exchange protocol. Computer scientists are already working on a
hardware realization of interacting Tree Parity Machines for
cryptographic purposes \cite{Volkmer:2004:LCS, Volkmer:2005:TPMa,
  Volkmer:2005:TPMb, Volkmer:2005:KEI, Volkmer:2005:LKE,
  Volkmer:2006:ICD}. They have especially found out that neural
synchronization only needs very basic mathematical operations and,
therefore, is very fast compared to algorithms based on number theory.
Consequently, one can use neural cryptography in small embedded
systems, which are unable to use RSA or other established methods
\cite{Beutelspacher:2002:K, Stinson:1995:CTP}. Here it does not matter
that the neural key-exchange protocol only reaches a moderate level of
security as long as one requires a small synchronization time. But
integrated circuits can achieve a very high frequency of key updates,
which compensates this disadvantage \cite{Volkmer:2004:LCS,
  Volkmer:2005:TPMa, Volkmer:2005:TPMb}.

Finally, these approaches indicate that further development of neural
cryptography is indeed possible. As mentioned before, there are, in
fact, two distinct directions: First, one can extend the neural
key-exchange protocol in order to improve the efficiency, security and
usefulness for specific cryptographic applications, e.~g.~embedded
systems. Second, one can replace the neural networks by other physical
systems, e.~g.~chaotic lasers, which have similar properties to those
identified in this thesis as essential for security.

\begin{appendix}
\chapter{Notation}

\begin{tabbing}
  \hspace{3cm}\=\hspace{10cm}\=\kill \\
  A \> sender   \\
  B \> receiver \\
  E \> attacker \\
  \\
  $K$ \> number of hidden units in a Tree Parity Machine \\
  $L$ \> synaptic depth of the neural networks           \\
  $N$ \> number of neurons per hidden unit               \\
  $M$ \> (maximum) number of attacking networks          \\
  \\
  $H$ \> absolute set value of the local field           \\
  $R$ \> threshold for the reset of the input vectors    \\
  $U$ \> minimal fitness                                 \\
  $V$ \> length of the output history                    \\
  \\
  $\mathbf{w}_i$          \> weight vector of the $i$-th hidden unit \\
  $\mathbf{x}_i$          \> input vector of the $i$-th hidden unit \\
  $w_{i,j}$               \> $j$-th element of $\mathbf{w}_i$       \\
  $x_{i,j}$               \> $j$-th element of $\mathbf{x}_i$       \\
  $\sigma_i$              \> output of the $i$-th hidden unit       \\
  $\tau$                  \> total output of a Tree Parity Machine  \\
  $h_i$                   \> local field of the $i$-th hidden unit  \\
  $\rho_i$                \> overlap of the $i$-th hidden unit      \\
  $\epsilon_i$            \> generalization error                   \\
  $\epsilon_i^\mathrm{p}$ \> prediction error                       \\
  \\
  $P_\mathrm{a}$                \> probability of attractive steps  \\
  $P_\mathrm{r}$                \> probability of repulsive steps   \\
  $\Delta \rho_\mathrm{a}$      \> step size of an attractive step  \\
  $\Delta \rho_\mathrm{r}$      \> step size of a repulsive step    \\
  $\langle \Delta \rho \rangle$ \> average change of the overlap    \\
  $\rho_\mathrm{f}$             \> fixed point of the dynamics      \\ 
  $\sigma_\mathrm{f}$           \> width of the $\rho$-distribution
                                   at the fixed point \\
  \\
  $I$               \> mutual information                         \\
  $S$               \> entropy of a weight distribution           \\
  $S_0$             \> maximal entropy of a single neural network \\ 
  $n_\mathrm{conf}$ \> number of possible weight configurations   \\
  $n_\mathrm{key}$  \> number of distinct keys                    \\
  $n_\mathrm{vs}$   \> size of the version space                  \\
  \\
  $T$               \> synchronization time for two random walks    \\
  $T_N$             \> synchronization time for $N$ pairs of random
                       walks \\ 
  $t_\mathrm{sync}$ \> synchronization time for two Tree Parity
                       Machines \\
  $P_E$ \> success probability of an attacker                      \\
  $y$   \> sensitivity of $P_E$ in regard to $L$                   \\
  $L_0$ \> minimal value of $L$ for the exponential decay of $P_E$ \\
  \\
  $\alpha$            \> rescaled local field $H / L$              \\
  $\alpha_\mathrm{c}$ \> minimum $\alpha$ for synchronization      \\
  $\alpha_\mathrm{s}$ \> maximum $\alpha$ for security             \\
  $\beta$             \> sensitivity of $P_E$ in regard to $H$     \\
  $\delta$            \> offset of $P_E(H)$                        \\
  \\
  $\gamma$ \> Euler-Mascheroni constant ($\gamma \approx 0.577$)
\end{tabbing}

\subsubsection{Auxiliary functions for the learning rules}

\begin{itemize}
\item control signal
  \begin{displaymath}
    f(\sigma, \tau^A, \tau^B) = 
    \Theta(\sigma \tau^A) \Theta(\tau^A \tau^B)
    \left\{
      \begin{array}{cl}
        \sigma  & \mbox{ Hebbian learning rule }      \\
        -\sigma & \mbox{ anti-Hebbian learning rule } \\
        1       & \mbox{ random walk learning rule }
      \end{array}
    \right.
  \end{displaymath}
\item boundary condition
  \begin{displaymath}
    g(w) = 
    \left\{
      \begin{array}{cl}
        \mathrm{sgn}(w) \, L & \mbox{ for $|w|>L$ } \\
        w                    & \mbox{ otherwise }   \\
      \end{array}
    \right.
  \end{displaymath}
\end{itemize}

\chapter{Iterative calculation}
\label{chap:icalc}

This appendix presents the algorithm, which is used to calculate the
time evolution of the weight distribution iteratively in the limit $N
\rightarrow \infty$ \cite{Rosen-Zvi:2002:MLT, Rosen-Zvi:2002:CBN,
  Ruttor:2004:NCF}. Compared to direct simulations $N$ weights are
replaced by $(2 L + 1) \times (2 L + 1)$ variables $p^i_{a,b}$, which
describe the probability that one finds a weight with $w_{i,j}^A=a$
and $w_{i,j}^B=b$. Consequently, one has to adapt both the calculation
of the output bits and the update of the weight configuration.

\section{Local field and output bits}

According to its definition (\ref{eq:field}) the local field $h_i$ of
a hidden unit is proportional to the sum over $N$ independent random
variables $w_{i,j} \, x_{i,j}$. Therefore the central limit theorem
applies and the probability to find certain values of $h_i^A$ and
$h_i^B$ in a time step is given by
\begin{equation}
  \label{eq:icalc}
  P(h_i^A,h_i^B) = \frac{e^{-(1/2)(h_i^A , h_i^B)
      \mathcal{C}_i^{-1} (h_i^A , h_i^B)^T}}{2 \pi \sqrt{
      \det \mathcal{C}_i}}\,.
\end{equation}
In this equation the covariance matrix $\mathcal{C}$ describes the
correlations between A's and B's Tree Parity Machines in terms of the
well-known order parameters $Q$ and $R$, which are functions of the
weight distribution according to (\ref{eq:qa}), (\ref{eq:qb}), and
(\ref{eq:rab}): 
\begin{equation}
  \mathcal{C}_k = \left(
    \begin{array}{cc}
      Q_i^A    & R_i^{AB} \\
      R_i^{AB} & Q_k^B                   
    \end{array}
  \right) \,.
\end{equation}
In order to generate local fields $h_i^A$ and $h_i^B$, which have the
correct joint probability distribution (\ref{eq:icalc}), the following
algorithm is used. A pseudo-random number generator produces two
independent uniformly distributed random numbers $z_1, z_2 \in [0,1[$.
Then the local fields are given by \cite{Knuth:1981:SA}
\begin{eqnarray}
  h_i^A &=& \sqrt{- 2 Q_i^A \ln(z_1)} \cos(2 \pi z_2) \,,\\
  h_i^B &=& \sqrt{- 2 Q_i^B \ln(z_1)} \left[ \rho \cos(2 \pi z_2) +
    \sqrt{1 - \rho^2} \sin(2 \pi z_2) \right] \,.
\end{eqnarray}
Afterwards one can calculate the outputs $\sigma_i$ and $\tau$ in the
same way as in the case of direct simulations. As the local fields are
known, it is possible to implement the geometric correction, too.
Therefore this method is also suitable to study synchronization by
unidirectional learning, e.~g.~for a geometric attacker. Additionally,
the algorithm can be extended to three and more interacting Tree
Parity Machines \cite{Ruttor:2004:NCF}.

\section{Equations of motion}

The equations of motion are generally independent of the learning
rule, because the behavior of Hebbian and anti-Hebbian learning
converges to that of the random walk learning rule in the limit $N
\rightarrow \infty$.  Consequently, the weights in both participating
Tree Parity Machines stay uniformly distributed, only the correlations
between $\mathbf{w}_i^A$ and $\mathbf{w}_i^B$ change.

\subsubsection{Attractive steps}

In an attractive step corresponding weights move in the same
direction. Thus the distribution of the weights changes according to
the following equations of motion for $-L < a,b < L$:
\begin{eqnarray}
  p^{i+}_{a,b}     &=& \frac{1}{2}
  \left(
    p^i_{a+1,b+1}   +
    p^i_{a-1,b-1}
  \right) \,, \\
  p^{i+}_{a,L}     &=& \frac{1}{2}
  \left(
    p^i_{a-1,L}     +
    p^i_{a-1,L-1}
  \right) \,, \\
  p^{i+}_{a,-L}    &=& \frac{1}{2}
  \left(
    p^i_{a+1,-L}    +
    p^i_{a+1,-L+1}
  \right) \,, \\
  p^{i+}_{L,b}     &=& \frac{1}{2}
  \left(
    p^i_{L,b-1}     +
    p^i_{L-1,b-1}
  \right) \,, \\
  p^{i+}_{-L,b}    &=& \frac{1}{2}
  \left(
    p^i_{-L,b+1}    +
    p^i_{-L+1,b+1}
  \right) \,, \\
  p^{i+}_{L,L}     &=& \frac{1}{2}
  \left(
    p^i_{L-1,L-1}   +
    p^i_{L-1,L}     +
    p^i_{L,L-1}     +
    p^i_{L,L}
  \right)\,, \\
  p^{i+}_{-L,-L}   &=& \frac{1}{2}
  \left(
    p^i_{-L+1,-L+1} +
    p^i_{-L+1,-L}   +
    p^i_{-L,-L+1}   +
    p^i_{-L,-L}
  \right)\,, \\
  p^{i+}_{L,-L}    &=& 0 \,, \\
  p^{i+}_{-L,L}    &=& 0 \,.
\end{eqnarray}

\subsubsection{Repulsive steps}

In a repulsive step only the weights in one hidden unit move, either
in A's or in B's Tree Parity Machine. However, the active hidden unit
is selected randomly by the output bits, so that both possibilities
occur with equal probability. Thus one can combine them in one set of
equations for $-L < a,b < L$:
\begin{eqnarray}
  p^{i+}_{a,b}     &=& \frac{1}{4}
  \left(
    p^i_{a+1,b}     +
    p^i_{a-1,b}     +
    p^i_{a,b+1}     +
    p^i_{a,b-1}
  \right) \,, \\
  p^{i+}_{a,L}     &=& \frac{1}{4}
  \left(
    p^i_{a+1,L}     +
    p^i_{a-1,L}     +
    p^i_{a,L}       +
    p^i_{a,L-1}
  \right) \,, \\
  p^{i+}_{a,-L}    &=& \frac{1}{4}
  \left(
    p^i_{a+1,-L}    +
    p^i_{a-1,-L}    +
    p^i_{a,-L+1}    +
    p^i_{a,-L}
  \right) \,, \\
  p^{i+}_{L,b}     &=& \frac{1}{4}
  \left(
    p^i_{L,b}       +
    p^i_{L-1,b}     +
    p^i_{L,b+1}     +
    p^i_{L,b-1}
  \right) \,, \\
  p^{i+}_{-L,b}    &=& \frac{1}{4}
  \left(
    p^i_{-L+1,b}    +
    p^i_{-L,b}      +
    p^i_{-L,b+1}    +
    p^i_{-L,b-1}
  \right) \,, \\
  p^{i+}_{L,L}     &=& \frac{1}{4}
  \left(
    2 p^i_{L,L}     +
    p^i_{L-1,L}     +
    p^i_{L,L-1}
  \right) \,, \\
  p^{i+}_{-L,-L}   &=& \frac{1}{4}
  \left(
    2 p^i_{-L,-L}   +
    p^i_{-L+1,-L}   +
    p^i_{-L,-L+1}
  \right) \,, \\
  p^{i+}_{L,-L}    &=& \frac{1}{4}
  \left(
    2 p^i_{L,-L}    +
    p^i_{L-1,-L}    +
    p^i_{L,-L+1}    
  \right) \,, \\
  p^{i+}_{-L,L}    &=& \frac{1}{4}
  \left(
    2 p^i_{-L,L}    +
    p^i_{-L+1,L}    +
    p^i_{-L,L-1}    
  \right) \,.
\end{eqnarray}

\subsubsection{Inverse attractive steps}

Inverse attractive steps are only possible if A and B do not interact
at all, but use common input vectors. In such a step corresponding
weights move in the opposite direction. Thus the distribution of the
weights changes according to the following equations of motion for $-L
< a,b < L$:

\begin{eqnarray}
  p^{i+}_{a,b}     &=& \frac{1}{2}
  \left(
    p^i_{a+1,b-1}   +
    p^i_{a-1,b+1}
  \right) \,, \\
  p^{i+}_{a,L}     &=& \frac{1}{2}
  \left(
    p^i_{a+1,L}     +
    p^i_{a+1,L-1}
  \right) \,, \\
  p^{i+}_{a,-L}    &=& \frac{1}{2}
  \left(
    p^i_{a-1,-L}    +
    p^i_{a-1,-L+1}
  \right) \,, \\
  p^{i+}_{L,b}     &=& \frac{1}{2}
  \left(
    p^i_{L,b+1}     +
    p^i_{L-1,b+1}
  \right) \,, \\
  p^{i+}_{-L,b}    &=& \frac{1}{2}
  \left(
    p^i_{-L,b-1}    +
    p^i_{-L+1,b-1}
  \right) \,, \\
  p^{i+}_{L,L}     &=& 0 \,, \\
  p^{i+}_{-L,-L}   &=& 0 \,, \\
  p^{i+}_{L,-L}    &=& \frac{1}{2}
  \left(
    p^i_{L-1,-L+1}  +
    p^i_{L-1,-L}    +
    p^i_{L,-L+1}    +
    p^i_{L,-L}
  \right)\,, \\
  p^{i+}_{-L,L}    &=& \frac{1}{2}
  \left(
    p^i_{-L+1,L-1}  +
    p^i_{-L+1,L}    +
    p^i_{-L,L-1}    +
    p^i_{-L,L}
  \right)\,.
\end{eqnarray}

\chapter{Generation of queries}
\label{chap:qgen}

This appendix describes the algorithm \cite{Ruttor:2005:NCQ} used to
generate a query $\mathbf{x}_i$, which results in a previously chosen
local field $h_i$. Finding an exact solution is similar to the
knapsack problem \cite{Schroeder:1986:NTS} and can be very difficult
depending on the parameters. Hence this is not useful for simulations,
as one has to generate a huge number of input vectors $\mathbf{x}_i$
in this case.  Instead a fast algorithm is employed, which gives an
approximate solution.

As both inputs $x_{i,j}$ and weights $w_{i,j}$ are discrete, there are
only $2 L + 1$ possible results for the product $w_{i,j} \, x_{i,j}$.
Therefore a set of input vectors consisting of all permutations, which
do not change $h_i$, can be described by counting the number $c_{i,l}$
of products with $w_{i,j} \, x_{i,j} = l$.  Then the local field is
given by
\begin{equation}
  h_i = \frac{1}{\sqrt{N}} \sum_{l=1}^{L} l (c_{i,l} - c_{i,-l}) \,,
\end{equation}
which depends on both inputs and weights. But the sum $n_{i,l} =
c_{i,l} + c_{i,-l}$ is equal to the number of weights with
$|w_{i,j}|=|l|$ and thus independent of $\mathbf{x}_i$. Consequently,
one can write $h_i$ as a function of only $L$ variables,
\begin{equation}
  h_i = \frac{1}{\sqrt{N}} \sum_{l=1}^{L} l (2 c_{i,l} - n_{i,l}) \,,
\end{equation}
as the values of $n_{i,l}$ are defined by the current weight vector
$\mathbf{w}_i$.

In the simulations the following algorithm \cite{Ruttor:2005:NCQ} is
used to generate the queries. First the output $\sigma_i$ of the
hidden unit is chosen randomly, so that the set value of the local
field is given by $h_i = \sigma_i H$. Then the values of $c_{i,L}$,
$c_{i,L-1}$, \dots, $c_{i,1}$ are calculated successively. For that
purpose one of the following equations is selected randomly with equal
probability, either
\begin{equation}
  \label{eq:qgenf}
  c_{i,l} = \left\lfloor \frac{n_{i,l} + 1}{2} + \frac{1}{2 l} \left(
      \sigma_i H \sqrt{N} - \sum_{j=l+1}^{L} j (2 c_{i,j} - n_{i,j})
    \right) \right\rfloor
\end{equation}
or
\begin{equation}
  \label{eq:qgenc}
  c_{i,l} = \left\lceil \frac{n_{i,l} - 1}{2} + \frac{1}{2 l} \left(
      \sigma_i H \sqrt{N} - \sum_{j=l+1}^{L} j (2 c_{i,j} - n_{i,j})
    \right) \right\rceil \,,
\end{equation}
in order to reduce the influence of rounding errors. Additionally, one
has to take the condition $0 \leq c_{i,l} \leq n_{i,l}$ into account.
If equation (\ref{eq:qgenf}) or equation (\ref{eq:qgenc}) yield a
result outside this range, $c_{i,l}$ is reset to the nearest boundary
value.

Afterwards the input vector $\mathbf{w}_i$ is generated. Those
$x_{i,j}$ associated with zero weights $w_{i,j}=0$ do not influence
the local field, so that their value is just chosen randomly. But the
other input bits $x_{i,j}$ are divided into $L$ groups according to
the absolute value $l = |w_{i,j}|$ of their corresponding weight. Then
$c_{i,l}$ input bits are selected randomly in each group and set to
$x_{i,j} = \mathrm{sgn}(w_{i,j})$, while the other $n_{k,l} - c_{k,l}$
inputs are set to $x_{i,j} = -\mathrm{sgn}(w_{i,j})$.

Simulations show that queries generated by this algorithm result in
local fields $h_i$ which match the set value $\sigma_i H$ on average
\cite{Ruttor:2005:NCQ}. Additionally, only very small deviations are
observed, which are caused by the restriction of inputs and weights to
discrete values. So this algorithm is indeed suitable for the
generation of queries.

\end{appendix}
\addcontentsline{toc}{chapter}{Bibliography}
\nocite{Bronstein:1999:TM}
\nocite{Hartmann:2001:PGC}
\bibliography{dissertation}

\begin{thebibliography}{59}
\providecommand{\natexlab}[1]{#1}
\providecommand{\url}[1]{\texttt{#1}}
\expandafter\ifx\csname urlstyle\endcsname\relax
  \providecommand{\doi}[1]{doi: #1}\else
  \providecommand{\doi}{doi: \begingroup \urlstyle{rm}\Url}\fi

\bibitem[Pikovsky et~al.(2001)Pikovsky, Rosenblum, and Kurths]{Pikovsky:2001:S}
A.~Pikovsky, M.~Rosenblum, and J.~Kurths.
\newblock \emph{Synchronization}.
\newblock Cambridge University Press, Cambridge, 2001.

\bibitem[Kim et~al.(2002)Kim, Rim, and Kye]{Kim:2002:SSC}
C.-M. Kim, S.~Rim, and W.-H. Kye.
\newblock Sequential synchronization of chaotic systems with an application to
  communication.
\newblock \emph{Phys. Rev. Lett.}, 88\penalty0 (1):\penalty0 014103, 2002.

\bibitem[Cuomo and Oppenheim(1993)]{Cuomo:1993:CIS}
K.~M. Cuomo and A.~V. Oppenheim.
\newblock Circuit implementation of synchronized chaos with applications to
  communications.
\newblock \emph{Phys. Rev. Lett.}, 71\penalty0 (1):\penalty0 65--68, 1993.

\bibitem[Pecora and Carroll(1990)]{Pecora:1990:SCS}
L.~M. Pecora and T.~L. Carroll.
\newblock Synchronization in chaotic systems.
\newblock \emph{Phys. Rev. Lett.}, 64\penalty0 (8):\penalty0 821--824, 1990.

\bibitem[Argyris et~al.(2005)Argyris, Syvridis, Larger, Annovazzi-Lodi, Colet,
  Fischer, Garc{\'i}a-Ojalvo, Mirasso, Pesquera, and Shore]{Argyris:2005:CBC}
A.~Argyris, D.~Syvridis, L.~Larger, V.~Annovazzi-Lodi, P.~Colet, I.~Fischer,
  J.~Garc{\'i}a-Ojalvo, C.~R. Mirasso, L.~Pesquera, and K.~A. Shore.
\newblock Chaos-based communications at high bit rates using commercial
  fibre-optic links.
\newblock \emph{Nature}, 437\penalty0 (7066):\penalty0 343--346, 2005.

\bibitem[Metzler et~al.(2000)Metzler, Kinzel, and Kanter]{Metzler:2000:INN}
R.~Metzler, W.~Kinzel, and I.~Kanter.
\newblock Interacting neural networks.
\newblock \emph{Phys. Rev. E}, 62\penalty0 (2):\penalty0 2555--2565, 2000.

\bibitem[Kinzel et~al.(2000)Kinzel, Metzler, and Kanter]{Kinzel:2000:DIN}
W.~Kinzel, R.~Metzler, and I.~Kanter.
\newblock Dynamics of interacting neural networks.
\newblock \emph{J. Phys. A: Math. Gen.}, 33\penalty0 (14):\penalty0 L141--L147,
  2000.

\bibitem[Hertz et~al.(1991)Hertz, Krogh, and Palmer]{Hertz:1991:ITN}
J.~Hertz, A.~Krogh, and R.~G. Palmer.
\newblock \emph{Introduction to the Theory of Neural Computation}.
\newblock Addison-Wesley, Redwood City, 1991.

\bibitem[Kinzel and Kanter(2003)]{Kinzel:2003:DGI}
W.~Kinzel and I.~Kanter.
\newblock Disorder generated by interacting neural networks: application to
  econophysics and cryptography.
\newblock \emph{J. Phys. A: Math. Gen.}, 36\penalty0 (43):\penalty0
  11173--11186, 2003.

\bibitem[Kinzel and Kanter(2002{\natexlab{a}})]{Kinzel:2002:INN}
W.~Kinzel and I.~Kanter.
\newblock Interacting neural networks and cryptography.
\newblock In B.~Kramer, editor, \emph{Advances in Solid State Physics},
  volume~42, pages 383--391. Springer, Berlin, 2002{\natexlab{a}}.

\bibitem[Kinzel(2002)]{Kinzel:2002:TIN}
W.~Kinzel.
\newblock Theory of interacting neural networks.
\newblock cond-mat/0204054, 2002.

\bibitem[Kinzel and Kanter(2002{\natexlab{b}})]{Kinzel:2002:NC}
W.~Kinzel and I.~Kanter.
\newblock Neural cryptography.
\newblock cond-mat/0208453, 2002{\natexlab{b}}.

\bibitem[Kanter et~al.(2002)Kanter, Kinzel, and Kanter]{Kanter:2002:SEI}
I.~Kanter, W.~Kinzel, and E.~Kanter.
\newblock Secure exchange of information by synchronization of neural networks.
\newblock \emph{Europhys. Lett.}, 57\penalty0 (1):\penalty0 141--147, 2002.

\bibitem[Kanter and Kinzel(2003)]{Kanter:2003:TNN}
I.~Kanter and W.~Kinzel.
\newblock The theory of neural networks and cryptography.
\newblock In I.~Antoniou, V.~A. Sadovnichy, and H.~Wather, editors,
  \emph{Proceedings of the XXII Solvay Conference on Physics on the Physics of
  Communication}, page 631. World Scientific, Singapore, 2003.

\bibitem[Klein et~al.(2005{\natexlab{a}})Klein, Mislovaty, Kanter, Ruttor, and
  Kinzel]{Klein:2005:SNN}
E.~Klein, R.~Mislovaty, I.~Kanter, A.~Ruttor, and W.~Kinzel.
\newblock Synchronization of neural networks by mutual learning and its
  application to cryptography.
\newblock In L.~K. Saul, Y.~Weiss, and L.~Bottou, editors, \emph{Advances in
  Neural Information Processing Systems}, volume~17, pages 689--696. MIT Press,
  Cambridge, MA, 2005{\natexlab{a}}.

\bibitem[Mislovaty et~al.(2002)Mislovaty, Perchenok, Kanter, and
  Kinzel]{Mislovaty:2002:SKE}
R.~Mislovaty, Y.~Perchenok, I.~Kanter, and W.~Kinzel.
\newblock Secure key-exchange protocol with an absence of injective functions.
\newblock \emph{Phys. Rev. E}, 66:\penalty0 066102, 2002.

\bibitem[Rosen-Zvi et~al.(2002{\natexlab{a}})Rosen-Zvi, Kanter, and
  Kinzel]{Rosen-Zvi:2002:CBN}
M.~Rosen-Zvi, I.~Kanter, and W.~Kinzel.
\newblock Cryptography based on neural networks---analytical results.
\newblock \emph{J. Phys. A: Math. Gen.}, 35:\penalty0 L707--L713,
  2002{\natexlab{a}}.

\bibitem[Rosen-Zvi et~al.(2002{\natexlab{b}})Rosen-Zvi, Klein, Kanter, and
  Kinzel]{Rosen-Zvi:2002:MLT}
M.~Rosen-Zvi, E.~Klein, I.~Kanter, and W.~Kinzel.
\newblock Mutual learning in a tree parity machine and its application to
  cryptography.
\newblock \emph{Phys. Rev. E}, 66:\penalty0 066135, 2002{\natexlab{b}}.

\bibitem[Ruttor et~al.(2004{\natexlab{a}})Ruttor, Kinzel, Shacham, and
  Kanter]{Ruttor:2004:NCF}
A.~Ruttor, W.~Kinzel, L.~Shacham, and I.~Kanter.
\newblock Neural cryptography with feedback.
\newblock \emph{Phys. Rev. E}, 69:\penalty0 046110, 2004{\natexlab{a}}.

\bibitem[Klimov et~al.(2003)Klimov, Mityaguine, and Shamir]{Klimov:2003:ANC}
A.~Klimov, A.~Mityaguine, and A.~Shamir.
\newblock Analysis of neural cryptography.
\newblock In Y.~Zheng, editor, \emph{Advances in Cryptology---ASIACRYPT 2002},
  page 288. Springer, Heidelberg, 2003.

\bibitem[Shacham et~al.(2004)Shacham, Klein, Mislovaty, Kanter, and
  Kinzel]{Shacham:2004:CAN}
L.~N. Shacham, E.~Klein, R.~Mislovaty, I.~Kanter, and W.~Kinzel.
\newblock Cooperating attackers in neural cryptography.
\newblock \emph{Phys. Rev. E}, 69\penalty0 (6):\penalty0 066137, 2004.

\bibitem[Ruttor et~al.(2006)Ruttor, Kinzel, Naeh, and Kanter]{Ruttor:2006:GAN}
A.~Ruttor, W.~Kinzel, R.~Naeh, and I.~Kanter.
\newblock Genetic attack on neural cryptography.
\newblock \emph{Phys. Rev. E}, 73\penalty0 (3):\penalty0 036121, 2006.

\bibitem[Ruttor et~al.(2005)Ruttor, Kinzel, and Kanter]{Ruttor:2005:NCQ}
A.~Ruttor, W.~Kinzel, and I.~Kanter.
\newblock Neural cryptography with queries.
\newblock \emph{J. Stat. Mech.}, 2005\penalty0 (01):\penalty0 P01009, 2005.

\bibitem[Mislovaty et~al.(2003)Mislovaty, Klein, Kanter, and
  Kinzel]{Mislovaty:2003:PCC}
R.~Mislovaty, E.~Klein, I.~Kanter, and W.~Kinzel.
\newblock Public channel cryptography by synchronization of neural networks and
  chaotic maps.
\newblock \emph{Phys. Rev. Lett.}, 91\penalty0 (11):\penalty0 118701, 2003.

\bibitem[Stinson(1995)]{Stinson:1995:CTP}
D.~R. Stinson.
\newblock \emph{Cryptography: Theory and Practice}.
\newblock CRC Press, Boca Raton, FL, 1995.

\bibitem[Beutelspacher(2002)]{Beutelspacher:2002:K}
A.~Beutelspacher.
\newblock \emph{Kryptologie}.
\newblock Vieweg \& Sohn Verlagsgesellschaft mbH, Braunschweig/Wiesbaden, 2002.

\bibitem[Ruttor et~al.(2004{\natexlab{b}})Ruttor, Reents, and
  Kinzel]{Ruttor:2004:SRW}
A.~Ruttor, G.~Reents, and W.~Kinzel.
\newblock Synchronization of random walks with reflecting boundaries.
\newblock \emph{J. Phys. A: Math. Gen.}, 37:\penalty0 8609--8618,
  2004{\natexlab{b}}.

\bibitem[Engel and {Van den Broeck}(2001)]{Engel:2001:SML}
A.~Engel and C.~{Van den Broeck}.
\newblock \emph{Statistical Mechanics of Learning}.
\newblock Cambridge University Press, Cambridge, 2001.

\bibitem[Cover and Thomas(1991)]{Cover:1991:EIT}
T.~M. Cover and J.~A. Thomas.
\newblock \emph{Elements of Information Theory}.
\newblock John Wiley \& Sons, New York, 1991.

\bibitem[Ein-Dor and Kanter(1999)]{Ein-Dor:1999:CPN}
L.~Ein-Dor and I.~Kanter.
\newblock Confidence in prediction by neural networks.
\newblock \emph{Phys. Rev. E}, 60\penalty0 (1):\penalty0 799--802, 1999.

\bibitem[Ruttor et~al.(2007)Ruttor, Kanter, and Kinzel]{Ruttor:2007:DNC}
A.~Ruttor, I.~Kanter, and W.~Kinzel.
\newblock Dynamics of neural cryptography.
\newblock \emph{Phys. Rev. E}, 75\penalty0 (5):\penalty0 056104, 2007.

\bibitem[Feller(1968)]{Feller:1968:IPT}
W.~Feller.
\newblock \emph{An Introduction to Probability Theory and Its Applications},
  volume~1.
\newblock John Wiley \& Sons, New York, 3rd edition, 1968.

\bibitem[Galambos(1940)]{Galambos:1940:ATE}
J.~Galambos.
\newblock \emph{The Asymptotic Theory of Extreme Order Statistics}.
\newblock John Wiley \& Sons, New York, 1940.

\bibitem[Reents and Urbanczik(1998)]{Reents:1998:SAL}
G.~Reents and R.~Urbanczik.
\newblock Self-averaging and on-line learning.
\newblock \emph{Phys. Rev. Lett.}, 80\penalty0 (24):\penalty0 5445--5448, 1998.

\bibitem[Urbanczik(2000)]{Urbanczik:2000:OLE}
R.~Urbanczik.
\newblock Online learning with ensembles.
\newblock \emph{Phys. Rev. E}, 62\penalty0 (1):\penalty0 1448--1451, 2000.

\bibitem[Watkin(1993)]{Watkin:1993:OLN}
T.~L.~H. Watkin.
\newblock Optimal learning with a neural network.
\newblock \emph{Europhys. Lett.}, 21\penalty0 (8):\penalty0 871--876, 1993.

\bibitem[Kang and Oh(1995)]{Kang:1995:LPP}
K.~Kang and J.-H. Oh.
\newblock Learning by a population of perceptrons.
\newblock In J.-H. Oh, C.~Kwon, and S.~Cho, editors, \emph{Neural Networks: The
  Statistical Mechanics Perspective}, volume~1 of \emph{Progress in Neural
  Processing}, pages 94--101. World Scientific, Singapore, 1995.

\bibitem[Eisenstein et~al.(1995)Eisenstein, Kanter, Kessler, and
  Kinzel]{Eisenstein:1995:GPT}
E.~Eisenstein, I.~Kanter, D.~Kessler, and W.~Kinzel.
\newblock Generation and prediction of time series by a neural network.
\newblock \emph{Phys. Rev. E}, 74\penalty0 (1):\penalty0 6--9, 1995.

\bibitem[Metzler et~al.(2001)Metzler, Kinzel, Ein-Dor, and
  Kanter]{Metzler:2001:GUT}
R.~Metzler, W.~Kinzel, L.~Ein-Dor, and I.~Kanter.
\newblock Generation of unpredictable time series by a neural network.
\newblock \emph{Phys. Rev. E}, 63:\penalty0 056126, 2001.

\bibitem[Bialek et~al.(2001)Bialek, Nemenman, and Tishby]{Bialek:2001:PCL}
W.~Bialek, I.~Nemenman, and N.~Tishby.
\newblock Predictability, complexity and learning.
\newblock \emph{Neural Computation}, 13\penalty0 (11):\penalty0 2409--2463,
  2001.

\bibitem[Zhu and Kinzel(1998)]{Zhu:1998:APS}
H.~Zhu and W.~Kinzel.
\newblock Anti-predictable sequences: Harder to predict than a random sequence.
\newblock \emph{Neural Comput.}, 10\penalty0 (8):\penalty0 2219--2230, 1998.

\bibitem[Volkmer and Schaumburg(2004)]{Volkmer:2004:ATP}
M.~Volkmer and A.~Schaumburg.
\newblock Authenticated tree parity machine key exchange.
\newblock cs/0408046, 2004.

\bibitem[Volkmer(2006)]{Volkmer:2006:EAA}
M.~Volkmer.
\newblock Entity authentication and authenticated key exchange with tree parity
  machines.
\newblock Cryptology ePrint Archive, Report 2006/112, 2006.

\bibitem[Kinzel and Rujan(1990)]{Kinzel:1990:ING}
W.~Kinzel and P.~Rujan.
\newblock Improving a network generalization ability by selecting examples.
\newblock \emph{Europhys. Lett.}, 13\penalty0 (5):\penalty0 473--477, 1990.

\bibitem[Maurer(1993)]{Maurer:1993:SKA}
U.~Maurer.
\newblock Secret key agreement by public discussion.
\newblock \emph{IEEE Trans. Inf. Theory}, 39\penalty0 (3):\penalty0 733--742,
  1993.

\bibitem[Gross et~al.(2005)Gross, Klein, Rosenbluh, Kinzel, Khaykovich, and
  Kanter]{Gross:2005:FPC}
N.~Gross, E.~Klein, M.~Rosenbluh, W.~Kinzel, L.~Khaykovich, and I.~Kanter.
\newblock A framework for public-channel cryptography using chaotic lasers.
\newblock cond-mat/0507554, 2005.

\bibitem[Klein et~al.(2005{\natexlab{b}})Klein, Mislovaty, Kanter, and
  Kinzel]{Klein:2005:PCC}
E.~Klein, R.~Mislovaty, I.~Kanter, and W.~Kinzel.
\newblock Public-channel cryptography using chaos synchronization.
\newblock \emph{Phys. Rev. E}, 72:\penalty0 016214, 2005{\natexlab{b}}.

\bibitem[Klein et~al.(2006{\natexlab{a}})Klein, Gross, Kopelowitz, Rosenbluh,
  Khaykovich, Kinzel, and Kanter]{Klein:2006:PCC}
E.~Klein, N.~Gross, E.~Kopelowitz, M.~Rosenbluh, L.~Khaykovich, W.~Kinzel, and
  I.~Kanter.
\newblock Public-channel cryptography based on mutual chaos pass filters.
\newblock \emph{Phys. Rev. E}, 74\penalty0 (4):\penalty0 046201,
  2006{\natexlab{a}}.

\bibitem[Klein et~al.(2006{\natexlab{b}})Klein, Gross, Rosenbluh, Kinzel,
  Khaykovich, and Kanter]{Klein:2006:SIS}
E.~Klein, N.~Gross, M.~Rosenbluh, W.~Kinzel, L.~Khaykovich, and I.~Kanter.
\newblock Stable isochronal synchronization of mutually coupled chaotic lasers.
\newblock \emph{Phys. Rev. E}, 73\penalty0 (6):\penalty0 066214,
  2006{\natexlab{b}}.

\bibitem[Volkmer and Wallner(2004)]{Volkmer:2004:LCS}
M.~Volkmer and S.~Wallner.
\newblock A low-cost solution for frequent symmetric key exchange in ad-hoc
  networks.
\newblock In P.~Dadam and M.~Reichert, editors, \emph{Proceedings of the 2nd
  German Workshop on Mobile Ad-hoc Networks, WMAN 2004}, volume P-50 of
  \emph{Lecture Notes in Informatics (LNI)}, pages 128--137, Ulm, 2004. Bonner
  K{\"o}llen Verlag.

\bibitem[Volkmer and Wallner(2005{\natexlab{a}})]{Volkmer:2005:TPMa}
M.~Volkmer and S.~Wallner.
\newblock Tree parity machine rekeying architectures.
\newblock \emph{IEEE Trans. Comput.}, 54\penalty0 (4):\penalty0 421--427,
  2005{\natexlab{a}}.

\bibitem[Volkmer and Wallner(2005{\natexlab{b}})]{Volkmer:2005:TPMb}
M.~Volkmer and S.~Wallner.
\newblock Tree parity machine rekeying architectures for embedded security.
\newblock Cryptology ePrint Archive, Report 2005/235, 2005{\natexlab{b}}.

\bibitem[Volkmer and Wallner(2005{\natexlab{c}})]{Volkmer:2005:KEI}
M.~Volkmer and S.~Wallner.
\newblock A key establishment ip-core for ubiquitous computing.
\newblock In \emph{Proceedings of the 1st International Workshop on Secure and
  Ubiquitous Networks, SUN'05}, pages 241--245, Copenhagen, 2005{\natexlab{c}}.
  IEEE Computer Society.

\bibitem[Volkmer and Wallner(2005{\natexlab{d}})]{Volkmer:2005:LKE}
M.~Volkmer and S.~Wallner.
\newblock Lightweight key exchange and stream cipher based solely on tree
  parity machines.
\newblock In \emph{ECRYPT (European Network of Excellence for Cryptology)
  Workshop on RFID and Lightweight Crypto}, pages 102--113, Graz,
  2005{\natexlab{d}}. Graz University of Technology.

\bibitem[Volkmer and Wallner(2006)]{Volkmer:2006:ICD}
M.~Volkmer and S.~Wallner.
\newblock Ein {IP}-{C}ore {D}esign f{\"u}r {S}chl{\"u}ssel{\-}austausch,
  {S}tromchiffre und {I}dentifikation auf ressourcenbeschr{\"a}nkten
  {G}er{\"a}ten.
\newblock In J.~Dittmann, editor, \emph{Workshop \dq{}Kryptographie in Theorie
  und Praxis\dq}, volume P-770 of \emph{Lecture Notes in Informatics (LNI)},
  pages 294--298, Magdeburg, 2006. Bonner K{\"o}llen Verlag.

\bibitem[Knuth(1981)]{Knuth:1981:SA}
D.~E. Knuth.
\newblock \emph{Seminumerical Algorithms}, volume~2 of \emph{The Art of
  Computer Programming}.
\newblock Addison-Wesley, Redwood City, second edition, 1981.

\bibitem[Schroeder(1986)]{Schroeder:1986:NTS}
M.~R. Schroeder.
\newblock \emph{Number Theory in Science and Communication}.
\newblock Springer, Berlin, second edition, 1986.

\bibitem[Bronstein et~al.(1999)Bronstein, Semendjajew, Musiol, and
  M{\"u}hlig]{Bronstein:1999:TM}
I.~N. Bronstein, K.~A. Semendjajew, G.~Musiol, and H.~M{\"u}hlig.
\newblock \emph{Taschenbuch der Mathematik}.
\newblock Verlag Harri Deutsch, Frankfurt am Main, 1999.

\bibitem[Hartmann and Rieger(2001)]{Hartmann:2001:PGC}
A.~K. Hartmann and H.~Rieger.
\newblock A practical guide to computer simulations.
\newblock cond-mat/0111531, 2001.

\end{thebibliography}
\bibliographystyle{unsrtnat}
\chapter*{Acknowledgment}

A lot of people have contributed to the success of this thesis in
different ways. Here I wish to express my gratitude to them:

\begin{itemize}
\item Prof. Dr. Wolfgang Kinzel for the excellent supervision. His
  proposals and tips regarding interesting questions have much
  influenced the direction of this thesis.
\item Prof. Dr. Haye Hinrichsen and Prof. Dr. Georg Reents for helpful
  advice on various problems appearing from time to time.
\item Prof. Ido Kanter and his work group for the interesting
  discussions, a lot of suggestions, and the fruitful teamwork leading
  to results, which we have published together.
\item Florian Grewe, Markus Volkmer, and Sebastian Wallner for their
  ideas concerning the realization of the neural key-exchange protocol
  in practise.
\item Markus Walther and Sebastian Weber for the diligent and
  attentive proof-reading of this thesis.
\item the system administrators Andreas Klein and Andreas Vetter for
  maintaining the computer system very well.
\item the Leibniz computing center in Munich for providing computing
  time on its high-performance Linux Cluster. Most simulations for
  this thesis have been done there.
\item the secretaries Bettina Spiegel, Brigitte Wehner and Nelia
  Meyer for their help with bureaucratic problems.
\item all members of the chair for Computational Physics for the 
  possibility to work in a constructive and relatively relaxed manner
\item the Deutsche Forschungsgemeinschaft for funding this thesis as
  part of the project Neural Cryptography.
\item Last but not least, I wish to thank my family for the
  encouragement and financial support during my studies.
\end{itemize}

\clearpage

\thispagestyle{empty}

\cleardoublepage

\end{document}